\documentclass[usegraphicx,usenatbib]{mn2e}
\usepackage{verbatim}
\usepackage{color}
\usepackage{aas_macros}

\usepackage[T1]{fontenc}
\usepackage{ae,aecompl}

\usepackage{hyperref}
\usepackage{amsmath} 
\usepackage{breqn}  

\usepackage{pdflscape} 

\newcommand{\beq}{\begin{equation}}
\newcommand{\eeq}{\end{equation}}

\newcommand{\hmpc}{\,$h^{-1}$Mpc }
\newcommand{\hmpcii}{\,$h^{-1}$Mpc}

\newcommand{\czHzrs}{\ensuremath{cz/H/r_{\mathrm s}} }
\newcommand{\czHzrsii}{\ensuremath{cz/H/r_{\mathrm s}}}
\newcommand{\Hzrs}{\,\ensuremath{H \cdot r_{\mathrm s}} }

\newcommand{\HrsfidHrs}{\,\ensuremath{(H r_{\rm s})^{\rm fid}/(H r_{\rm s})} }
\newcommand{\HrsfidHrsii}{\,\ensuremath{(H r_{\rm s})^{\rm fid}/(H r_{\rm s})}}
\newcommand{\HzfidHz}{\,\ensuremath{H^{\rm fid}/H} }
\newcommand{\HzfidHzii}{\,\ensuremath{H^{\rm fid}/H}}

\newcommand{\Dazrs}{\ensuremath{D_{\rm A}/r_{\mathrm s}} }
\newcommand{\Dazrsii}{\ensuremath{D_{\rm A}/r_{\mathrm s}}}
\newcommand{\DazDazfid}{\,\ensuremath{D_{\rm A}/D_{\rm A}^{\rm fid}} }
\newcommand{\DazDazfidii}{\,\ensuremath{D_{\rm A}/D_{\rm A}^{\rm fid}}}
\newcommand{\DrsDrsfid}{\,\ensuremath{(D_{\rm A}/r_{\rm s})/(D_{\rm A}/r_{\rm s})^{\rm fid}} }
\newcommand{\DrsDrsfidii}{\,\ensuremath{(D_{\rm A}/r_{\rm s})/(D_{\rm A}/r_{\rm s})^{\rm fid}}}

\newcommand{\baf}{\,baryonic acoustic feature }
\newcommand{\bafii}{\,baryonic acoustic feature}

\newcommand{\czHrscmass}{\ensuremath{12.28}}  
\newcommand{\czHrsunc}{\ensuremath{0.82} }  
  
\newcommand{\czHrsperc}{\ensuremath{6.7\%} }

\newcommand{\Hzcmasswmapii}{\ensuremath{90.8}}
\newcommand{\Hzcmasswmapunc}{\ensuremath{6.2} }

\newcommand{\Darscmass}{\,\ensuremath{9.05}} 
\newcommand{\Darsunc}{\,\ensuremath{0.27} }  
\newcommand{\Darsperc}{\ensuremath{3.0\%} } 
\newcommand{\Darspercii}{\ensuremath{3.0\%}}

\newcommand{\Dacmasswmapii}{\ensuremath{1386}}
\newcommand{\Dacmasswmapunc}{\ensuremath{45} }

\newcommand{\crosscorrczHrsDars}{\,\ensuremath{-0.5} } 
\newcommand{\crosscorrczHrsDarsii}{\,\ensuremath{-0.5}}

\newcommand{\avg}[1]{\ensuremath{\langle{#1}\rangle}}

\begin{document}


\title[BOSS DR9 $H(z)$ and $D_{\rm A}(z)$ from Clustering Wedges]
{The Clustering of Galaxies in the SDSS-III Baryon Oscillation Spectroscopic Survey: 
Measuring $H(z)$ and $D_{\rm A}(z)$ at $z=0.57$ with Clustering Wedges}

\author[Kazin E., S\'anchez A. et al.]
{\parbox[t]{\textwidth}{
Eyal A. Kazin$^{1,2}$\thanks{E-mail: eyalkazin@gmail.com}, 
Ariel G. S\'anchez$^{3}$, 
Antonio J. Cuesta$^{4}$,   
Florian Beutler$^{5}$,  
Chia-Hsun Chuang$^{6}$, 
Daniel J. Eisenstein$^{7}$, 
Marc Manera$^{8}$, 
Nikhil Padmanabhan$^{4}$, 
Will J. Percival$^{8}$, 
Francisco Prada$^{6,9,10}$, 
Ashley J. Ross$^{8}$,  
Hee-Jong Seo$^{5}$, 
Jeremy Tinker$^{11}$, 
Rita Tojeiro$^{8}$, 
Xiaoying Xu$^{12}$, 
J. Brinkmann$^{13}$,
Brownstein Joel$^{14}$,
Robert C. Nichol$^{8}$, 
David J Schlegel$^{5}$,
Donald P. Schneider$^{15,16}$ and 
Daniel Thomas$^{8}$}
\vspace*{6pt} \\ 
$^{1}$ Centre for Astrophysics \& Supercomputing, Swinburne University of Technology, PO Box 218, Hawthorn, VIC 3122, Australia.\\
$^{2}$ ARC Centre of Excellence for All-sky Astrophysics (CAASTRO). \\
$^{3}$ Max-Planck-Institut f\"ur Extraterrestrische Physik, Giessenbachstra\ss e, 85748 Garching, Germany.\\
$^{4}$Department of Physics, Yale University, 260 Whitney Ave, New Haven, CT 06520, USA.\\
$^{5}$Lawrence Berkeley National Lab, 1 Cyclotron Rd, Berkeley CA 94720, USA.\\
$^{6}$Instituto de F\'{\i}sica Te\'orica, (UAM/CSIC), Universidad Aut\'onoma de Madrid,  Cantoblanco, E-28049 Madrid, Spain. \\
$^{7}$Harvard-Smithsonian Center for Astrophysics, 60 Garden St., Cambridge, MA 02138, USA.\\
$^{8}$Institute of Cosmology \& Gravitation, University of Portsmouth, Dennis Sciama Building, Portsmouth PO1 3FX, UK.\\
$^{9}$Instituto de Astrof\'{\i}sica de Andaluc\'{\i}a (CSIC), Glorieta de la Astronom\'{\i}a, E-18080 Granada, Spain. \\
$^{10}$Campus of International Excellence UAM+CSIC, Cantoblanco, E-28049 Madrid, Spain. \\
$^{11}$Center for Cosmology and Particle Physics, New York University, New York, NY 10003, USA. \\
$^{12}$Department of Physics, Carnegie Mellon University, 5000 Forbes Ave., Pittsburgh, PA 15213, USA.\\
$^{13}$Apache Point Observatory.\\
$^{14}$Department of Physics and Astronomy, University of Utah, 115 S 1400 E, Salt Lake City, UT 84112, USA. \\
$^{15}$ Department of Astronomy and Astrophysics, The Pennsylvania State University, University Park, PA 16802, USA. \\
$^{16}$Institute for Gravitation and the Cosmos, The Pennsylvania State University,  University Park, PA 16802, USA.}
\date{Submitted to MNRAS}
\maketitle

\begin{abstract}
We analyze the 2D correlation function of the 
SDSS-III 
Baryon Oscillation Spectroscopic Survey (BOSS) 
CMASS  sample of massive galaxies of the ninth data release to 
measure cosmic expansion $H$ and the angular diameter distance  $D_{\rm A}$ 
at a mean redshift of $\avg{z}=0.57$.  
We apply, for the first time, a new correlation function technique called 
{\it clustering wedges}  $\xi_{\Delta \mu}(s)$. 
Using a physically motivated model, the anisotropic \baf in the galaxy sample 
is detected at a significance level of $4.7\sigma$ 
compared to a featureless model.
The \baf is used to obtain  
model independent constraints 
\czHzrsii$\ = \ $\czHrscmass$\ \pm \ $\czHrsunc (\czHrsperc accuracy)
and 
\Dazrsii$\ = \ $\Darscmass$\ \pm \ $\Darsunc (\Darspercii)
with a correlation coefficient of \crosscorrczHrsDarsii,  
where $r_{\rm s}$ is the sound horizon scale at the end of the baryonic drag era. 
We conduct thorough tests on the data and 600 simulated realizations, 
finding robustness of the results regardless 
of the details of the analysis method. 
Combining with $r_{\rm s}$ constraints from the Cosmic Microwave Background 
we obtain 
$H(0.57)=\ $\Hzcmasswmapii$\ \pm \ $\Hzcmasswmapunc kms$^{-1}$Mpc$^{-1}$ 
and  $D_{\rm A}(0.57)=\ $\Dacmasswmapii$\ \pm \ $\Dacmasswmapunc Mpc. 
We use simulations to forecast results of the 
final BOSS CMASS data set.  
We apply the reconstruction technique on the simulations 
demonstrating that the sharpening of the anisotropic \baf 
should improve the detection 
as well as tighten constraints of 
$H$ and $D_{\rm A}$ by $\sim 30\%$ on average.   
\end{abstract}

\begin{keywords}
cosmological parameters, large scale structure of the universe, distance scale
\end{keywords}
\section{Introduction}\label{intro_section}
One of the most exciting recent observations 
is the 
acceleration 
of the expansion of the Universe 
since redshift $z=1$ (\citealt{riess98,perlmutter99a}).  
The origin of this phenomenon 
is thought to be 
an energy component with negative pressure 
(e.g, the so-called dark energy or a cosmological constant), 
or otherwise a break-down of General Relativity (\citealt{einstein16a}) on cosmic scales.
For an in-depth summary of the observed acceleration and its possible interpretations, 
see \cite{weinberg12a}.

One method of measuring geometry 
from a 3D map of cosmological objects is through a geometric  
technique called the Alcock-Paczynski test (\citealt{alcock79}) along with a standard ruler 
known as the baryonic acoustic feature.
\cite{alcock79} demonstrated that 
by assuming an incorrect cosmology when converting 
observed redshifts  $z_{\rm obs}$ to comoving distances $\chi$, 
a spherical cosmological body 
will appear deformed  
due to geometrical distortions.  
In the context of galaxy maps, 
this would cause a coherent distortion of the apparent 
radial positions and an anisotropic signature in clustering probes. 
Reproducing an isotropic clustering signal would 
result in obtaining the true cosmology. 
The observable of this process is $HD_{\rm A}$, 
where $H$ is the expansion factor and $D_{\rm A}$ is the angular diameter distance. 
To break this 
degeneracy, 
and hence improve constraining power, 
one can apply the technique on a standard ruler, 
such as the \bafii.  

Early Universe plasma-photon waves propagated 
at close to the speed of sound 
from over dense regions, 
and came to a near halt at 
the era of decoupling of photons 
from baryons at $z_{*}\sim 1100$ 
at a characteristic comoving distance of $r_{\rm s}\sim 150$ Mpc 
from the originating over-density. 
This process left a 
distinctive signature in CMB anisotropies 
and in the large-scale structure of galaxies (\citealt{peebles70a}). 
\cite{hu97a} review how the CMB 
anisotropies can be used to constrain 
fundamental cosmological parameters.  
\cite{bassett10a} review 
the baryonic acoustic signature in the  
clustering of matter 
and its usage as a standard ruler.

Following first \baf measurements in 
the clustering of galaxies 
by \cite{eisenstein05b} and \cite{cole05a},  
two recent surveys, the WiggleZ Dark Energy Sky Survey (\citealt{drinkwater10a}), 
and the Sloan Digital Sky Survey (SDSS-III; \citealt{york00a,eisenstein11a}) 
Baryonic Oscillation Spectroscopic Survey (BOSS; \citealt{dawson13a}), 
have reported detections  
of the \baf  at $z>0.5$ (\citealt{blake11b,blake11c,anderson12a, sanchez12a}), 
as well as the 6dF Galaxy Survey (\citealt{jones09a}) at $z<0.2$ (\citealt{beutler11a}). 
\cite{busca12a} and \cite{slosar13a} also detect the \bafii, 
for the first time, 
in the Lyman-alpha forest of BOSS quasars between $2<z<3$.  

The focus of 
of most of these studies has been on the angle-averaged 
signal which constrains $(D_{\mathrm A}^2/H)^{\frac{1}{3}}/r_{\mathrm s}$, 
where $r_{\rm s}$ is the sound horizon at the end of the baryon drag era $z_{\rm d}$ 
(see \S\ref{geometry_section}).
This degeneracy originates because for every line-of-sight 
clustering mode (which constrains $ H r_{\rm s}$), 
there are two transverse modes that constrain $D_{\rm A}/r_{\rm s}$. 

The subject of this study is breaking the $D_{\rm A}^2/H$ degeneracy 
by using the Alcock-Paczynski effect through anisotropic clustering. 
This approach was first suggested by 
\cite{hu03a} by using 
the two-dimensional power spectrum $P({\bf k})$. 
\cite{wagner08a} used mock catalogues at $z=1$ and 3 to demonstrate
the usefulness of the technique. 
\cite{shoji09a} argued that $H$ and $D_{\rm A}$ information is encoded in the 
full 2D shape, 
and presented a generic algorithm that takes into account dynamic
distortions on all scales, assuming all non-linear effects are understood. 
First attempts to apply these techniques on 2D $P({\bf k})$ and  $\xi({\bf s})$ clustering planes  
were performed by  \cite{okumura08}, \cite{chuang11a}, and \cite{blake11d}. 

\cite{padmanabhan08a} suggested 
decomposing the 2D correlation function into Legendre moments. 
They argue that the monopole ($\xi_0$ angle averaged signal) 
and the quadrupole components ($\xi_2$, see Equation \ref{xiell_equation}) 
contain most of the relevant constraining information. 
 \cite{taruya11a} and \cite{kazin11a} show that the hexadecapole term $\xi_4$ 
contains extra constraining power, which could be harnessed  
in the future with higher S/N than that currently available.

The advantage of analyzing 1D projections over 
the 2D plane is 
the relative simplicity of building a stable covariance matrix.

The first analyses of 
the anisotropic \baf using 
$\xi_0$ and $\xi_2$  have been 
performed on the SDSS-II luminous red galaxy sample 
($z\sim 0.35$; \citealt{xu12b,chuang12c,chuang12b}) 
and the DR9-CMASS sample tested here ($z\sim 0.57$; \citealt{reid12a}).

We analyze, for the first time, an alternative 1D basis suggested by \cite{kazin11a}, 
called  {\it clustering wedges} $\xi_{\Delta \mu}(s)$.
\cite{gaztanaga08iv} focused on a narrow clustering cylinder $\xi(s_{||},s_{\perp}<5$\hmpcii). 
In a subsequent analysis, \cite{kazin10b} proposed using wider  
clustering wedges $\xi_{\Delta \mu}(s)$ to improve S/N of the measurements. 
\cite{kazin11a} 
analyzed the constraining power of $H$ and $D_{\rm A}$ 
of $\xi_{\Delta \mu=0.5}(s)$ on mock catalogues. 
They concluded that these statistics should be comparable in performance 
to the multipoles ($\xi_0$, $\xi_2$) and provide a useful tool to test for systematics. 
The current study is the first analysis to perform such 
a thorough comparison on both data and mock galaxy catalogs. 

Our analysis differs from the previous ones in a few other aspects.  
First, we compare results 
both before and after reconstruction.  
Reconstruction is a technique which corrects for the damping of the 
\baf due to the large-scale coherent motions of galaxies. 
The baryonic acoustic feature is sharpened by calculating the displacement field 
and shifting galaxies to their near-original positions (\citealt{eisenstein07}). 
Second, we follow a similar approach as in \cite{xu12b}, 
by focusing on   
\czHzrs and \Dazrs and marginalizing over shape information. 
One notable difference from \cite{xu12b}, however, is that they apply 
a linear approximation of the Alcock-Paczynski test, 
where here we use the full non-linear equations. 
We compare both methods in Appendix \ref{linear_ap_appendix}, 
and show that the linear approach under-estimates 
the uncertainties of the obtained 
constraints. 
Finally, 
we compare between two independent theoretical $\xi$ templates.

This study 
is part of a series of 
papers 
analyzing the anisotropic clustering signal 
of the DR9 CMASS galaxy sample, containing $264,283$ massive galaxies between $0.43<z<0.7$. 
Here we measure $H$ and $D_{\rm A}$ 
in a model independent fashion 
through $\Delta \mu=0.5$  clustering wedges. 
\cite{sdss3dr9aniso} 
uses ``consensus" values of clustering wedges 
and multipoles to infer cosmological implications. 
Both of these studies focus on the information contained within the 
anisotropic \bafii. 
Two further studies analyze the information from the full shape of $\xi({\bf s})$: 
\cite{sanchez13a} use the $\xi_{\Delta \mu=0.5}$, 
and \cite{chuang13a} focus on the multipoles $\xi_{0,2}$. 

This study is constructed as follows: 
in \S\ref{geometry_section} we explain 
in detail the geometric information 
encoded in redshift maps. 
In \S\ref{wedgesdef_section} we define 
the clustering wedges and 
in \S\ref{data_section} we present 
the data and mock catalogs. 
In \S\ref{methodlogy_section} 
we  describe 
the method used in our analysis;   
\S\ref{results_section} describes our results. 
We discuss the results in \S\ref{discussion_section} 
and summarize in \S\ref{summary_section}.


To avoid semantic confusion, we briefly explain here the terminology of the different
spaces mentioned throughout the text. 
First, all analyses are based on two-point correlation functions, 
which we refer to as  
configuration-space, as opposed to 
the Fourier domain called $k-$space. 
Second, when referring to a 
space affected by redshift distortions,  
we call it redshift space,
and when there are none we refer to it 
as real space.


All the fiducial values 
calculated here are based on using 
the WMAP7 flat $\Lambda$CDM cosmology (\citealt{komatsu11a}).  
To calculate comoving distances we assume 
the matter density $\Omega_{\rm M}=0.274$. 
Assuming $h=0.7$ this yields fiducial values: 
$H^{\rm f}=93.57 $ kms$^{-1}$Mpc$^{-1}$, 
$D_{\rm A}^{\rm f}=1359.6$ Mpc at $z=0.57$. 
Throughout, we also use 
derived unit-less relationships 
(\czHzrsii$)^{\rm f}=11.94$, (\Dazrsii$)^{\rm f}=8.88$, 
where $r_{\rm s}^{\rm f}=153.1$ Mpc 
(at $z_{\rm d}^{\rm f}=1020$). 
For these we assume the baryon density 
$\Omega_{b}h^2=0.0224$, 
radiation density $10^{5}\Omega_{r}h^2=4.17$ 
and photon density of $10^{5}\Omega_{\gamma}h^2=2.47$.

\section{Cosmic Geometry from Galaxy Maps}\label{geometry_section}
Although galaxy distributions in real-space  
are assumed to be statistically isotropic, 
measured clustering signals from galaxies from redshift maps 
are anisotropic. 
This is a result of  
two physical effects that are at play when 
converting observed redshifts $z_{\rm obs}$ 
to comoving distances $\chi$: 
\beq\label{comoving_equation}
\chi(z_{\rm obs})=c\int_0^{z_{\rm obs}}{\frac{dz}{H(z)}}. 
\eeq 

The first, which we refer to as  redshift-distortions, 
stems from the fact that 
$z_{\rm obs}$ 
is a degenerate combination 
of the cosmological flow 
and the radial component 
of the peculiar velocity.  
This results in anisotropic clustering components 
due to large-scale coherent flows (\citealt{kaiser87}), 
and velocity dispersion effects within galaxy clusters. 
For a  detailed introduction on dynamical redshift-distortions see 
\cite{hamilton98a}.

On large scales,  
these effects can be used to test 
for deviations from General Relativity 
(\citealt{kaiser87,linder08}; see \citealt{guzzo07,samushia11a,blake11a,samushia13a,beutler12a} for the most recent measurements).
The observable in this test is 
$f\sigma_8$, 
where $b$ is the linear tracer to matter density bias,  
$f\equiv {\rm d}D_1/{\rm d}\ln a$ is the 
rate of change of growth 
of structure, 
$D_1$ is the linear growth of structure, 
and $\sigma_8$ is the linear r.m.s of density fluctuations 
averaged in spheres of radii $8$\hmpcii. 
This study focuses on a second more subtle effect 
which involves geometric distortions. 

Comoving separations between two nearby points in space 
depend both on $z$ 
and the observer angle between them $\Theta$. 
Assuming the plane-parallel approximation between galaxy pairs, 
radial separations are defined as $s_{||}\equiv c\Delta z/H(z)$, where $c$ 
is the speed of light, 
and transverse distances $s_{\perp}\equiv\Theta (1+z) D_{\rm A}$,  
where the proper (physical) angular diameter distance $D_{\rm A}$ is defined as: 
\beq\label{da_equation}
D_{\rm A}=\frac{1}{1+z}\frac{c}{H_0} \frac{1}{\sqrt{-\Omega_{\rm K}}} \sin\left(   \sqrt{-\Omega_{\rm K}} \frac{\chi}{c/H_0}  \right),
\eeq
where $H_0\equiv H(0)$ 
and $\Omega_{\rm K}=1-\sum_{\rm X} \Omega_{\rm X}$ is the representation 
of the curvature, 
and $\Omega_{\rm X}$ are the energy densities of 
compoenents X (matter, radiation, etc.).\footnote{Note that this is generic because $i\sin(ix)=-\sinh(x)$.} 
Hence, assuming an incorrect cosmology in Equation (\ref{comoving_equation}) 
would cause a spherical body (meaning $s_{||}=s_{\perp}$) 
to be deformed.  
For example, 
a lower $H(z)$ than the true one would cause an elongation 
along the line-of-sight 
due to an increased $s_{||}$,
where a lower $D_{\rm A}(z)$ than the true value would cause a 
transverse squashing, 
because of a decrease of $s_{\perp}$.
Therefore, by fixing the observables $\Theta$ and $\Delta z$, 
retrieving a spherical shape constrains the $H D_{\rm A}$ combination. 

Various techniques have been suggested 
to measure $H  D_{\rm A}$  
through this Alcock-Paczynski test (AP henceforth; \citealt{alcock79,phillipps94a,lavaux12a}). 
Here we focus on clustering of galaxies, 
where 
line-of-sight clustering modes depend 
on $s_{||}$ ($1/H$) 
and transverse modes on $s_{\perp}$ ($D_{\rm A}$), 
and hence the anisotropies due the AP effect. 

It has been pointed out that the anisotropies 
from this geometric effect are degenerate with 
those from redshift-effects (\citealt{ballinger96a}). 
Various studies, such as \cite{blake11d} and \cite{reid12a},  
show the degeneracy between  $H  D_{\rm A}$ 
and $f\sigma_8$. In this study 
we marginalize over the  redshift distortion information and focus 
on the geometric distortions. 

In practice, when converting redshifts to comoving distances, 
the $H_0$ 
factors out trivially and thus we express comoving distance in units of \hmpcii, 
where $h\equiv H_0/(100\,{\rm km}\,{\rm s}^{-1}{\rm Mpc}^{-1}$). 
The rest of the parameters in $H(z)$ 
($\Omega_{\rm X}$ and their equation of states $w_{\rm X}$)
have more important, and potentially measurable, effects. 

One way of overcoming these effects is to recalculate $\chi$ 
and then the 
clustering statistics for every set of parameters when 
determining cosmological constraints. 
However, that approach is currently not practical;  
instead, we vary a fixed clustering template,  
as described below. 

Although the \baf comoving scale 
is fixed, the apparent position measured 
in the correlation function depends 
on $Hr_{\rm s}$ and $D_{\rm A}/r_{\rm s}$. 
As demonstrated in \cite{eisenstein05b},  
the \baf in the angle average signal 
is sensitive, to first order, to 
$\left(D_{\rm A}^2/H\right)^{1/3}/r_{\rm s}$. 
\cite{padmanabhan08a} show that analysis of the anistropic signal 
adds  $HD_{\rm A}$ information, and hence breaks 
the degeneracy. 
To break the degeneracy with $r_{\rm s}$ 
one needs to add additional information  
from the CMB anisotropies. 

When relating $r_{\rm s}$ 
measured from the CMB to that 
in the large-scale structure,  
one must 
take into account that these two definitions correspond to 
slightly different sound horizon radii (see Equation 1 in \citealt{blake03}). 
Because the baryons have momentum at decoupling $z_*$, 
the baryonic acoustic signature in the distribution of matter is related to 
$r_{\rm s}(z_{\rm d})>r_{\rm s}(z_*)$,   
where $z_{\rm d}$ is the epoch when the baryonic drag effectively ended (\citealt{eisenstein98}). 
The baryonic acoustic signature in the CMB anisotropies corresponds to $z_*$. 
For current $r_{\rm s}(z_{*})$  measurements 
see \cite{hinshaw12a}, 
and for 
$r_{\rm s}(z_{\rm d})$ predictions from the CMB, 
see Table 3 of \cite{komatsu09a}.

Conservation of the observer angle $\Theta$ 
means that true separations transverse to the line-of-sight component $s_{\perp}^{\rm t}$
will be related to an apparent ``fiducial" component $s_{\perp}^{\rm f}$ by:\footnote{Here we assume the plane-parallel approximation for each pair.}
\beq\label{strv_shift_eq}
s_{\perp}^{\rm t}=s_{\perp}^{\rm f}\cdot \alpha_{\perp}, 
\eeq
where 
\beq\label{alphatrv_eq}
\alpha_{\perp}\equiv \frac{D_{\rm A}^{\rm t}}{D_{\rm A}^{\rm f}}\cdot\frac{r_{\rm s}^{\rm f}}{r_{\rm s}^{\rm t}}.
\eeq
where the ``f" subscript indicates the fiducial cosmology when calculating $\chi(z)$, 
and ``t" indicates the true cosmology. 
 
Similarly, 
the true line-of-sight separation component is related to the fiducial by:
\beq\label{slos_shift_eq}
s_{||}^{\rm t}=s_{||}^{\rm f}\cdot \alpha_{||}, 
\eeq
with  
\beq\label{alphalos_eq}
\alpha_{||}\equiv \frac{H^{\rm f}}{H^{\rm t}}\cdot\frac{r_{\rm s}^{\rm f}}{r_{\rm s}^{\rm t}}.
\eeq
The sound horizon $r_{\rm s}(z_{\rm d})$ terms 
appear due to the degeneracy with $D_{\rm A}$ and $H$, 
when applied to the \baf as a standard ruler. 
Here we quote the rescaling in the position 
of the peak of the $\xi$. 
The purely geometrical effect of changing the 
cosmology does not 
depend 
on $r_{\rm s}(z_{\rm d})$. 

In Appendix \ref{ap_inpractice_section} we explain how we apply  
the AP test in practice through the mapping of 
$\xi$ between 
these coordinates systems. 


We also make use of an alternative representation 
of $\alpha_{||}$ and $\alpha_{||}$ through the 
isotropic dilation parameter $\alpha$ (\citealt{eisenstein05b})
and the anisotropic warping parameter $\epsilon$ (\citealt{padmanabhan08a}): 
\beq\label{alpha_equation}
\alpha\equiv \left(\frac{D_{\rm A}}{D_{\rm A}^{\rm f}}\right)^{\frac{2}{3}} \left(\frac{H^{\rm f}}{H}\right)^{\frac{1}{3}}\frac{r_{\rm s}^{\rm f}}{r_{\rm s}}=\alpha_{\perp}^{2/3}\alpha_{||}^{1/3};
\eeq
\beq\label{epsilon_equation}
1+\epsilon= \left( \frac{D_{\rm A}^{\rm f}H^{\rm f}}{D_{\rm A}H} \right)^{\frac{1}{3}}=\left( \frac{\alpha_{||}}{\alpha_{\perp}}\right)^{1/3}.
\eeq

\section{Clustering Wedges}\label{wedgesdef_section}

Assuming azimuthal statistical symmetry around 
the line-of-sight\footnote{The assumption of azimuthal statistical symmetry around the line-of-sight is true even with geometrical distortions.}
the 3D correlation function $\xi({\bf s})$ 
can be projected into 2D polar coordinates: 
the comoving separation $s$ and 
the cosine of the angle from the line-of-sight $\mu$, 
where the line-of-sight direction is $\mu=1$. 


The 2D plane of $\xi(\mu,s)$ can then be projected to 
{\it clustering wedges} $\Delta \mu$ as:
\beq\label{wedges_from_ximus_equation}
\xi_{\Delta\mu}(s)=\frac{1}{\Delta\mu}\int_{\mu_{\rm min}}^{\mu_{\rm min}+\Delta \mu}\xi(\mu,s)d\mu.
\eeq
For the purpose of this study, 
we focus on two clustering wedges of 
$\Delta\mu=0.5$, which we call  
line-of-sight $\xi_{\rm ||}(s)\equiv\xi_{0.5}(\mu_{\rm min}=0.5,s)$ and 
transverse $\xi_{\rm \perp}(s)\equiv\xi_{0.5}(\mu_{\rm min}=0,s)$. 
For consistency we compare all results to the 
multipole statistics defined as: 
 \beq\label{xiell_equation}
 \xi_{\ell}(s)=\frac{2\ell+1}{2}\int_{-1}^{+1}\xi(\mu,s)\mathcal{L}_{\ell}(\mu)d\mu,
 \eeq
where 
$\mathcal{L}_{\ell}(x)$ are the standard Legendre polynomials. 

The clustering wedges and multipoles are 
complementary bases of similar information. 
As shown by \cite{kazin11a} up to order $\ell=4$ they are related by: 
\beq
\xi_{||}(s)= \xi_0(s) + \frac{3}{8} \xi_2(s) - \frac{15}{128}\xi_4(s), 
\eeq

\beq
\xi_{\perp}(s)= \xi_0(s) - \frac{3}{8} \xi_2(s) + \frac{15}{128}\xi_4(s). 
\eeq

A useful relationship is the fact that the average of the $\Delta \mu=0.5$ clustering wedges 
results in $\xi_0$. 

In real space, where there are no anisotropies, 
all $\ell>0$ components are nulled, and clustering wedges of any $\Delta\mu$ width correspond to the monopole signal.\footnote{Homogeneity and isotropy are assumed here.}
The AP effect 
breaks this symmetry, causing $\ell>0$ components 
due to geometric distortions. 


\section{Data}\label{data_section}
We base our measurements of 
\czHzrs and \Dazrs on the large-scale 
anisotropic correlation function 
of the BOSS DR9-CMASS galaxy sample.
Here we give a brief description of the sample, 
and the calculated $\xi(\mu,s)$. 

\subsection{The DR9-CMASS galaxy sample}\label{galaxysample_section}
We use data from the SDSS-III BOSS survey \citep{eisenstein11a,dawson13a}.
The galaxy targets for BOSS are divided into two samples, LOWZ and CMASS. These 
are selected on the basis of photometric observations 
carried out with a drift-scanning mosaic CCD camera \citep{gunn98a,gunn05a} 
on the Sloan Foundation telescope at the Apache Point observatory.
Spectra of these galaxies are obtained using the double-armed BOSS spectrographs \citep{smee12a}.
Spectroscopic redshifts are then measured by means of the minimum-$\chi^2$ template-fitting procedure 
described in \citet{aihara11a}, with templates and methods updated for BOSS data as described in
\citet{bolton12a}.

Our analysis is based on the CMASS galaxy sample of SDSS Data Release 9 (DR9, \citealt{ahn12a}). 
This sample was designed to cover the redshift ranges
$0.43< z< 0.7$ down to a limiting stellar mass, resulting on a roughly constant
number density of $n \simeq 3 \times 10^{-4} h^3{\rm Mpc}^{-3}$
\citep[][Padmanabhan et al. in preparation]{eisenstein11a,dawson13a}.
This sample contains mostly central galaxies, with a $\sim 10\%$ satellite fraction
\citep{white10a,nuza12a} and it is dominated by early type galaxies, although it contains a significant
fraction of massive spirals \citep[$\sim26\%$,][]{masters11a}.

\citet{anderson12a} present a detailed description of the construction of a CMASS catalogue for LSS studies.
We follow the procedure detailed there and refer the reader to this article for more details.

\subsection{PTHalo mock catalogues}\label{mocks_section}
Mock catalogues play 
a major role in the analysis and interpretation of large-scale structure information, 
as they offer a useful tool to test for systematics 
and provide the means with which to estimate statistical errors.
In this analysis we use 600 PTHalo mock galaxy realizations to 
test our analysis pipeline 
and 
construct a covariance matrix of our measurements. 
Full details of the mock catalogues are given in \cite{manera12a}.
Briefly, the mocks are based on dark matter
2LPT simulations ($2^{\rm nd}$ order Lagrangian Perturbation Theory) , that were populated 
with mock galaxies within dark matter haloes. 
The halo occupation distribution (\citealt{peacock00a,seljak00a,scoccimarro01a,berlind02a,cooray02})
parameters are determined by comparing  
the correlation function 
to that of the data in the scale range of $[30,80]$\hmpcii.

To match the selection function of the data, 
the mock data is split by 
the Northern and Southern CMASS angular geometry 
and galaxies were excluded to match the radial profile. 

\subsection{The Anisotropic Correlation Function}\label{anixi_section}
\begin{figure*}
\begin{center}
\includegraphics[width=0.49\textwidth]{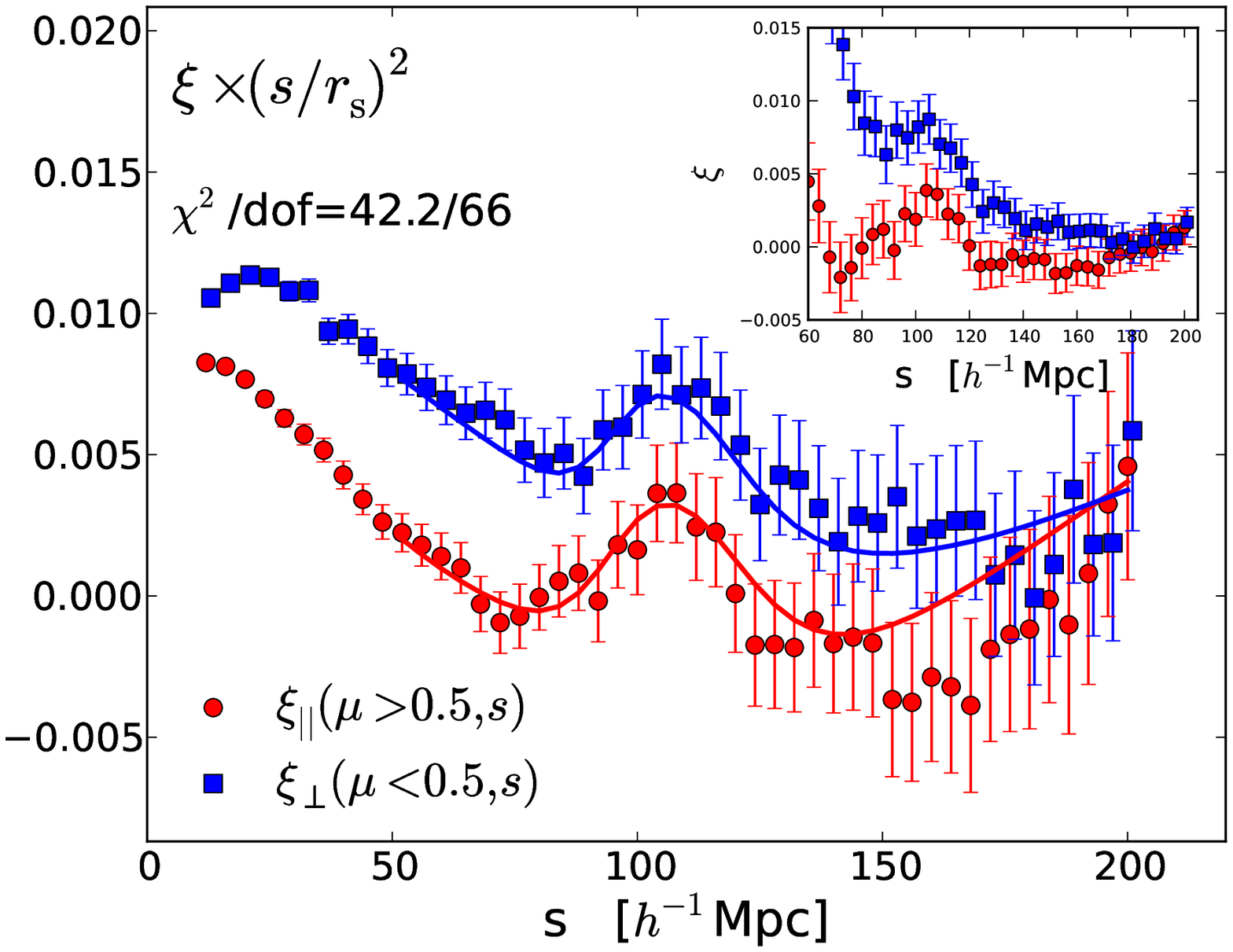}
\includegraphics[width=0.49\textwidth]{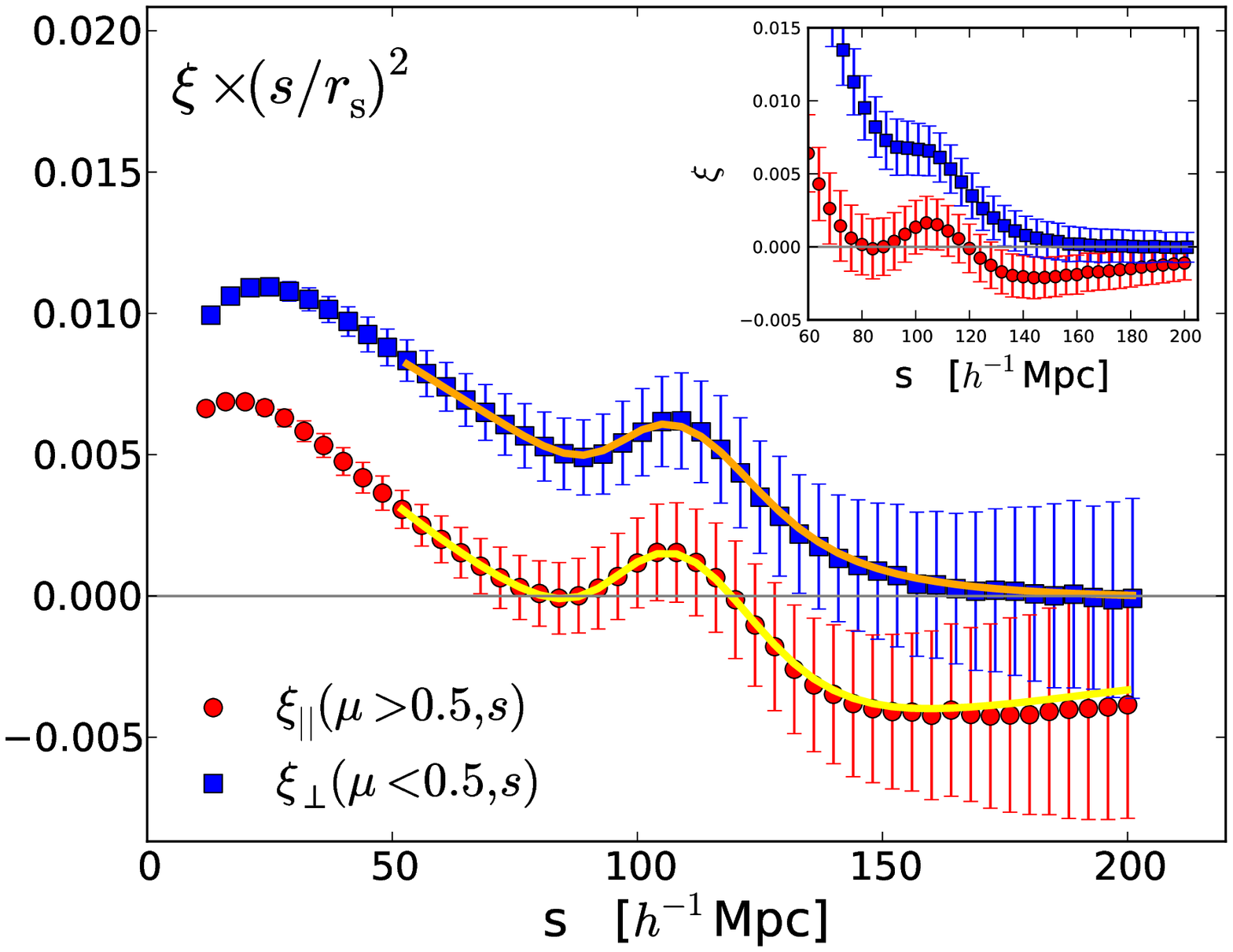}
\includegraphics[width=0.49\textwidth]{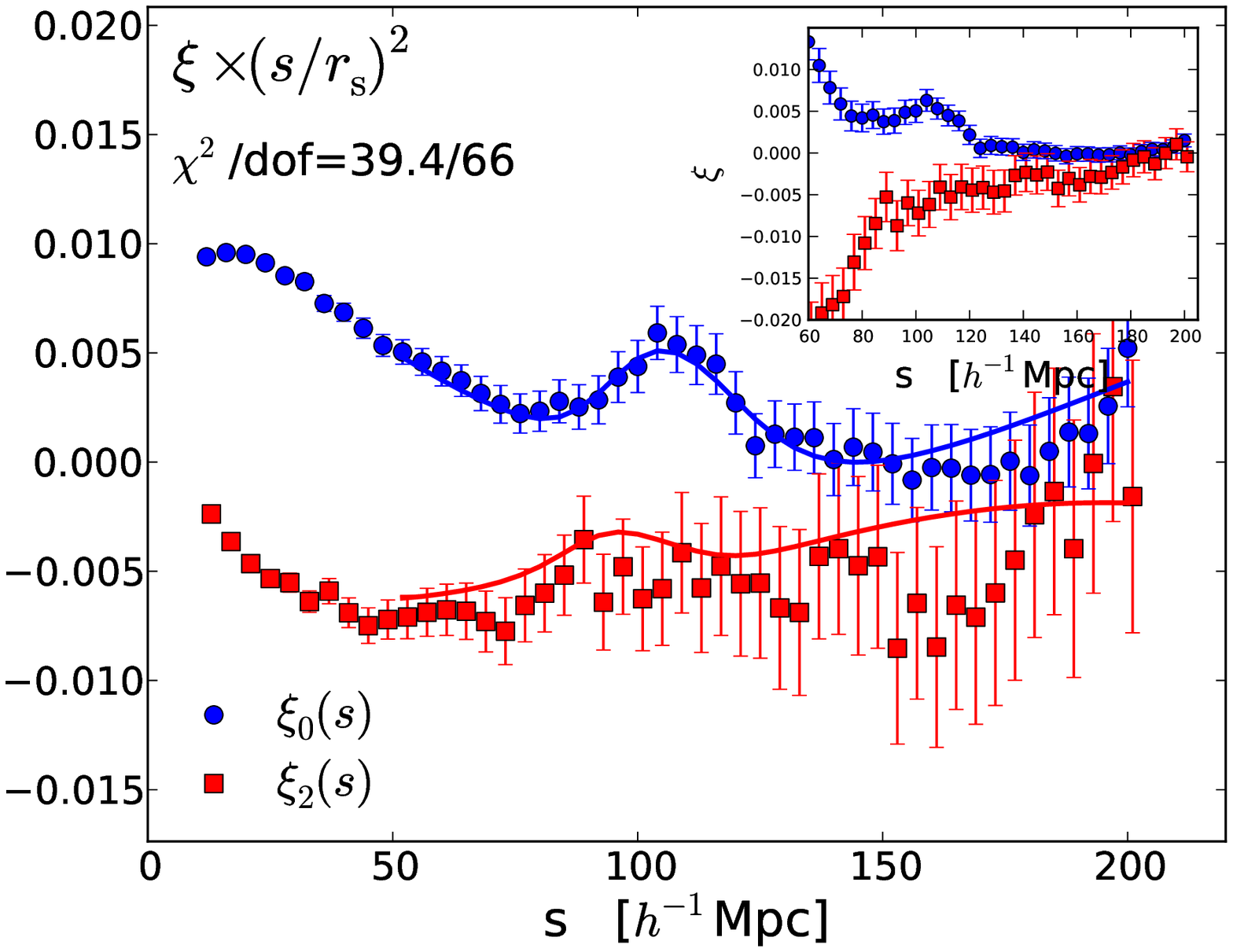}
\includegraphics[width=0.49\textwidth]{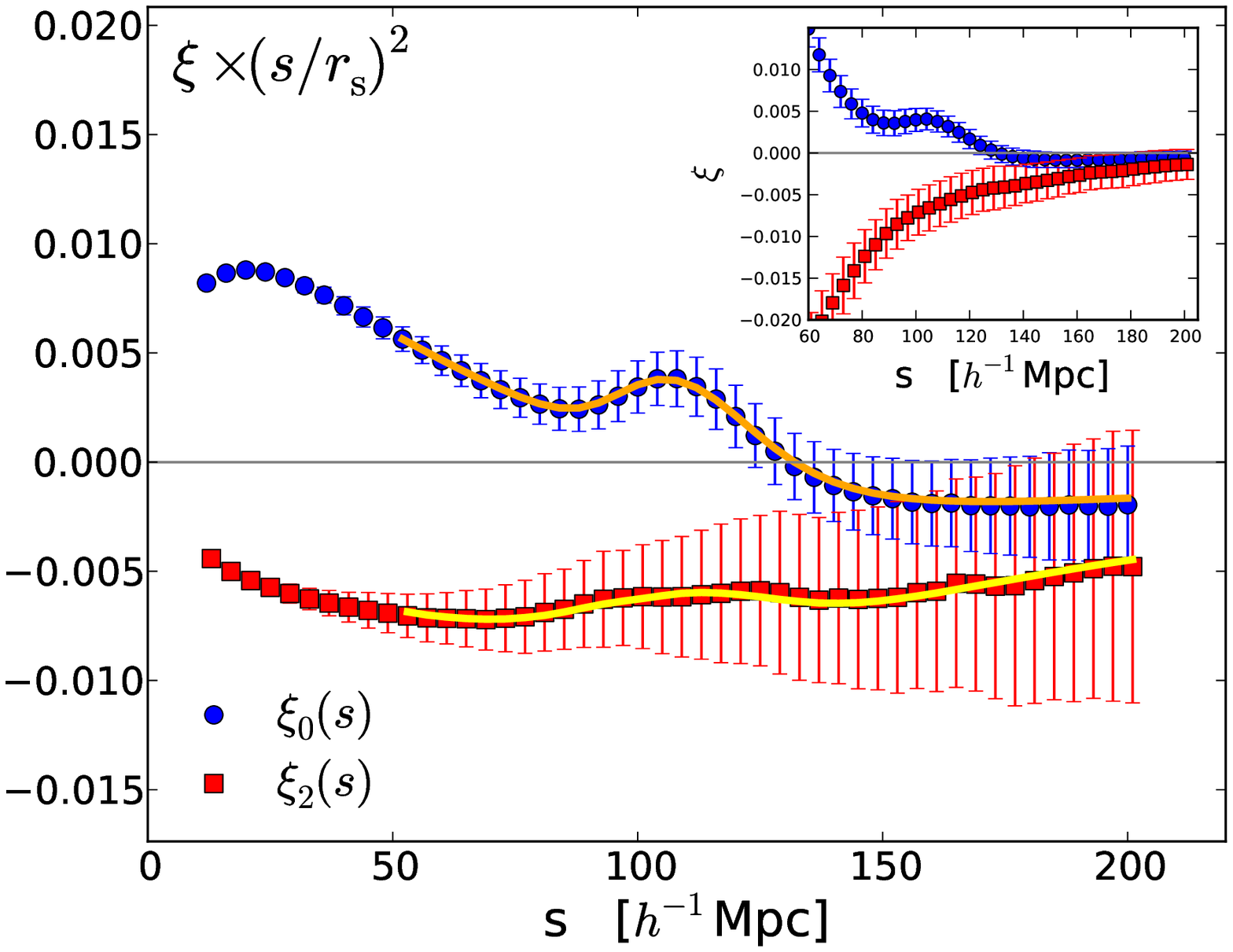}
\caption{
Top Left: the pre-reconstruction DR9-CMASS $\avg{z}=0.57$ clustering wedges are displayed 
multiplied by $\left(s/r_{\rm s}\right)^2$ in the main plot, and without in the inset. 
Bottom Left: the CMASS monopole and quadrupole.
The solid lines in the left hand panels are the RPT-based best fit models (Renormalized Perturbation Theory; see \S\ref{xitemplates_section}). 
The $\chi^2$ and degrees of freedom are indicated. 
Right: The same statistics using the mean signal of $600$ PTHalo mock realizations. 
The uncertainty estimates are the square root of the diagonal elements of the covariance matrix.
The solid lines in the right hand panels are the RPT-based templates.
(For clarity the transverse wedge and quadrupole are shifted by 1 \hmpcii.)
The line-of-sight wedge ($\mu>0.5$ red circles) is clearly weaker 
than the transverse wedge ($\mu<0.5$ blue squares). 
In both clustering wedges there is a clear 
signature of the \bafii. 
}
\label{cmass_xi1d_stats_figure}  
\end{center}
\end{figure*}

\begin{figure*}
\begin{center}
\includegraphics[width=0.49\textwidth]{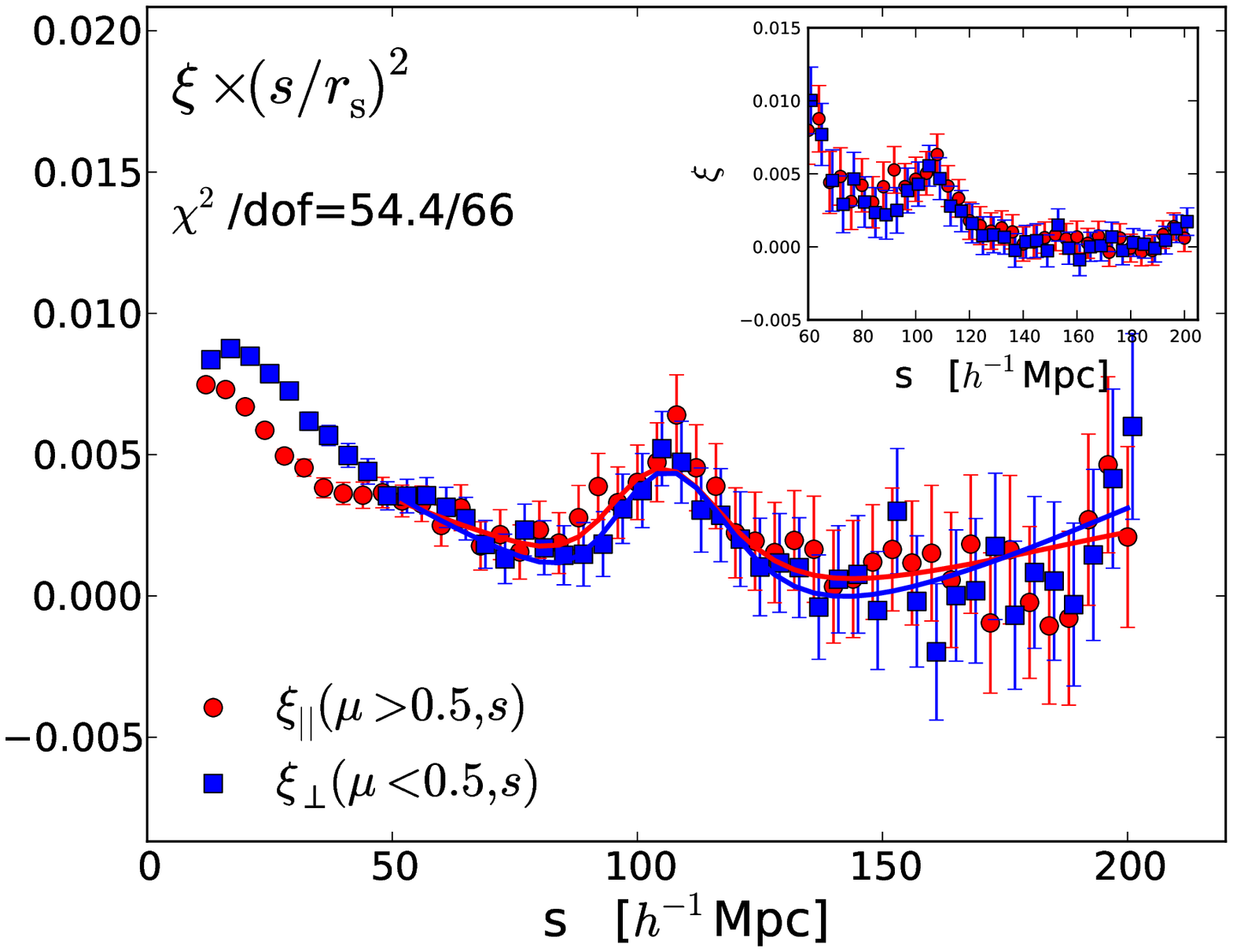}
\includegraphics[width=0.49\textwidth]{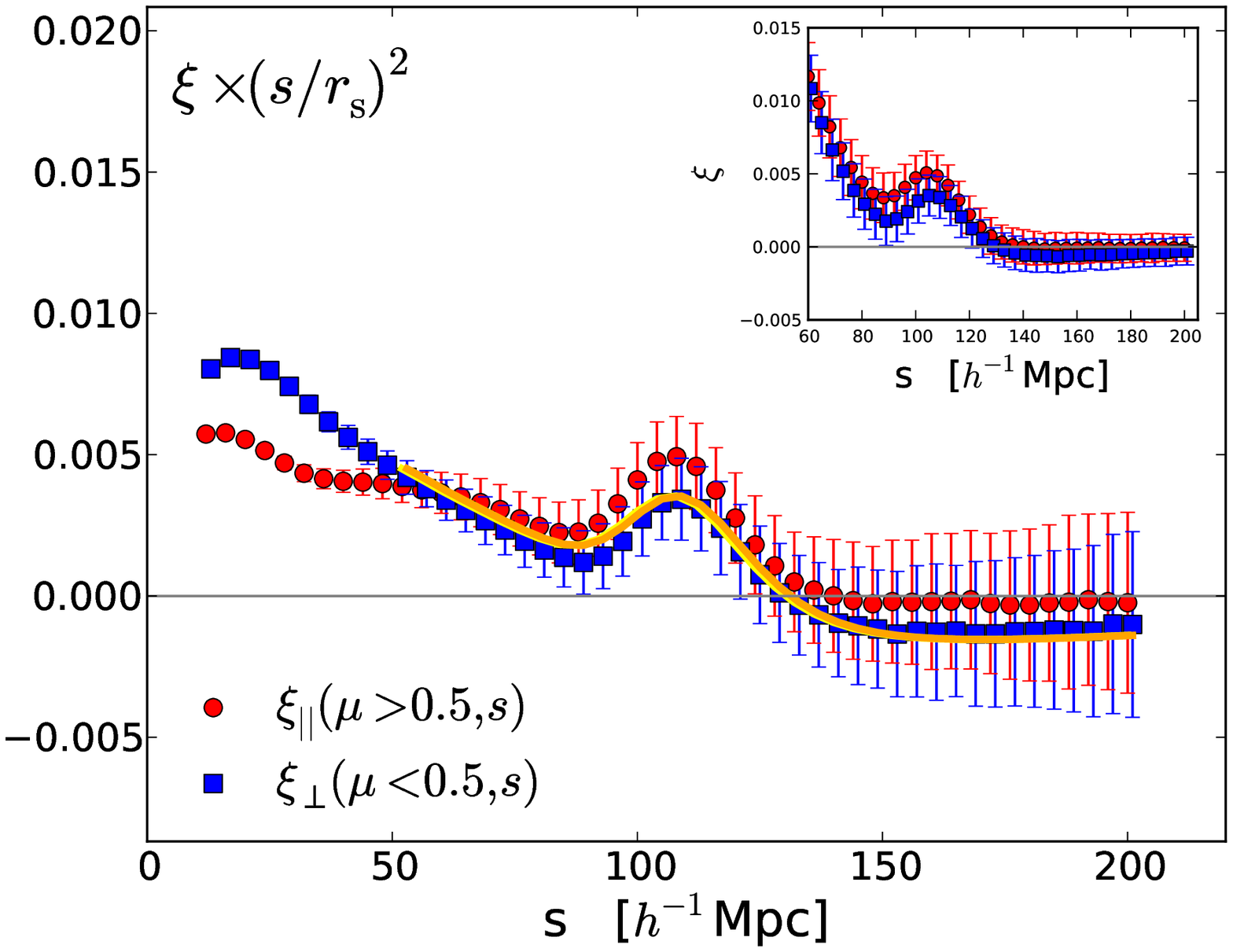}
\includegraphics[width=0.49\textwidth]{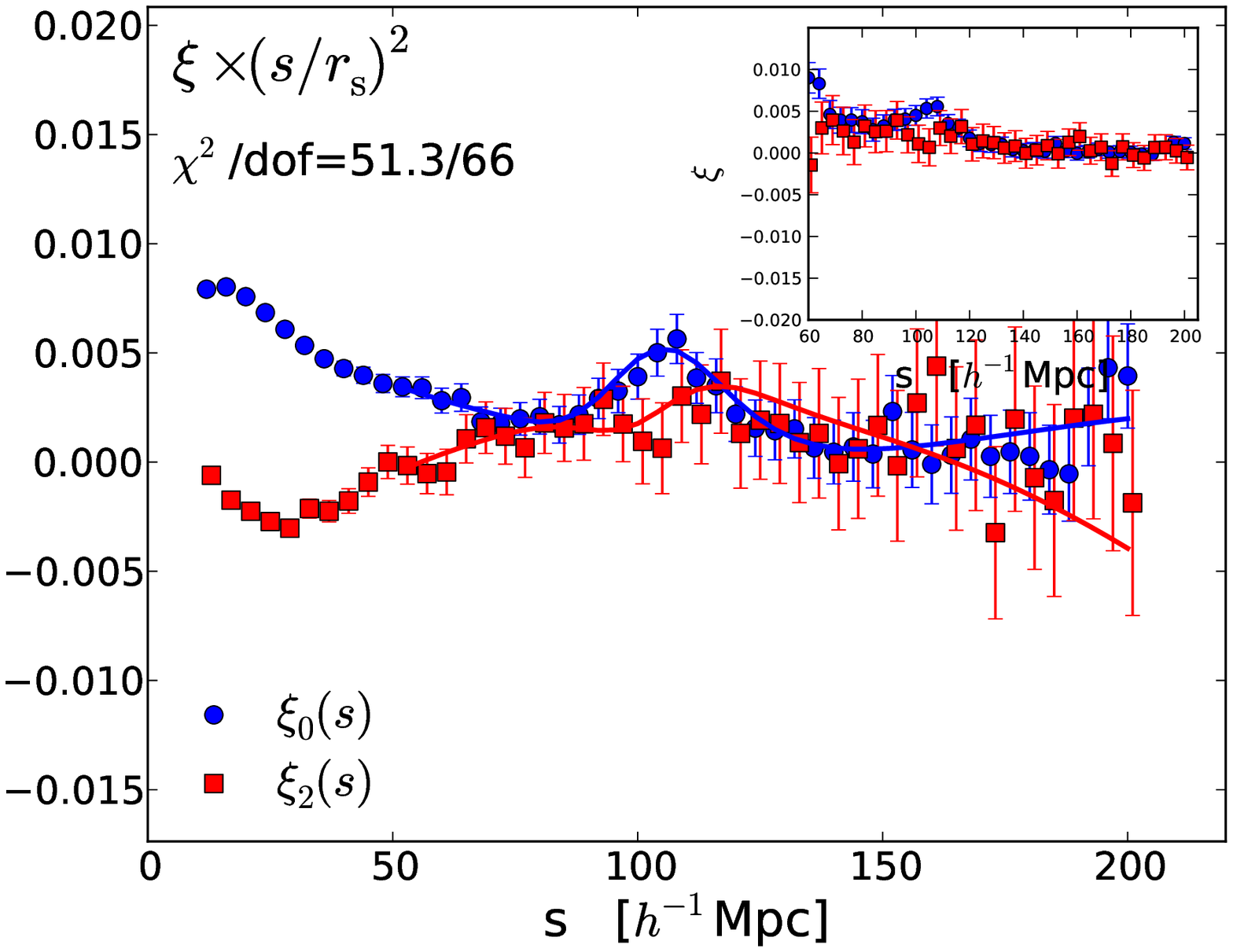}
\includegraphics[width=0.49\textwidth]{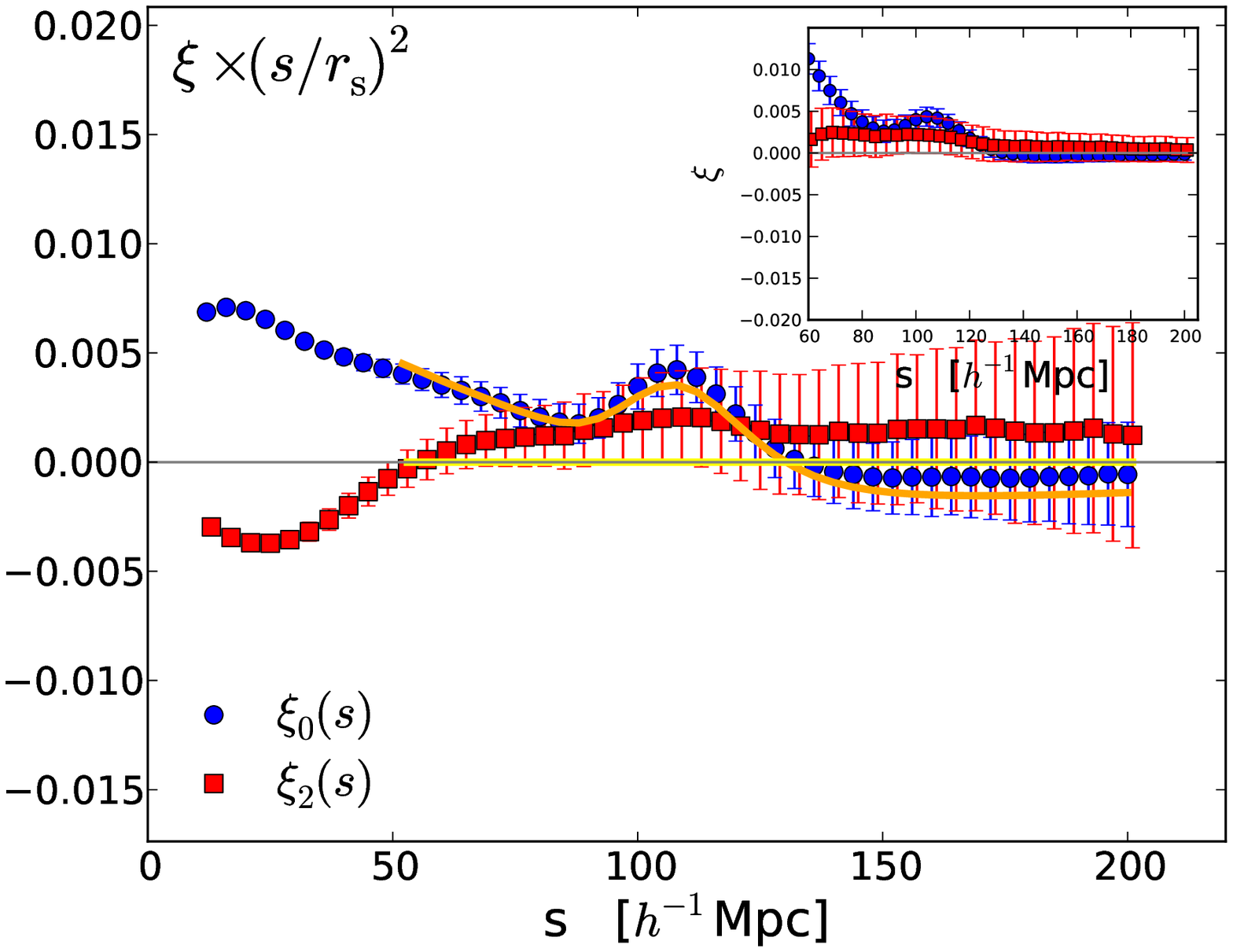}
\caption{
Top Left: the post-reconstruction DR9-CMASS $\avg{z}=0.57$ clustering wedges are displayed 
multiplied by $\left(s/r_{\rm s}\right)^2$ in the main plot, and without in the inset. 
Bottom Left: the CMASS monopole and quadrupole.
The solid lines in the left hand panels are the RPT-based best fit models (Renormalized Perturbation Theory; see \S\ref{xitemplates_section}). 
The $\chi^2$ and degrees of freedom are indicated. 
Right: The same statistics using the mean signal of $600$ PTHalo mock realizations. 
The uncertainty estimates are the square root of the diagonal elements of the covariance matrix.
The solid lines in the right hand panels are the RPT-based templates, 
where we set $\xi_2=0$, as explained in \S\ref{postrec_templates_section}. 
(For clarity the transverse wedge and quadrupole are shifted by 1 \hmpcii.)
The line-of-sight wedge ($\mu>0.5$ red circles) is similar to  
the transverse wedge ($\mu<0.5$ blue squares). 
This result shows that reconstruction substantially reduces effects of redshift-distortions.
In both clustering wedges there is a clear 
signature of the \bafii. 
}
\label{cmass_xi1d_stats_figure_postrec}  
\end{center}
\end{figure*}

To compute the correlation function, we use the  
\cite{landy93a} estimator with an angular dependence: 
\beq
\xi(\mu,s)= \frac{DD(\mu,s)+RR(\mu,s)-2DR(\mu,s)}{RR(\mu,s)},
\eeq
We calculate 
the normalized data-data pair counts in bins of evenly separated 
$\mu$ and $s$,  
$DD(\mu,s)$,
and similarly for data-random pairs, $DR$, and random-random, $RR$, 
where for each pair $\mu=1$ is defined as the direction 
in which a vector from the observer bisects ${\bf s}$. 
The $\mu$ values of each bin are the flat mean value. 
Our choice of binning is $\Delta \mu=1/100$ and $\Delta s=4$\hmpcii.

To obtain the clustering wedges we 
use Equation (\ref{wedges_from_ximus_equation}), 
where for $\xi_{\perp}(s)$ we use the $\mu$ range [0,0.5] and for $\xi_{||}(s)$  [0.5,1]. 
The resulting pre-reconstruction clustering wedges 
and multipoles 
are presented in top and bottom panels of Figure \ref{cmass_xi1d_stats_figure}, respectively.
The line-of-sight wedge $\xi_{||}(\mu>0.5,s)$ is clearly weaker 
than the transverse wedge $\xi_{\perp}(\mu<0.5,s)$. 
This large difference in amplitudes on large-scales 
is due to 
redshift-distortions. 

For comparison in the right panels of 
Figure \ref{cmass_xi1d_stats_figure} 
we show the mock-mean signals, i.e., the mean $\xi_{\Delta\mu}$ and $\xi_\ell$ 
of 600 mock catalogs. 

\subsection{Reconstructing the \baf}\label{data_reconstruction_section}
\cite{eisenstein07} showed  
that large-scale coherent motions, 
which cause a damping 
of the \bafii, can be ameliorated  
by using the gravitational potential estimated from the 
large-scale galaxy distribution to predict the bulk flows, 
and undo their non-linear effect on the density field. 
First studies focusing on periodic boxes shown 
that this reconstruction technique sharpens 
the \bafii, and hence improves its usage as a standard ruler  
(\citealt{padmanabhan09a,noh09a,seo10a,mehta11a}). 
We follow the procedure in \cite{padmanabhan12a}, 
which takes into account practical issues 
as edge effects by applying a Weiner filter (\citealt{hoffman91a,zaroubi95a}). 
We apply the reconstruction procedure on both the DR9-CMASS 
data, as well as on the mocks. 

Figure \ref{cmass_xi1d_stats_figure_postrec} 
displays the post-reconstruction results 
for $\xi_{||,\perp}(s)$ (top left) and the $\xi_{0,2}(s)$ (bottom left). 
We clearly see that the amplitudes of the clustering wedges are aligned 
at the scales of the \baf and larger. 
This is due to the fact that reconstruction not only 
corrects for large-scale coherent motions, 
but also corrects, to a certain extent, 
for redshift-distortions, as is seen 
by the near nullifying of the $\xi_2(s)$.

For comparison, 
the right panels Figure \ref{cmass_xi1d_stats_figure_postrec} 
show results of the post-reconstruction mock-mean signal. 
We clearly see that the $\xi_2(s)$ reverses 
from negative at \baf scales from the pre-reconstruction signal 
to positive.  
This change might be attributed to an over compensation 
of the redshift-distortions. 
In other words, 
throughout the reconstruction process,  
we estimate $f$ to 
shift galaxies in the radial direction, 
with the aim to reduce the Kaiser effect. 
An over-estimation could potentially put 
field galaxies a bit further away from high 
dense regions, and hence reverse the $\xi_2(s)$ signal,  
yielding a $\xi_{\perp}(s)$ that is slightly weaker than the $\xi_{||}(s)$.
We are not concerned with this issue, 
because we do not expect redshift-distortions 
to shift the position of the anisotropic \bafii.


In both clustering wedges, in pre- and post-reconstruction, 
there is a clear 
signature of the \bafii. 
We quantify the significance of the detection in \S\ref{significance_section}. 

\section{Analysis Methodology}\label{methodlogy_section}
\subsection{Statistics used}\label{statistics_section}
When computing likelihoods of a model $M$ 
with a variable parameter space $\Phi$
to fit data $D$, 
we calculate the $\chi^2$: 
\beq
\chi^2(\Phi)= \sum_{i,j}\left(M_i\left(\Phi\right)-D_i\right)C_{ij}^{-1}(M_j(\Phi)-D_j), 
\eeq
where $i,j$ are the bins tested.
The likelihood is then assumed to be Gaussian $L(\Phi)\propto\exp\left(-\frac{1}{2}\chi^2(\Phi)\right).$ 

Throughout this analysis we run Monte Carlo Markov Chains (MCMC) 
nominally for nine or ten parameters as described in 
\S\ref{paramspace_section}.
We quote  
the mode of the posterior as our {\it measurement}  
and half the $68\%$ CL region (68CLr henceforth) for the {\it uncertainty}, 
because these are well defined 
regardless of asymmetries in likelihood profiles. 

\subsubsection{Covariance matrix}\label{cij_section}
We construct the covariance matrix $C_{ij}$ 
from the $N_{\rm mocks}=600$ mock catalogues.
(For a description of the mocks used  see \S\ref{mocks_section}.)
The $\xi_{||,\perp}$ signals are not independent 
but have significant cross-correlations. 
To take these correlations into account, when 
constructing the $C_{ij}$, we treat the mocks signals  
in an array of the form $\xi_{\rm [2s]}=[\xi_{||},\xi_{\perp}]$, 
meaning a 1D array with twice the length of the separation range 
of analysis. When analyzing the multipoles we apply a similar 
convention $\xi_{\rm [2s]}=[\xi_{0},\xi_{2}]$.
We then construct 
a  covariance matrix of  $\xi_{\rm [2s]}$  defined as:
\begin{dmath}\label{cij_equation}
C_{ij}= \frac{1}{N_{\rm mocks}-1} \sum_{m=1}^{N_{\rm mocks}} \left(   {\overline{\xi}_{\rm [2s]}}_i - {\xi_{\rm [2s]}^m}_i \right) \left(   {\overline{\xi}_{\rm [2s]}}_j - {\xi_{\rm [2s]}^m}_j \right).
\end{dmath}

\begin{figure*}
\begin{center}
\includegraphics[width=0.49\textwidth]{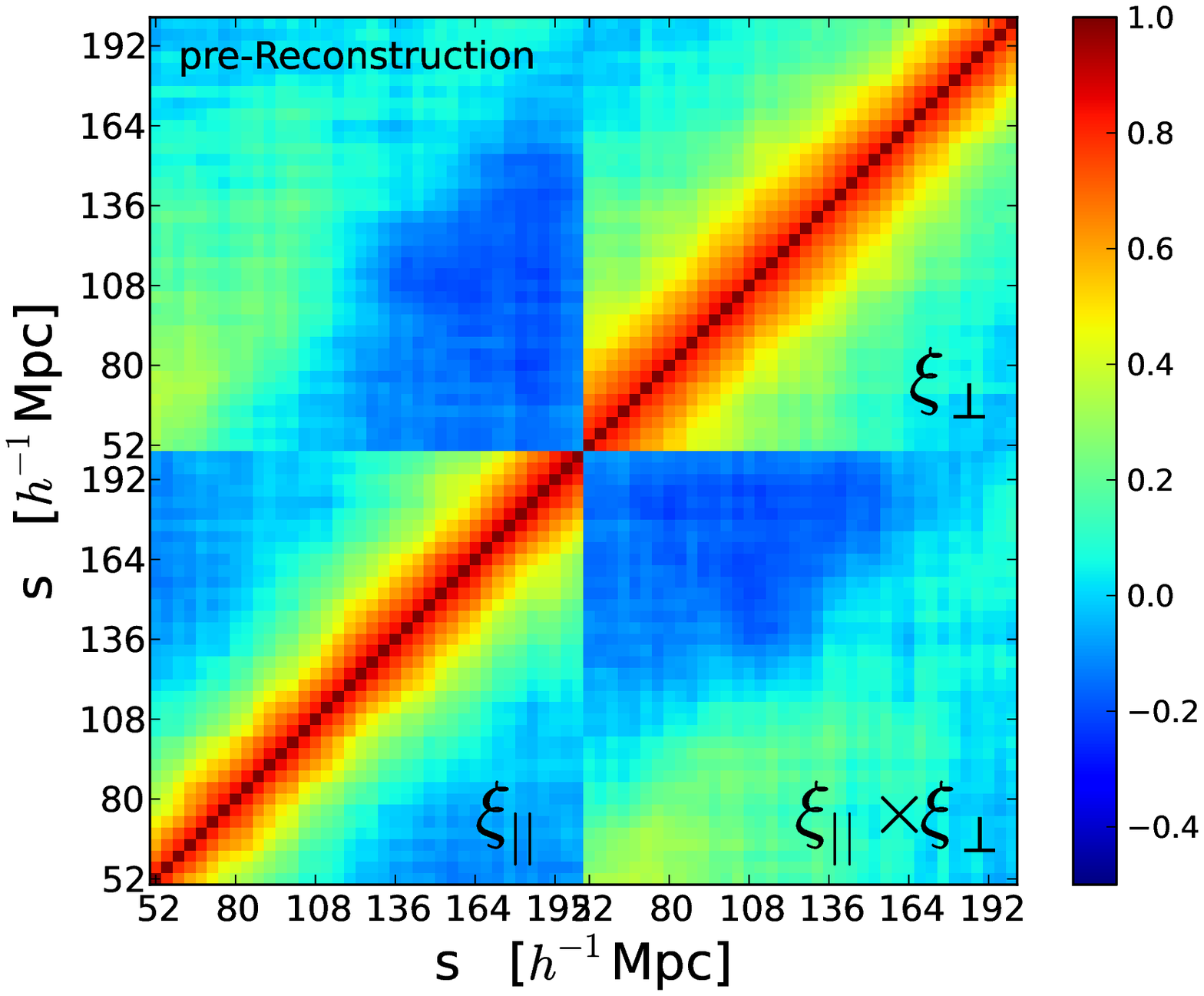}
\includegraphics[width=0.49\textwidth]{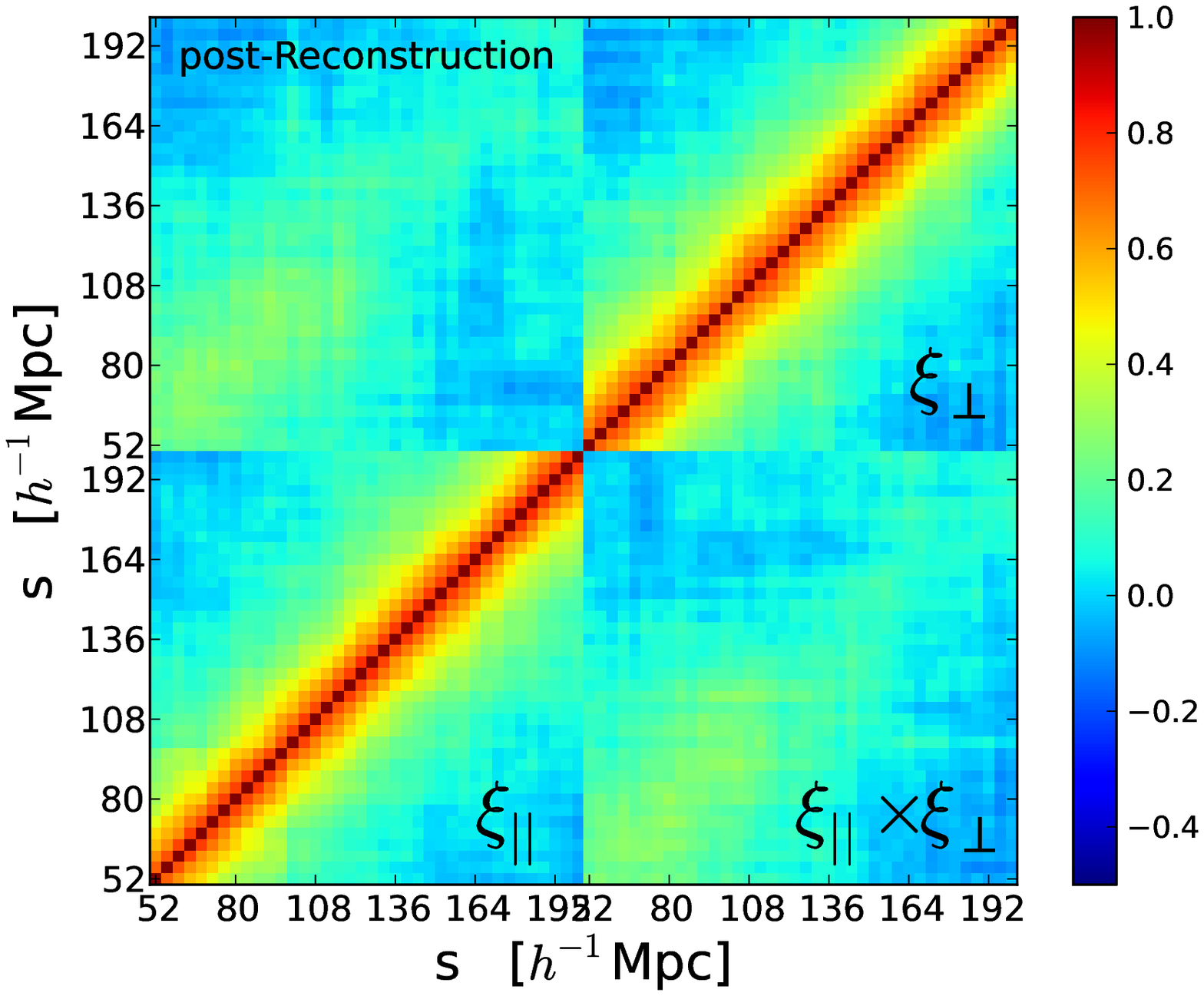}
\caption{
We use a suite of 600 PTHalo pre-reconstruction 
mock catalogs (Left) 
and post-reconstruction (Right) 
to construct the covariance matrix of the clustering wedges, 
displayed here in correlation matrix form 
$C_{i,j}/\sqrt{C_{i,i}C_{j,j}}$.  
The bottom left quadrant is that of $\xi_{||}$, 
the top right quadrant is that of $\xi_{\perp}$. 
The other quadrants, which are mirrored, are the 
cross-correlation between the bins of $\xi_{||}$ and $\xi_{\perp}$. 
}
\label{NCij_50-200_figure}  
\end{center}
\end{figure*}

Figure \ref{NCij_50-200_figure} shows  
the correlation matrix $C_{i,j}/\sqrt{C_{i,i}C_{j,j}}$ 
of $\xi_{||,\perp}$ pre- and post-reconstrucion.
The $\xi_{\perp}$ 
quartile has slightly larger (normalized) off-diagonal  
terms than in the $\xi_{||}$ quartile,  
demonstrated by the less steep gradient. 
There are also non-trivial positive and negative 
covariance cross-terms 
between the $\xi_{\perp}$ and $\xi_{||}$.  
In the reconstructed $C_{i,j}$ we notice a sharper gradient, 
and a shallower negative region, 
indicating less dominance of the off-diagonal terms. 
This means that the reconstruction procedure 
reduces the covariance between the data points. 
Examining $C_{i,j}$ of the $\xi_{0,2}$ we 
find similar trends. 


\begin{figure*}
\begin{center}
\includegraphics[width=0.49\textwidth]{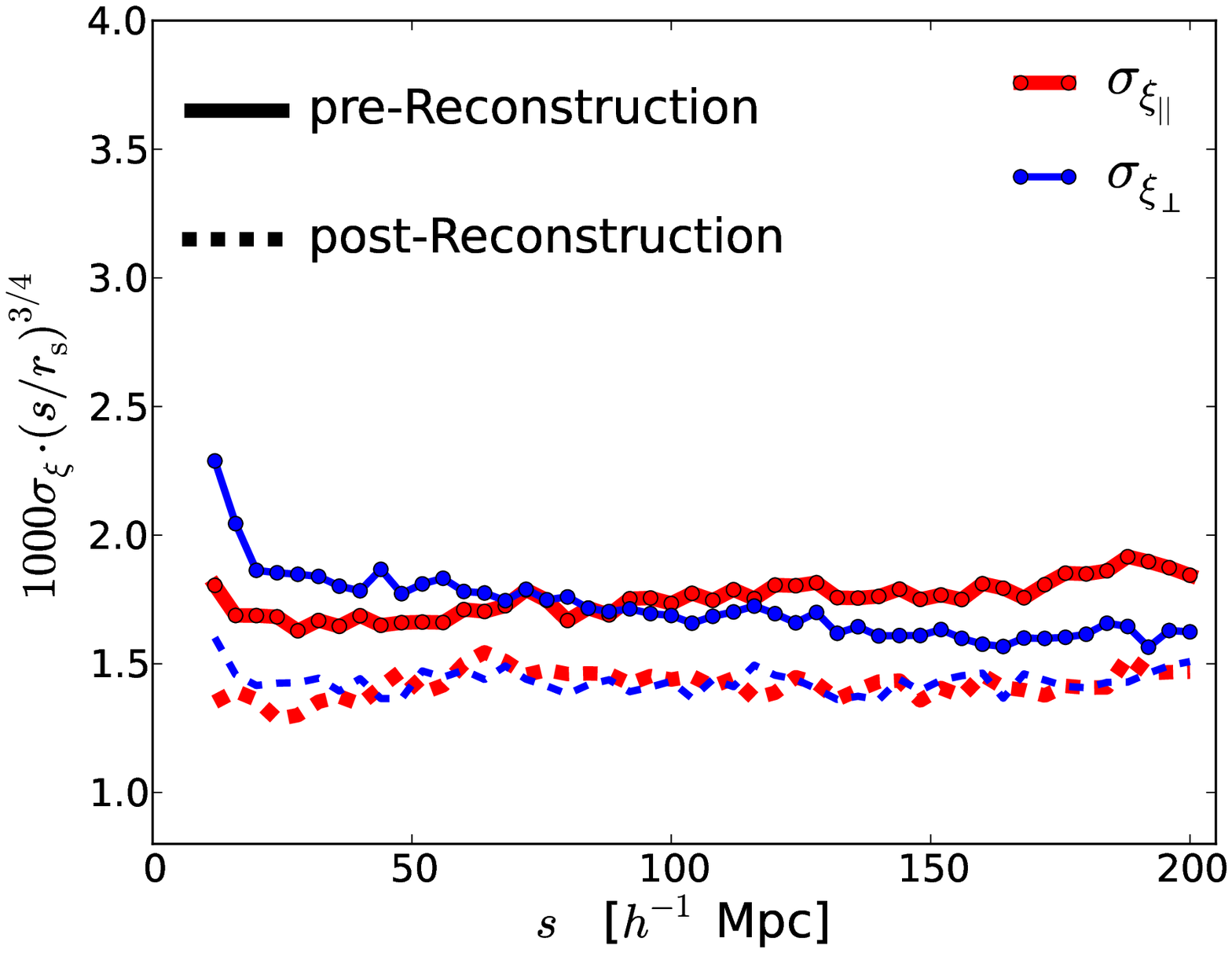}
\includegraphics[width=0.49\textwidth]{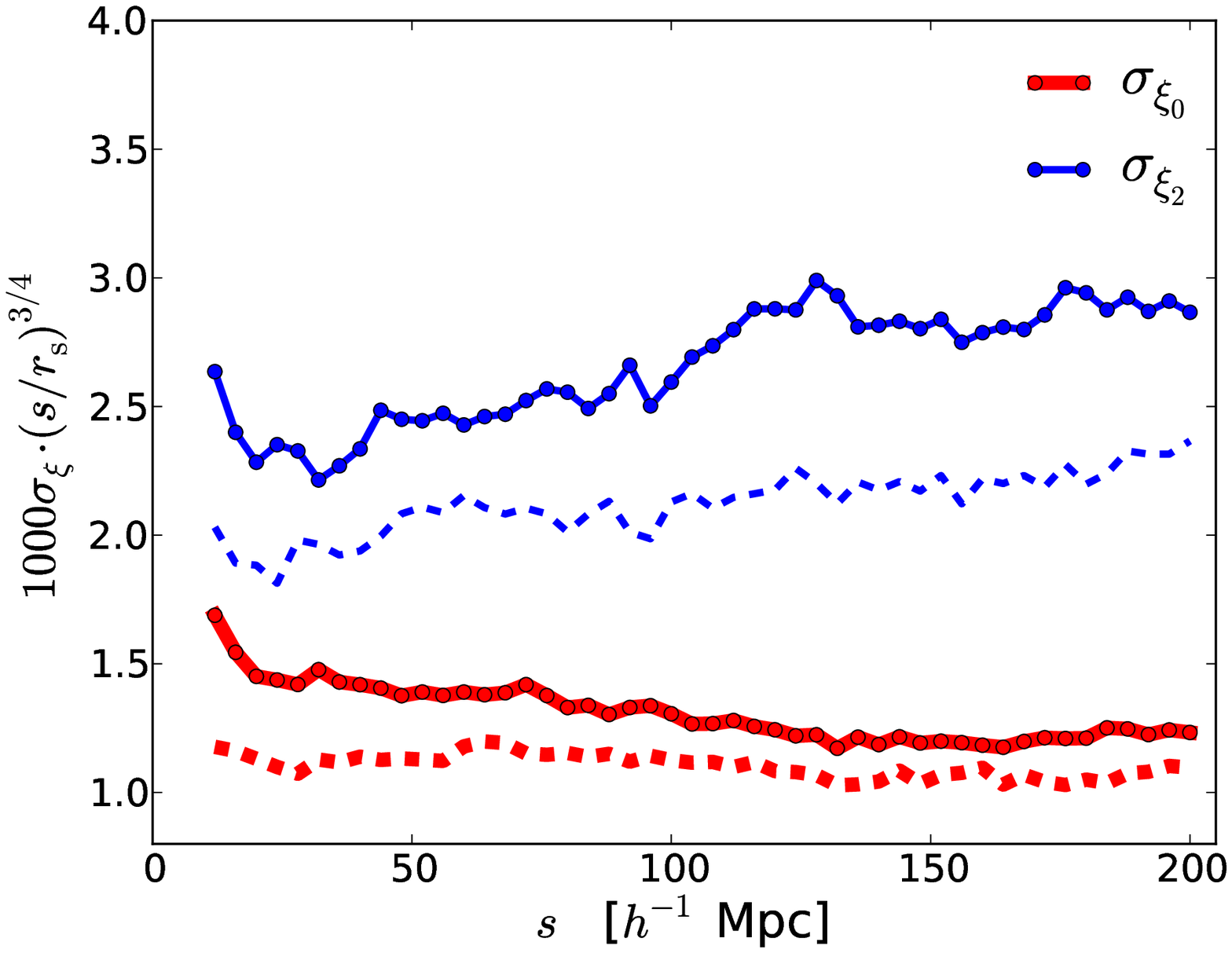}
\caption{
The $\sqrt{C_{ii}}$ values 
constructed from  
the pre- (solid) 
and post-reconstruction (dashed) mocks. 
Left plot: results of the clustering wedges 
$\xi_{||}$ (thick red), $\xi_{\perp}$ (thin blue). 
Right plot: 
similar for the $\xi_0$ (thick red), $\xi_2$ (thin blue). 
Reconstruction substantially reduces the covariance in 
these measurements. 
}
\label{sigma_50-200_figure}  
\end{center}
\end{figure*}


Figure \ref{sigma_50-200_figure} displays 
the square root of the diagonal elements. 
It is clear that pre-reconstruction the scatter 
in the two clustering wedges is slightly different, 
where post-reconstruction they are similar, 
and less than that of pre-reconstruction. 
We clearly see that $\xi_{0}$ yields 
the lowest scatter of all the $\xi$ statistics, 
and $\xi_{2}$ the highest, 
both pre- and post-reconstruction. 

To correct for 
the bias due to the finite number 
of realizations used to estimate the 
covariance matrix 
and avoid underestimation of 
the parameter constraining region,  
after inverting the matrix to $C^{-1}_{\rm original}$, 
we multiply 
it by the correction factors 
given in \cite{hartlap07a}:
\beq
C^{-1}=C^{-1}_{\rm original}\cdot\frac{(N_{\rm mocks}-N_{\rm bins}-2)}{(N_{\rm mocks}-1)}. 
\eeq
In our analysis $N_{\rm mocks}=600, N_{\rm bins}=76$ (2$\times$38) 
(when analyzing region [50,200]\hmpcii), yielding a factor 
of $0.87$. 

\subsection{Non-linear $\xi$ templates}\label{xitemplates_section}
The modeling is split 
into two parts: 
inclusion of redshift-distortions 
and modeling for non-linearities. 
Here we describe the former, 
and later consider two procedures 
for defining non-linearities. 

Once the non-linear $P_{\rm NL}$ 
is defined (see \S\ref{nonlinear_section}),  
redshift-distortions are added such that the 
non-linear $z$-space power spectrum is:
\beq\label{prerec_pkmu_eq}
P_{\rm NL}^{\rm S}(k,\mu_k)= \frac{1}{(1 + (k f \sigma_{\rm V} \mu_k)^2)^2} (1+\beta\mu_k^2)^2 P_{\rm NL}, 
\eeq
where $\beta\equiv f/b$. 

\begin{table*} 
\begin{minipage}{172mm}
\caption{Non linear anisotropic $\xi$ templates}
\label{xitemplate_tabel}
\begin{tabular}{@{}cllccc@{}}
 \hline
 Template Name                  &     Equation Base     &    Fixed parameter values                                  &Comment\\
 \hline
 $\xi^{\rm RPT-based}$ pre-rec        & (\ref{pkrpt_equation})&$k_{\rm NL}=0.19h$Mpc$^{-1}$, $A_{\rm MC}=2.44$, $\sigma_{\rm V}=5.26$\hmpcii&    \\
 $\xi^{\rm RPT-based}$ post-rec       & (\ref{pkrpt_equation})&$k_{\rm NL}=0.50h$Mpc$^{-1}$, $A_{\rm MC}=2.44$, $\sigma_{\rm V}=0$\hmpcii& $\xi_2=0$   \\
 $\xi^{\rm dewiggled}$ pre-rec  & (\ref{pkdw_equation}) &$\Sigma_{||}=11$\hmpcii, $\Sigma_{\perp}=6$\hmpcii, $\sigma_{\rm V}=1$\hmpcii &    \\
 $\xi^{\rm dewiggled}$ post-rec & (\ref{pkdw_equation}) &$\Sigma_{||}=\Sigma_{\perp}=3$\hmpcii, $\sigma_{\rm V}=1$\hmpcii & $\xi_2\ne 0$ but small   \\
 \hline
\end{tabular}

\medskip
 The RPT-based templates are shown in Figures \ref{cmass_xi1d_stats_figure} and \ref{cmass_xi1d_stats_figure_postrec}.\\  
 The dewiggled templates are shown in Figure 1 of \cite{sdss3dr9aniso}. \\
 After the base equation is calculated, Equation (\ref{prerec_pkmu_eq}) includes redshift-distortions (post-rec assumes $\beta=0$)
\end{minipage}
\end{table*}

Although the velocity dispersion parameter $\sigma_{\rm V}$ 
appears to be an unresolved subject of investigation (\citealt{taruya09a}), 
we find that applying it in the above 
Lorentzian format yields a good agreement with    
the mock-mean signals $\xi_{||,\perp}$ and $\xi_{0,2}$ 
down to $s>50$\hmpcii. 
We find the Lorentzian format is preferred over the 
popular Gaussian.  

The conversion to configuration space is accomplished by 
means of Equations (4.8) and (4.17)  
in \cite{taruya09a}. 
As described in Appendix \ref{ap_inpractice_section}, 
we apply this calculation once to obtain 
the $\xi_0$, $\xi_2$ templates, 
which are stored during the MCMC calculations. 
This approach means that we fix  
the parameters $f$, $\beta$, $\sigma_{\rm V}$ constant, 
and allow for their effective changes 
through the $a_{0 \ \rm stat}$ and $A(s)$ shape parameters, 
as described in \S\ref{shapeparams_section}, \S\ref{cmass_nuisance_section}. 
The values assumed for these parameters are summarized in Table \ref{xitemplate_tabel}.

\subsubsection{Non-linear $P(k)$}\label{nonlinear_section}

We use two anisotropic templates. 
The primary focus is on a physically motivated model based 
on Renormalized Perturbation Theory (RPT), 
which takes into account first order corrections 
of ${\bf k}$ mode coupling. 
Throughout this study we compare performance of this template 
to one that 
describes the effect of non-linearities 
in the \baf 
through the ``dewiggling" procedure. 
Both templates assume an exponential damping 
of the \baf due to large-scale coherent motions, 
where in RPT-based this is assumed to be isotropic 
and in dewiggled anisotropic.

For the RPT-based template we write: 
\beq\label{pkrpt_equation}
P_{\rm RPT}(k)= P_{\rm Linear}(k)\exp\left({-\left(\frac{k}{k_{\rm NL}}\right)^2}\right) + A_{\rm MC} P_{1\rm loop}(k), 
\eeq
where
\beq
 P_{1\rm loop}(k)=\frac{1}{4\pi^3}\int d{\bf q} | F_2({\bf k}-{\bf q},{\bf q})|^2 P_{\rm Linear}(|{\bf k}-{\bf q}|)P_{\rm Linear}({\bf q}). 
\eeq
The mode coupling term $F_2$  
is given by Equation (45) in \cite{bernardeau02a}.
Pre-reconstruction we fix $k_{\rm NL}=0.19h$Mpc$^{-1}$,  
which causes damping of the \bafii, 
and $A_{\rm MC}=2.44$ which takes into account mode coupling. 
These values are determined by analyzing the mean 
signal of the mocks whilst fixing 
\czHzrs and \Dazrs to the true values (and not using shape parameters). 
See \S 3 of \cite{sanchez13a} for a thorough discussion 
of the template and a summary earlier investigations (e.g, \citealt{crocce08,sanchez08}). 


We compare the results obtained by means of the RPT-based model to 
a popular model denoted as dewiggled, 
which also includes a Gaussian damping of the \bafii:
\begin{dmath}\label{pkdw_equation}
P_{\rm dewiggled}(k,\mu_{k})= \left( P_{\rm Linear}- P_{\rm No Wiggle}\right)\cdot  {\mathcal D}(k,\mu_{k}) + P_{\rm No Wiggle},
\end{dmath}
where the anisotropic damping is defined by:
\beq
{\mathcal D}(k,\mu_{k}) \equiv \exp{\left[-\frac{1}{2}k^2 \left(\mu_{k}^2\Sigma_{||}^2  + (1-\mu_k^2)\Sigma_{\perp}^2\right)\right]}.
\eeq
The $P_{\rm No Wiggle}(k)$ is the no-wiggle model given in \cite{eisenstein98}.
For a full description of this model, 
the reader is referred to \cite{eisenstein07b}, \cite{xu12b}, 
and \S 4.3 in \cite{sdss3dr9aniso}. 
Here we use values $\Sigma_{\perp},\Sigma_{||}=6,11$\hmpc 
for the pre-reconstruction case, 
and  
$\Sigma_{||}=\Sigma_{\perp}=3$\hmpc post-reconstruction. 

\subsubsection{Post-reconstruction templates}\label{postrec_templates_section}
Equation (\ref{prerec_pkmu_eq}) is used 
for both the RPT-based and dewiggled templates 
pre-reconstruction. 
Post-reconstruction templates  
are described in this section. 

Assuming that the reconstruction procedure works correctly, 
one expects,  
in addition to the sharpening of the \bafii, 
a correction for redshift-distortions, 
yielding an isotropic $\xi(s)$. 
We apply this approach in the RPT-based modeling. 
Due to the sharpening of the \bafii, 
we set $k_{\rm NL}=0.50h$Mpc$^{-1}$, 
which effectively yields the linear $\xi$. 
The isotropy in the post-reconstruction template  
is introduced by setting $\sigma_{\rm V},\beta=0$ 
and hence $\xi_2=0$. 

When applying reconstruction we also 
expect, ideally, no need for the coupling term. 
However, when analyzing the mocks, we find that setting 
$A_{\rm MC}$ to zero, 
yields a 
$0.7\%-1\%$ bias in \czHzrsii.  
For this reason we fix $A_{\rm MC}=2.44$ as the pre-reconstruction 
template, which produces lower bias 
($<0.5\%$ see Tables \ref{hda_dr9mock_table}, \ref{hda_highsn_table}).  

For the dewiggled post-reconstruction template  
we assume an isotropic $P_{\rm NL}$ model,  
but do include $\sigma_{\rm V}=1$\hmpc contributions, 
which are small at the \baf scale.
This is the same template used in the analysis 
of \cite{sdss3dr9aniso}. 

As clearly seen in Figure \ref{cmass_xi1d_stats_figure_postrec}, 
the reconstruction procedure, 
as applied on the PTHalos, 
yields a systematic effect in $\xi_2$, 
which reverses sign 
at scales of the \bafii, 
suggesting there might be an over-compensation 
of the Kaiser effect. 
We are not concerned by this fact, 
as we are interested in the 
peak positions to extract 
\czHzrs and \Dazrsii, and not $\beta$. 
When using each of the templates, linear and non linear systematics of 
the reconstruction procedure are 
corrected by the shape parameters 
as described in \S\ref{shapeparams_section}. 
In \S\ref{testingmethodology_section} we demonstrate that for the 
RPT-based model the post-reconstruction 
results are essentially unbiassed.

The RPT-based templates used are plotted 
in Figures \ref{cmass_xi1d_stats_figure} (pre-reconstruction) 
and \ref{cmass_xi1d_stats_figure_postrec}  (post-reconstruction).  
In the pre-reconstruction case 
we see a clear agreement with the $\xi_{\Delta\mu}$ and $\xi_{0,2}$. 
The dewiggled templates are plotted in Figure 1 of \cite{sdss3dr9aniso}.

\subsection{The model tested}\label{paramspace_section}
In this study we focus on the geometric
information \czHzrs and \Dazrs contained in $\xi$ 
in a model-independent fashion. 
This is done by focusing on the information 
contained in the anisotropic \bafii, 
and hence marginalize over the shape effects, 
in a similar approach to that used  in \cite{xu12b,xu12a} and \cite{anderson12a}. 

For each statistic analyzed  
we define a model based on a 
template 
using the following prescription:
\beq\label{model_equation}
 \xi_{\rm stat}^{\rm model}(s_{\rm f}) =  a_{0 \ \rm{stat}}\cdot \xi_{\rm stat}^{\rm AP \ template}(s_{\rm f}) + A_{\rm stat}(s_{\rm f}), 
\eeq
where ($ \xi_{\rm stat}=\xi_{||}$,$\xi_{\perp}$,$\xi_{0}$ or $\xi_{2}$).  
The \czHzrs and \Dazrs parameters are varied within $\xi_{\rm stat}^{\rm AP \ template}$ 
by application of the non-linear AP effect, 
as described in \S\ref{geometry_section} and Appendix \ref{ap_inpractice_section}. 
In Appendix \ref{linear_ap_appendix}, 
we also compare the non-linear to the linear AP shift  
and conclude 
that for DR9-CMASS the linear method underestimates constraints by 
$\sigma^{\rm linear}_{\rm X}/\sigma^{\rm non-linear}_{\rm X}\sim 0.8$, 
where $\sigma^{\rm method}_{\rm X}$ is the 1D marginalized 68CLr 
of X=$H, D_{\rm A}$.\footnote{Results from tests on the DR9-CMASS pre-reconstructed $\xi_{0,2}$.}

\subsubsection{The shape parameters}\label{shapeparams_section}



As indicated in Equation (\ref{model_equation}), 
each statistic ``stat" 
is multiplied by 
its own independent amplitude factor $a_{0 \ \rm{stat}}$. 
These factors take into account effective 
variations of $\sigma_8$, galaxy-to-matter linear bias, 
and the effective linear Kaiser boost. 

For each clustering wedge $\xi_{||,\perp}$ model 
we add three additional non-linear parameters according to: 
\beq\label{As_equation}
A_{\rm ||}(s) = \frac{a_{1 \ ||}}{s^2} + \frac{a_{2 \ ||}}{s} + a_{3 \rm \ ||},  
\eeq
\beq\label{As_equation2}
A_{\rm \perp}(s) = \frac{a_{1 \ \perp}}{s^2} + \frac{a_{2 \ \perp}}{s} + a_{3 \rm \ \perp}. 
\eeq  
When testing for the $\xi_{0,2}$ we apply a similar approach. 
These $A(s)$ terms are applied to the model only after the original template is shifted 
by the AP mapping. 
Hence, 
the parameter space used contains ten parameters: 
\begin{dmath}
\Phi_{10}=[\czHzrsii,\Dazrsii,{\mathcal S}],
\end{dmath}
where 
\begin{dmath}
{\mathcal S}=[a_{0 \ \rm stat1},a_{1 \ \rm stat1},a_{2 \ \rm stat1},a_{3 \ \rm stat1}, \\ a_{0 \ \rm stat2}, a_{1 \ \rm stat2},a_{2 \ \rm stat2},a_{3 \ \rm stat2}], 
\end{dmath}
where $a_{i \ {\rm stat}j}$ is the $i^{\rm th}$ 
shape parameter for the $j^{\rm th}$ $\xi-$statistic, 
as described in Equations (\ref{As_equation},\ref{As_equation2}). 

In our analysis 
we find, 
however, 
that $a_{0 \ \xi_2}$ 
is not well constrained 
both pre- and post-reconstruction 
(this is not the case for the rest of $a_{0 \ \rm stat}$). 
We decide to fix 
this parameter, 
and hence are left 
with a nine parameter 
space $\Phi_9$, 
when analyzing  $\xi_{[0,2]}$. 
In Appendix \ref{stackmock_section} we verify 
that the results obtained with $\xi_{0,2}$ using $\Phi_9$ yield similar 
results (modes and uncertainties) to those 
obtained with $\xi_{\Delta\mu}$ using $\Phi_{10}$  
both pre- and post-reconstruction. 
In \S\ref{cmass_nuisance_section}   
we describe degeneracies of the shape 
parameters with \czHzrs and \Dazrs constraints.

\subsubsection{Priors}\label{priors_section}
We limit \DrsDrsfid and \HrsfidHrs 
each to the region [0.5,1.5]. 
As suggested by \cite{xu12b}, 
we test the effect of applying a Gaussian prior 
on the warping parameter $\epsilon$. 
We also examine applying a flat prior on $\epsilon$. 

For most of this analysis, 
we do not 
use these priors, 
but we do examine using various $\epsilon$ prior 
values, and report a few results with flat prior $|\epsilon|\leq 0.15$. 
These priors 
are
physically motivated. 
First,  
most reasonable cosmologies  
would find $|\epsilon|>0.07$ highly improbable. 
Second, the covariance matrix is limited to some  
extent, and reliability is questionable at high deviations 
from the fiducial cosmology (e.g, see Figures 16,17 in \citealt{samushia11a}). 
Third, 
if the results yield a high $\epsilon$, 
the fiducial cosmology should be revisited. 
Overall we find CMASS results are not sensitive 
to the choice of prior. 
\section{Results}\label{results_section}
In this section, we 
determine the significance with which
the DR9-CMASS anisotropic \baf is  
detected, and compare this to simulated realizations.
We later describe the measurements of \czHzrs and \Dazrsii.  

\subsection{Significance of the detection of the anisotropic \bafii}\label{significance_section}
We generalize the standard technique 
of determining the significance of the 
detection of the \baf 
to the 2D anisotropic case by usage of the clustering wedges, 
and apply this to the DR9-CMASS and the 600 mock realizations. 

The method involves comparing the lowest $\chi^2$ result 
of a chosen physical model to a no-wiggle model.
For a no-wiggle model we use the \cite{eisenstein98} formalism 
(see their \S 4.2),
and derive monopole and quadrupole components using of Equation (\ref{xiell_equation}). 


Using this approach as a template, we run the same 
modeling and AP mapping (Equation \ref{model_equation})
with the same parameter space $\Phi_{10}$ 
as the physically motivated templates.  
In the procedure we do not attempt 
to analyze the clustering wedges separately from each other, 
i.e, we do not attempt 
to quantify significance of detection of the \baf 
only in $\xi_{||}$ or $\xi_{\perp}$. 
Instead, we quantify the significance of the 
detection of the anisotropic \baf in the 
$\xi(\bf s)$ by using both $\xi_{\Delta\mu}$. 
This is due to the co-variance between the clustering wedges, 
as well as the strong correlation between 
\HrsfidHrs and \DrsDrsfidii. 
All the following results are similar when 
using the RPT-based or the dewiggled template. 

\begin{figure*}
\begin{center}
\includegraphics[width=0.49\textwidth]{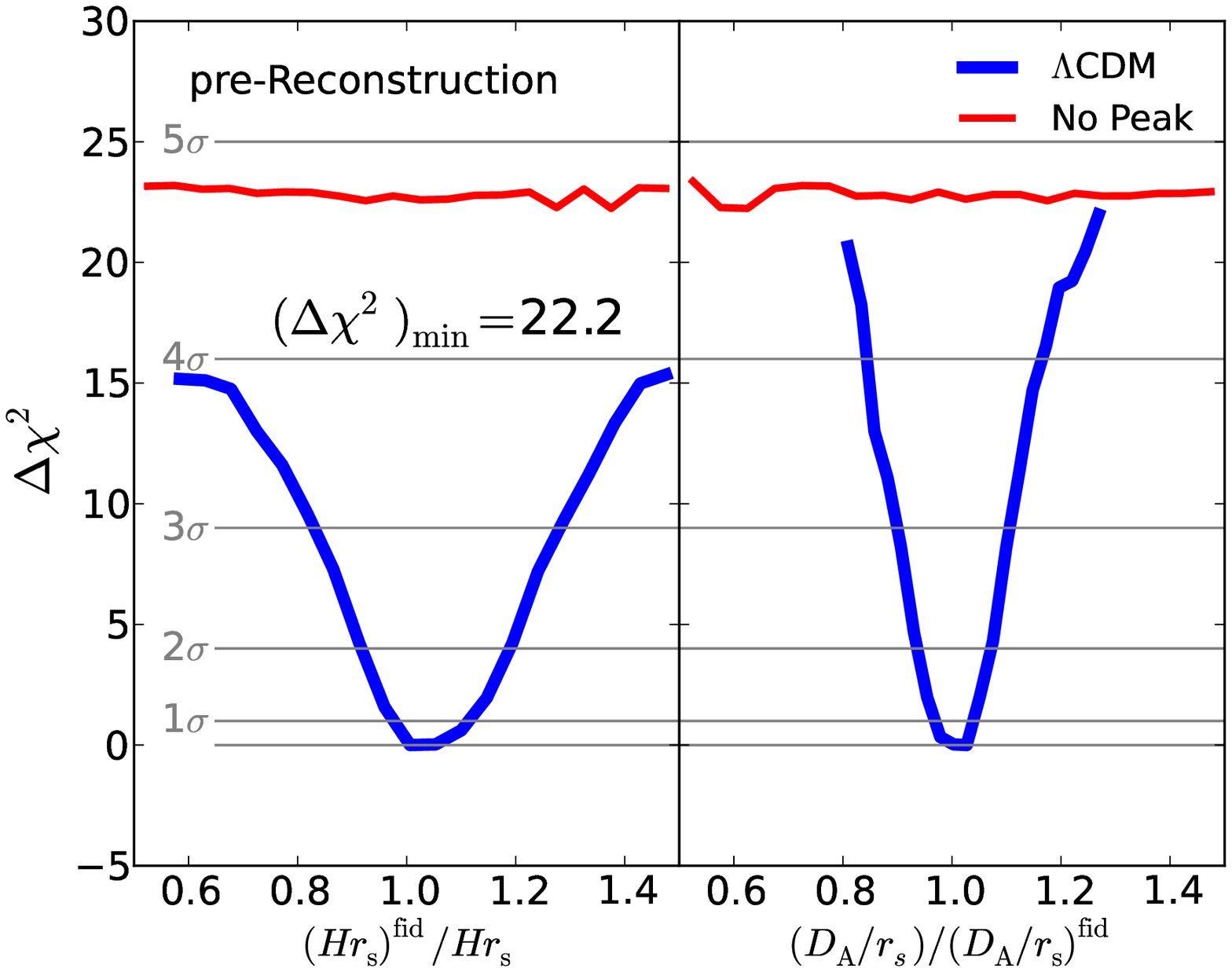}
\includegraphics[width=0.49\textwidth]{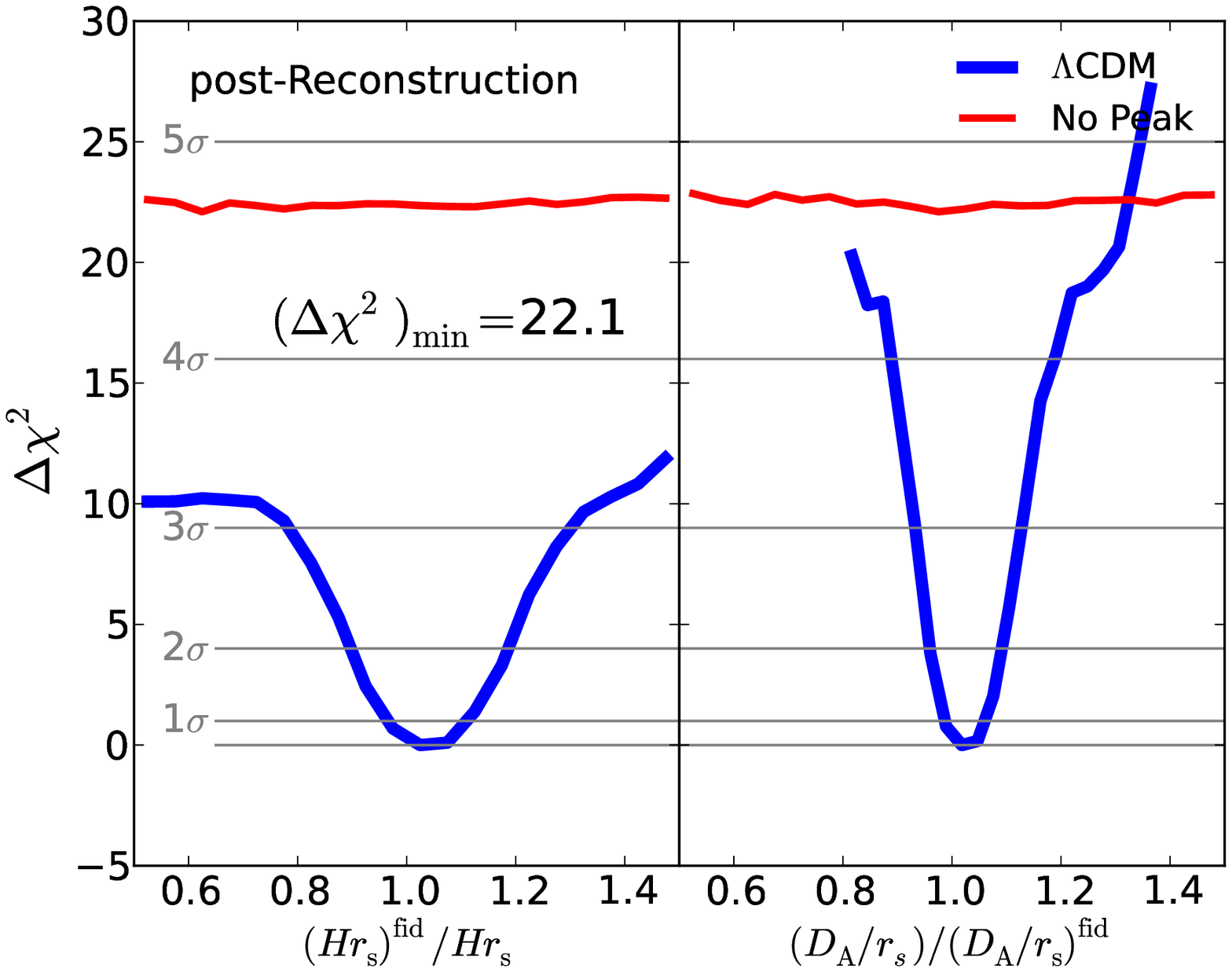}
\includegraphics[width=0.49\textwidth]{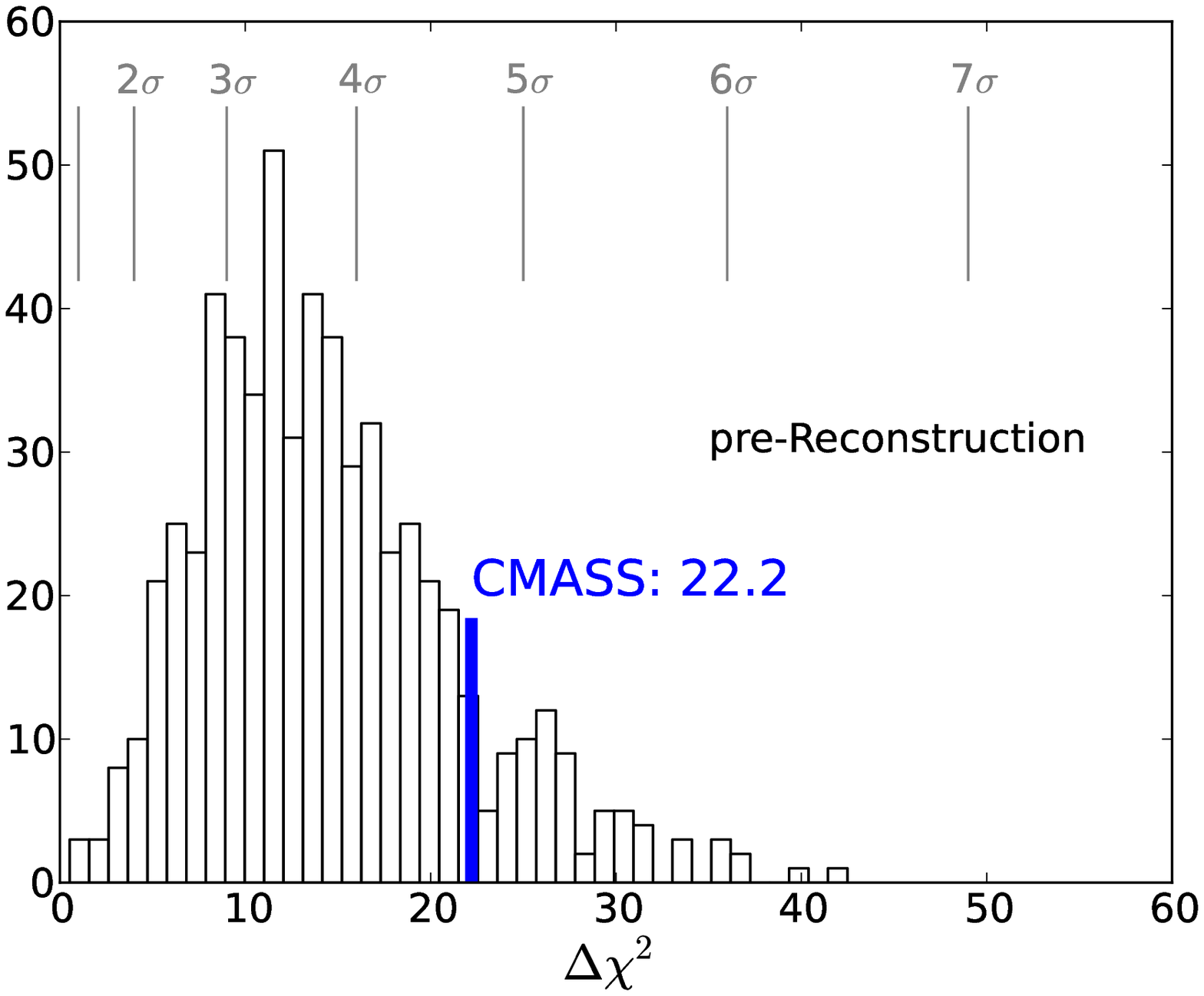}
\includegraphics[width=0.49\textwidth]{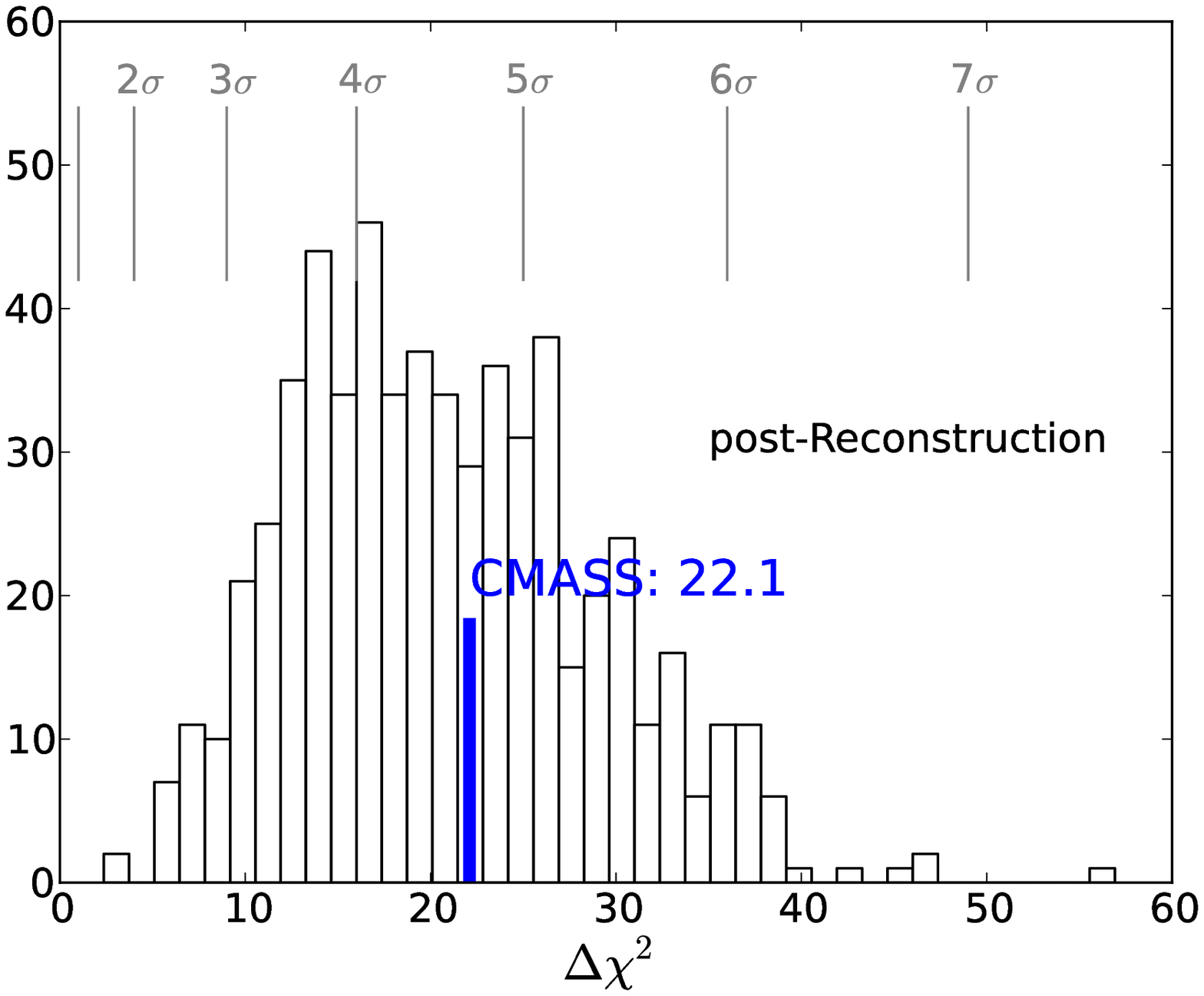}
\caption{
In the top plots we examine the significance of the detection of the anisotropic \baf 
in the CMASS clustering wedges by comparing 
$\chi^2$ results of two templates:  
a physical $\Lambda$CDM template (thick blue) and one with no baryonic peak (thin red). 
In each panel in the plots 
we display the the minimum $\chi^2$ surface for the marginalized \HrsfidHrs (left), 
and \DrsDrsfid (right). 
The reference $\chi^2$ from which each binned result is  
compared to is that of the best fit of the $\Lambda$CDM model. 
The left plots correspond to the data pre-reconstruction 
and the right to post-reconstruction. 
In CMASS we find the significance of the detection of 
the anisotropic \baf to be 
$\sqrt{\Delta\chi^2_{\rm min}}=4.7 \ \sigma$ for both 
the pre- and post-reconstruction cases. 
In the bottom plots 
we run the same procedure on $600$ mock catalogs and 
present the histogram of the distribution while indicating 
the CMASS result. 
}
\label{significance_plots}  
\end{center}
\end{figure*}

We apply this procedure on both the CMASS and the mock catalogs.
The results are summarized in Figure \ref{significance_plots}, 
where the left panels correspond to pre-, 
and the right post-reconstruction. 
The top two panels correspond to the CMASS $\Delta \chi^2 \equiv \chi^2_{\rm ref}-\chi^2$ 
results as a function of 
\HrsfidHrs and \DrsDrsfidii. 
The thick blue lines show the minimum $\chi^2$ surface 
of the RPT-based model compared to its minimum $\chi^2_{\rm ref}$. 
The thin red line corresponds to the no-wiggle (no-peak) $\chi^2$ surface minimum 
model compared with $\chi^2_{\rm ref}$. 
The bottom two panels are histogram 
results of the mock realizations, 
where the CMASS results are indicated with the thick vertical line. 
No priors on $\epsilon$ have been applied. 

The pre-reconstruction CMASS 
clustering wedges yield a result of 
$(\Delta \chi^2)_{\rm min}\equiv {\rm min}(\chi^2_{\rm ref}-\chi^2)=22.2$, 
meaning a 4.7$\sigma$ detection of the anisotropic \bafii, 
and we obtain a similar result after applying reconstruction. 
This result appears to be consistent with 
the isotropic \baf detection of CMASS-DR9 
as reported by \cite{anderson12a},  
who showed a $5\sigma$ detection that did not 
improve with reconstruction. 

In the pre-reconstruction case, 
the CMASS sample appears to be on the fortunate side of the mock distribution 
of the detection of the anisotropic \bafii, where 
$68\%$ of the mocks lie between 
$2.8-4.6 \ \sigma$. 
In the post-reconstruction case 
we see a clear shift of the mocks between $3.6-5.4 \ \sigma$. 

\begin{figure}
\begin{center}
\includegraphics[width=0.49\textwidth]{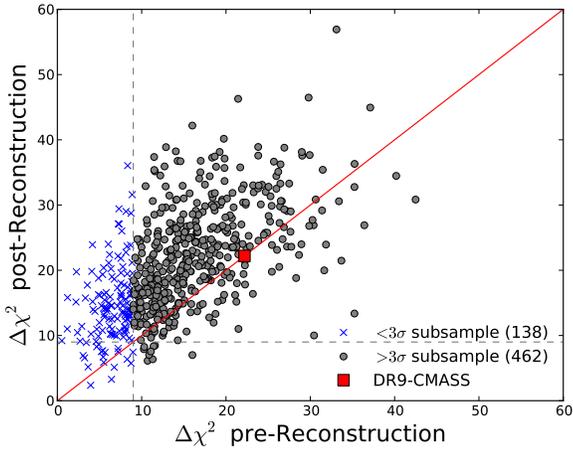}
\caption{
Here we show a quantification 
of the detection of the 
anisotropic \baf for all 600 mocks, 
and the data both pre- and post-reconstruction. 
We define the $\geq 3\sigma$ subsample 
as 462 realizations 
for which the anisotropic \baf is 
detected with at least $\Delta\chi^2=9$  
in the pre-reconstruction case (grey circles). 
The complementary are defined 
as the $< 3\sigma$ subsample (blue x's).
}
\label{subsample_by_significance_plots}  
\end{center}
\end{figure}

For later reference we  define 
a subsample of 462  
realisations with a $\geq 3\sigma$ detection 
as the ``$\geq3\sigma$ subsample", 
and its complement the  ``$<3\sigma$" subsample. 
For a consistent comparison between the various methods 
this subsample is defined when using the 
pre-reconstruction wedges RPT-based method. 
In the context of the DR9-CMASS volume, 
we find this separation useful for interpretation 
of the \HrsfidHrs and \DrsDrsfid results. 
For a visual of the subsamples in 
terms of $\Delta\chi^2$, 
please see Figure \ref{subsample_by_significance_plots}. 

In the following section 
we analyze how well we expect to measure 
\HrsfidHrs and \DrsDrsfid 
both pre- and post-reconstruction.

\subsection{Measuring $H$, $D_{\rm A}$: testing methodology on mocks}\label{testingmethodology_section}
To test the various assumptions made throughout 
the analysis 
we first apply the pipeline to our mock catalogs. 
To differentiate between systematic effects 
and peculiarities due to mocks with low \baf signal, 
in Appendix \ref{stackmock_section} we investigate high 
S/N mocks to answer the following questions (answers based on results in Table \ref{hda_highsn_table}): \\ \\
{\it Does the method outlined in \S\ref{paramspace_section} affect the AP test?\\}
The RPT-based result entries show that 
the marginalization 
over the shape information 
yields small biases ($<0.5\%$) in the geometric information measured. \\ \\
{\it Is one $\xi$ template preferred over the other?} \\
We find that although the RPT-based and dewiggled templates 
yield similar constraints and strong mode correlations, 
the dewiggled one yields a $\sim 1\%$ bias in measuring \HzfidHz
(Appendix \ref{rptdiwiggled_section}). 
The dewiggled template does not 
have a mode coupling term, 
which might explain tendencies to yield more biased mock 
results than the RPT-based template.
In \S\ref{cmass_nuisance_section} we report 
results with varied $A_{\rm MC}$, but
defer a more intensive investigation 
of possible effects 
for future studies (e.g. the \citealt{taruya10a} model).\\ \\
{\it Is one $\xi$ combination preferred over the other?} \\
We find that $\xi_{\Delta\mu}$ and $\xi_{\ell}$ 
contain similar constraining power (Appendix \ref{wedgesmultipoles_section}). \\ \\ 
{\it Are the resulting distributions of the \HrsfidHrs and \DrsDrsfid Gaussian?} \\
We find that results of high S/N mocks 
yield close to Gaussian (or symmetric) 
posterior distributions but DR9-volume mocks do not. 
This result is probably due to the fact 
that the DR9 mock volumes 
contain a large fraction of 
mocks with low S/N anisotropic \bafii. \\ \\
{\it Does reconstruction improve/bias the above?}\\
We find that the reconstructed RPT-based template 
yields a good description of the PTHalo mocks  
and, on average, improves constraints of \czHzrs and \Dazrs by $\sim 30\%$ (Appendix \ref{recimprove_section}). \\

These tests show that the methods applied 
work well on high S/N mocks. 
Analyzing 600 PTHalo DR9-volumes, we find that a non-negligible 
amount of realizations 
yield low anisotropic \baf signals. 

\begin{figure*}
\begin{center}
\includegraphics[width=0.49\textwidth]{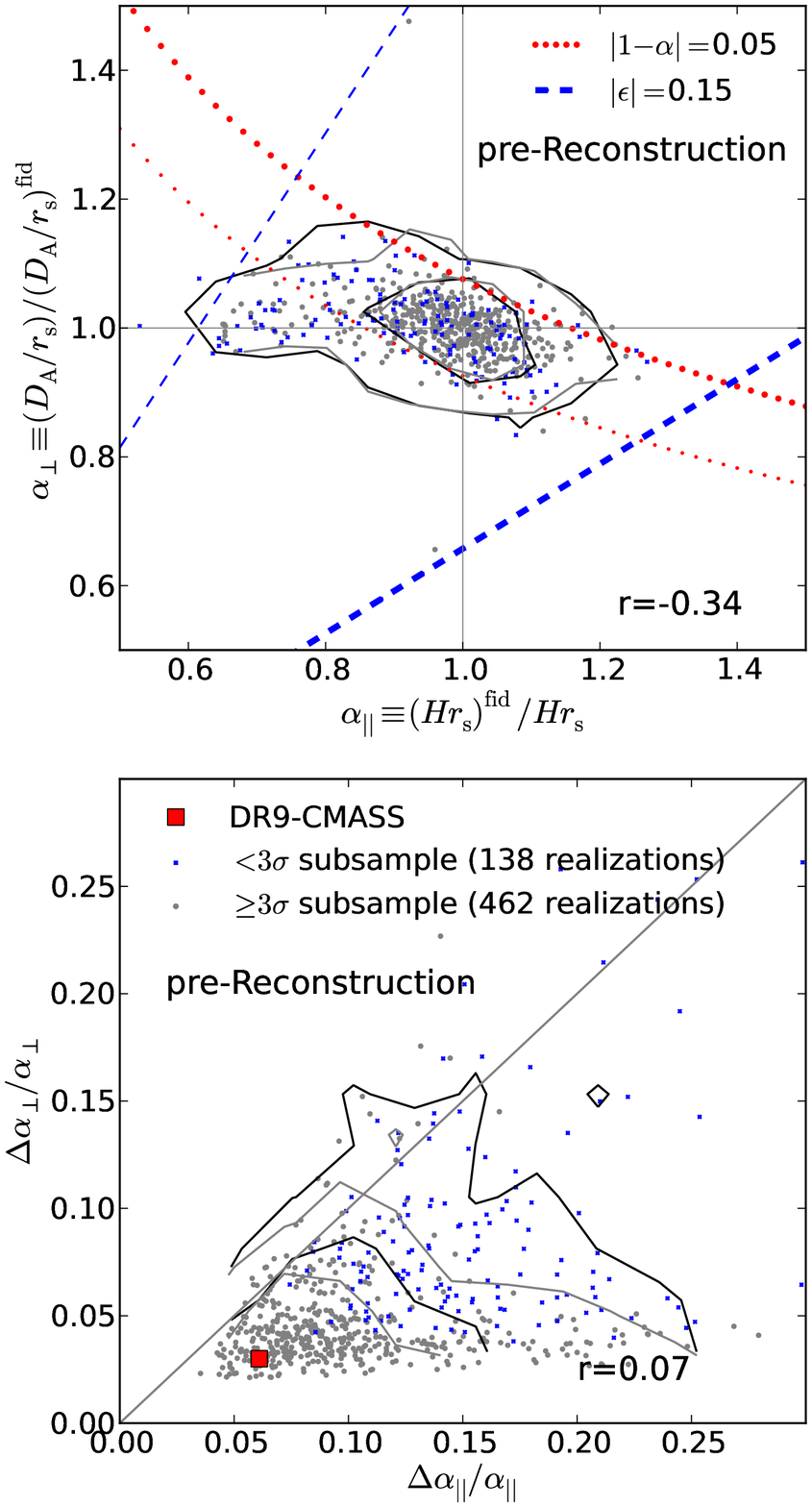}
\includegraphics[width=0.49\textwidth]{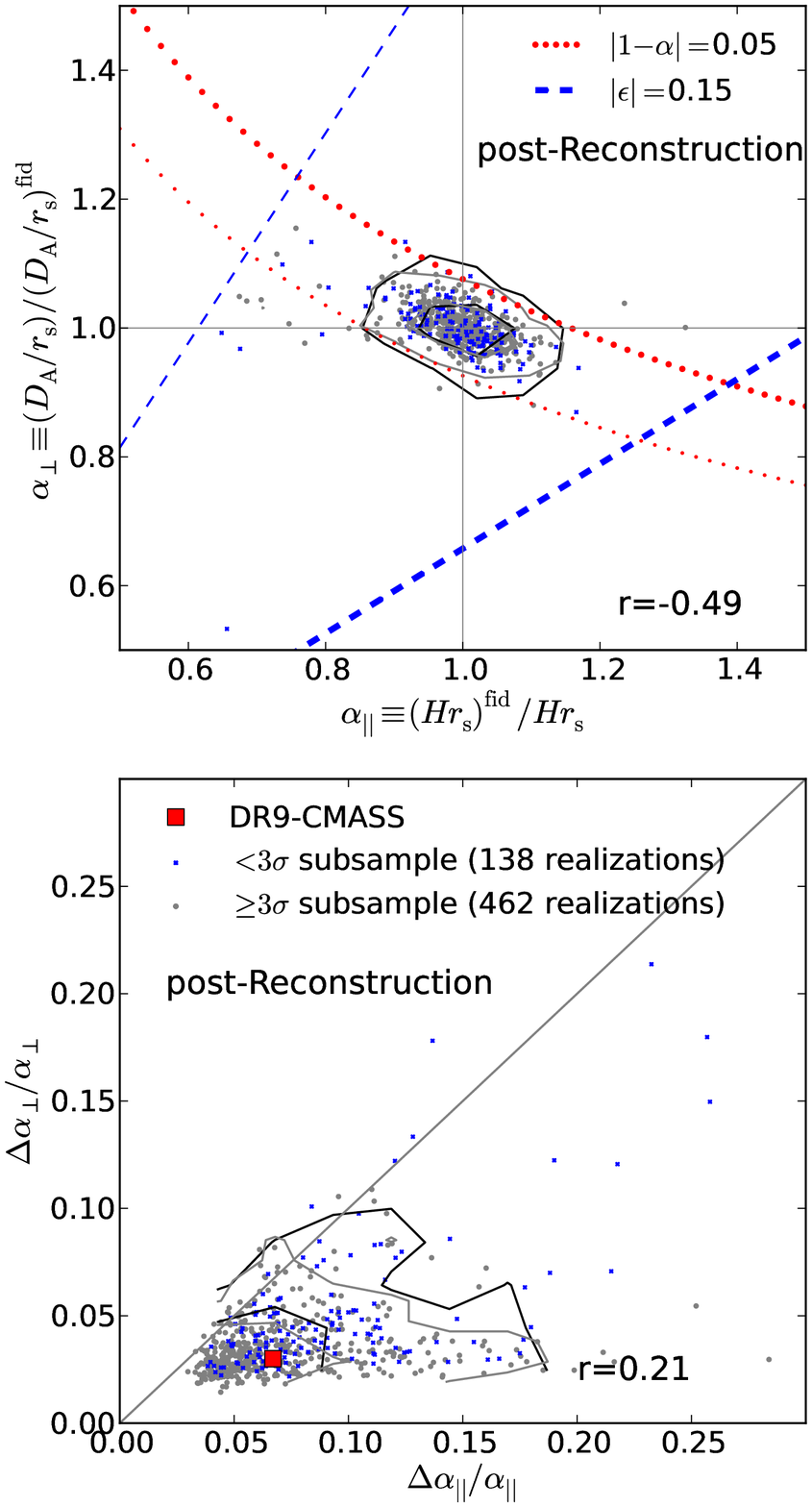} 
\caption{
Pre-(Left) and post-reconstruction (Right) distributions of 
$\alpha_{||}=$\HrsfidHrs and $\alpha_{\perp}=$\DrsDrsfid modes and their uncertainties 
of the mock PTHalos 
using the RPT-based $\xi_{||},\xi_{\perp}$.  
The top panels show the scatter of mode measurements; 
the bottom presents the scatter of uncertainties.  
Each panel presents the results of all 600 mock 
realizations, where the grey dots are  
the $\geq 3\sigma$ subsample (462 realizations), 
and blue for the complementary $<3\sigma$ subsample. 
The solid contours in each panel are the $68,95\%$ CL regions  
for the $\geq 3\sigma$ subsample (gray) 
and the full sample (black). 
The cross-correlation coefficient for the $\geq 3\sigma$ subsample
in each panel is indicated by $r$. 
Numerical results are summarized in Table \ref{hda_dr9mock_table}. 
In the top panels we emphasize the constant $\alpha$ 
and $\epsilon$ lines, as indicated (where the thicker line of each indicates the larger value). 
In the bottom panels we mark the DR9-CMASS uncertainty measurement (red filled squares). 
For plotting purposes we apply a prior of $|\epsilon|<0.15$. 
}
\label{scatterplots_valunc_prepostrec}  
\end{center}
\end{figure*}

Figure \ref{scatterplots_valunc_prepostrec} shows 
correlations between \HzfidHz and \DazDazfid modes (top) 
and their uncertainties (bottom). 
As explained in Appendix \ref{stackmock_section} (and apparent in Figure \ref{ks_plots}), 
these distributions 
of $\Delta(1/H)/(1/H)$ and  $\Delta  D_{\rm A}/D_{\rm A}$ are not Gaussian. 
A visual inspection of various individual mocks 
reveals some cases  
with weak line-of-sight and/or 
transverse baryonic acoustic features. 
This is quantified in \S\ref{significance_section}, 
where we find that $\sim 23\%$ of the realizations 
have  an anisotropic baryonic feature 
with a significance of less than $3\sigma$. 
For this reason, 
we separate the results 
to the $\geq 3\sigma$ subsample (gray points) 
and its complementary $<3\sigma$ subsample (blue points) 
Note that in both pre- and post-reconstruction 
the subsamples are the same as that in 
pre-reconstruction 
(for a visual see Figure \ref{subsample_by_significance_plots}). 
This separation points 
to interesting trends in  
\HzfidHz and \DazDazfid modes and uncertainties.

Most of the outliers that 
measure \HzfidHz and \DazDazfid modes at $>10\%$ 
from the true values   
tend to be from 
the  $<3\sigma$ subsample 
in both pre- 
and post-reconstruction.  
The plot clearly shows that reconstruction substantially improves 
both mode and uncertainty scatters and constraints.

The uncertainty-uncertainty plots  
also show 
that most of the extremely  
large uncertainties are in the 
$<3\sigma$  subsample.
Although post-reconstruction 
removes 
the trend differences in the uncertainties, 
we clearly see that the tightest constraints 
are on the $\geq 3\sigma$ subsample. 

For clarity of the plots  
and interpretation of results, 
we have applied a  
$|\epsilon|<0.15$ prior on the 
MCMC propositions. 
In the mode-mode plots this limit 
is shown by the dashed lines. 
The motivation behind this choice is given 
in \S\ref{priors_section}. 
Without this prior, 
we find a systematic ``pile-up" on the 
flat prior limit of 
\HzfidHzii=0.5, 
which is dominated by the 
$<3\sigma$  subsample.
We verify that these mocks 
have line-of-sight baryonic acoustic 
features that are either washed out, 
or contain a $\xi_{||}$ with a spurious strong 
clustering measurement at $110<s<200$\hmpcii.
For some of the ``double-mode" realizations 
(meaning with both at line-of-sight \baf signal 
and a spurious strong feature) the 
$\epsilon$ prior strengthens 
the true mode. For realizations 
with strong spurious features 
the $\epsilon$ prior causes them 
to move from $\HzfidHzii=0.5$ 
closer to the $\epsilon=-0.15$ boundary. 

All the above trends appear 
in both templates examined (RPT-based, dewiggled), 
and in both clustering wedges and multipoles. 

In Table  \ref{hda_dr9mock_table} we summarize the mock results 
of \HzfidHz and \DazDazfid modes and uncertainties 
and their scatter. 
Most entries are for the RPT-based clustering wedges pre- and post-reconstruction.  
For completeness, 
the first and last entries include the dewiggled templates 
as well as including results of multipoles in all templates. 
The sample examined is indicated (e.g., full sample or the $\geq 3\sigma$ subsample) 
as well as 
if a prior on $\epsilon$ is used. 
For example, we investigate various $|\epsilon|$ priors, 
or restricting to realizations 
with \HzfidHz and \DazDazfid 
modes within $14\%$ from the true values, or both. 

\begin{table*} 
\begin{minipage}{172mm}
\caption{Mock DR9 PTHalo results}
\label{hda_dr9mock_table}
\begin{tabular}{@{}ccccc@{}}
\hline\hline
$\xi$ ($\#$ of realizations) & $\alpha_{||}$ & $\Delta \alpha_{||}$/$\alpha_{||}$ &$\alpha_{\perp}$ &  $\Delta \alpha_{\perp}$/$\alpha_{\perp}$ \\
\hline
{\bf Full sample, no priors:}\\ 
RPT-based wedges pre-Rec (600)&$ 0.970\pm 0.188$&$ 0.132\pm 0.075$&$ 1.006\pm 0.087$&$ 0.048\pm 0.061$  \\
RPT-based wedges post-Rec (600)&$ 0.997\pm 0.101$&$ 0.068\pm 0.065$&$ 1.000\pm 0.042$&$ 0.034\pm 0.033$  \\
RPT-based multipoles pre-Rec (600)&$ 0.986\pm 0.194$&$ 0.102\pm 0.076$&$ 1.001\pm 0.082$&$ 0.050\pm 0.053$  \\
RPT-based multipoles post-Rec (600)&$ 0.992\pm 0.176$&$ 0.077\pm 0.083$&$ 1.002\pm 0.052$&$ 0.037\pm 0.024$  \\ \\

dewiggled wedges pre-Rec (600)&$ 0.983\pm 0.190$&$ 0.129\pm 0.075$&$ 1.014\pm 0.086$&$ 0.047\pm 0.060$  \\
dewiggled wedges post-Rec (600)&$ 0.999\pm 0.106$&$ 0.065\pm 0.067$&$ 1.002\pm 0.047$&$ 0.033\pm 0.036$  \\
dewiggled multipoles pre-Rec (600)&$ 0.990\pm 0.183$&$ 0.100\pm 0.072$&$ 1.008\pm 0.086$&$ 0.049\pm 0.047$  \\
dewiggled multipoles post-Rec (600)&$ 1.002\pm 0.134$&$ 0.056\pm 0.077$&$ 1.000\pm 0.045$&$ 0.030\pm 0.025$  \\
\hline
{\bf Full sample, $|\epsilon<0.15|$:} \\
RPT-based wedges pre-Rec (600)&$ 0.982\pm 0.114$&$ 0.098\pm 0.049$&$ 1.002\pm 0.050$&$ 0.044\pm 0.034$  \\
RPT-based wedges post-Rec (600)&$ 0.998\pm 0.067$&$ 0.064\pm 0.038$&$ 0.999\pm 0.038$&$ 0.033\pm 0.020$  \\

\\{\bf $\geq 3\sigma$  subsample, no priors:} \\
RPT-based wedges pre-Rec (462)&$ 0.983\pm 0.146$&$ 0.103\pm 0.065$&$ 1.003\pm 0.064$&$ 0.042\pm 0.044$  \\
RPT-based wedges post-Rec (462)&$ 0.997\pm 0.086$&$ 0.061\pm 0.061$&$ 1.000\pm 0.034$&$ 0.032\pm 0.026$  \\

\\{\bf $\geq 4.0\sigma$  subsample, no priors:} \\
RPT-based wedges pre-Rec (208)&$ 0.990\pm 0.111$&$ 0.074\pm 0.054$&$ 1.001\pm 0.043$&$ 0.034\pm 0.027$  \\
RPT-based wedges post-Rec (208)&$ 0.996\pm 0.079$&$ 0.053\pm 0.051$&$ 0.999\pm 0.030$&$ 0.028\pm 0.020$  \\

\\{\bf $\geq 4.5\sigma$  subsample, no priors:} \\
RPT-based wedges pre-Rec (104)&$ 0.992\pm 0.099$&$ 0.065\pm 0.050$&$ 1.000\pm 0.035$&$ 0.031\pm 0.021$  \\
RPT-based wedges post-Rec (104)&$ 0.999\pm 0.045$&$ 0.052\pm 0.040$&$ 0.997\pm 0.024$&$ 0.028\pm 0.009$  \\

\\{\bf $\geq 3\sigma$  subsample, $|\epsilon<0.15|$:} \\
RPT-based wedges pre-Rec (462)&$ 0.988\pm 0.102$&$ 0.089\pm 0.042$&$ 1.001\pm 0.048$&$ 0.040\pm 0.023$  \\
RPT-based wedges post-Rec (462)&$ 0.997\pm 0.061$&$ 0.059\pm 0.035$&$ 0.999\pm 0.031$&$ 0.031\pm 0.014$  \\

\hline
{\bf $\geq 3\sigma$  subsample, $|\epsilon<0.15|$, $|1-{\rm mode}|<0.14$:} \\
RPT-based wedges pre-Rec (394)&$ 0.996\pm 0.060$&$ 0.087\pm 0.035$&$ 1.000\pm 0.037$&$ 0.040\pm 0.021$  \\
RPT-based wedges post-Rec (450)&$ 0.998\pm 0.046$&$ 0.059\pm 0.032$&$ 0.999\pm 0.029$&$ 0.031\pm 0.014$  \\
RPT-based multipoles pre-Rec (374)&$ 0.999\pm 0.061$&$ 0.079\pm 0.030$&$ 0.997\pm 0.038$&$ 0.044\pm 0.011$  \\
RPT-based multipoles post-Rec (434)&$ 1.001\pm 0.047$&$ 0.062\pm 0.030$&$ 0.998\pm 0.031$&$ 0.033\pm 0.011$  \\ \\
dewiggled wedges pre-Rec (392)&$ 1.003\pm 0.063$&$ 0.087\pm 0.036$&$ 1.009\pm 0.038$&$ 0.039\pm 0.019$  \\
dewiggled wedges post-Rec (445)&$ 1.003\pm 0.047$&$ 0.056\pm 0.034$&$ 1.000\pm 0.031$&$ 0.030\pm 0.016$  \\
dewiggled multipoles pre-Rec (371)&$ 1.003\pm 0.061$&$ 0.077\pm 0.030$&$ 1.006\pm 0.037$&$ 0.042\pm 0.019$  \\
dewiggled multipoles post-Rec (434)&$ 1.006\pm 0.043$&$ 0.051\pm 0.033$&$ 0.999\pm 0.028$&$ 0.028\pm 0.007$  \\
\hline\hline
\end{tabular}

\medskip
 * The $\alpha_{||}$ and $\alpha_{\perp}$ columns show the median and rms of the modes. \\
 * The $\Delta \alpha_{||}/\alpha_{||}$ and $\Delta \alpha_{\perp}$/$\alpha_{\perp}$ columns show the median and rms of the fractional uncertainties.
\end{minipage}
\end{table*}

Regarding the post-reconstruction RPT-based $\xi_{||,\perp}$ 
we notice that 
\HzfidHz
has in all cases a median mode bias of $\leq0.3\%$, 
and 
\DazDazfid $\leq 0.1\%$. 
Pre-reconstruction mode results, 
on the other hand, 
improve substantially when 
applying the various priors and cuts 
($\geq 3\sigma$  subsample, $|\epsilon|<0.15$,
mode limitation). 
These results show the effects of 
mocks with low anisotropic \baf signal. 
For example 
when limiting the sample 
to the most constrained 2/3 of the realizations (meaning 394/600),  
the bias on \HzfidHz improves from $3\%$ to $0.4\%$,  
and of \DazDazfid from $0.6\%$ to $\leq 0.1\%$.  

The \HzfidHz and \DazDazfid uncertainties improve in different 
manners when applying the various 
priors and cuts. 
The most noticeable trend, 
which is common for both parameter results, 
is the reduction of the scatter on the 
uncertainty when 
applying the $|\epsilon|<0.15$ prior.


For ill-constrained DR9 volumes 
the median uncertainties vary with choice of $\epsilon$. 
On the other hand, for well-constrained realizations, such as CMASS-DR9, 
results  do not depend on the $\epsilon$ prior (see \S\ref{cmass_results_section}). 

We also find that the dewiggled pre-reconstruction 
template yields similar 
\HzfidHz and \DazDazfid constraints as the RPT-based ones, 
although the dewiggled pre-reconstruction template shows a systematic bias of $\sim 1\%$ on \DazDazfidii. 
This effect is not apparent in the high S/N mocks (Appendix \ref{wedgesmultipoles_section}), 
which yield a median $1.4\%$ bias on \HzfidHzii, which is not apparent here. 
The post-reconstruction dewiggled wedges results 
are in line with the RPT-based. 

Perhaps the most notable feature in Table \ref{hda_dr9mock_table} 
is that the scatter in the \HzfidHz modes  
is different from the median of the uncertainties. 
Focusing on the most constrained subsample (the bottom entry), 
we see that the scatter in the \HzfidHz modes 
is smaller than the median of the uncertainties in all cases. 
In \S\ref{dr12_forecast_section} and Appendix \ref{stackmock_section}, 
we show that this should improve with higher S/N samples. 
For \DazDazfid we see that the scatter of the modes 
and median of the uncertainties are fairly similar.

The {\it fiducial} cosmology 
of the analyses 
is the 
{\it true} cosmology of the mocks. 
We defer testing possible effects of using an incorrect 
fiducial cosmology (for preliminary tests on mocks see \citealt{kazin11a}).

To summarize, 
we find that a significant minority of DR9-CMASS pre-reconstruction realizations 
yield unreliable results. However, 
the majority 
$>3\sigma$ subsample  
yields a low bias result ($<0.5\%$). 
Moreover, the results show that the 
post-reconstruction wedges results yield 
low bias ($<0.3\%$) with both RPT-based and dewiggled.
We also find that for a DR9 volume we expect non-Gaussian 
likelihood profiles of \czHzrs and \Dazrs in both pre- and post-reconstruction.
We next turn to apply the same method 
used here on the data both pre- and post-reconstruction.

\subsection{DR9-CMASS $H$, $D_{\rm A}$ results}\label{cmass_results_section}
In this section we present our measurements 
of \czHzrs and \Dazrs in the DR9 CMASS dataset.

\begin{figure*}
\begin{center}
\includegraphics[width=0.49\textwidth]{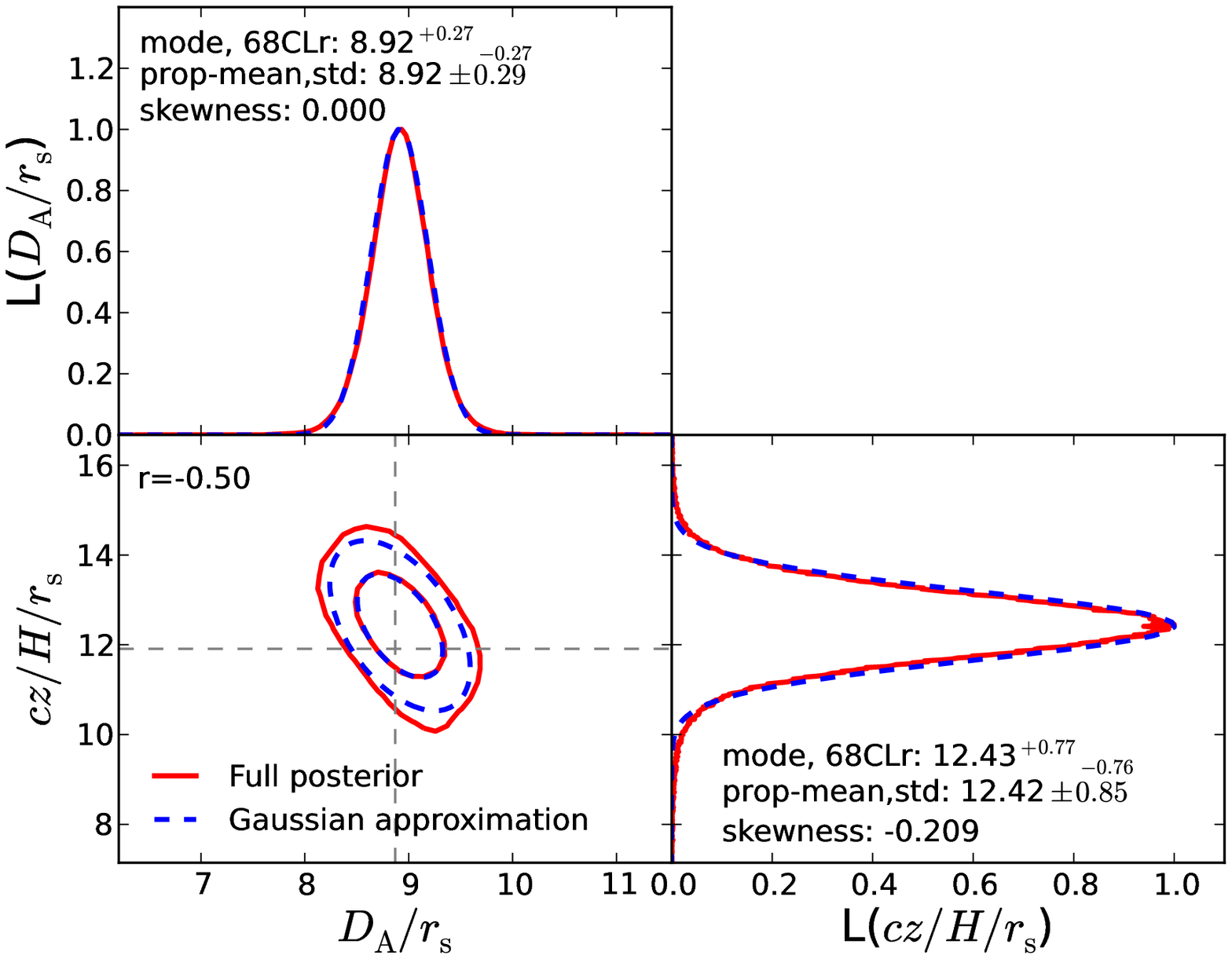} 
\includegraphics[width=0.49\textwidth]{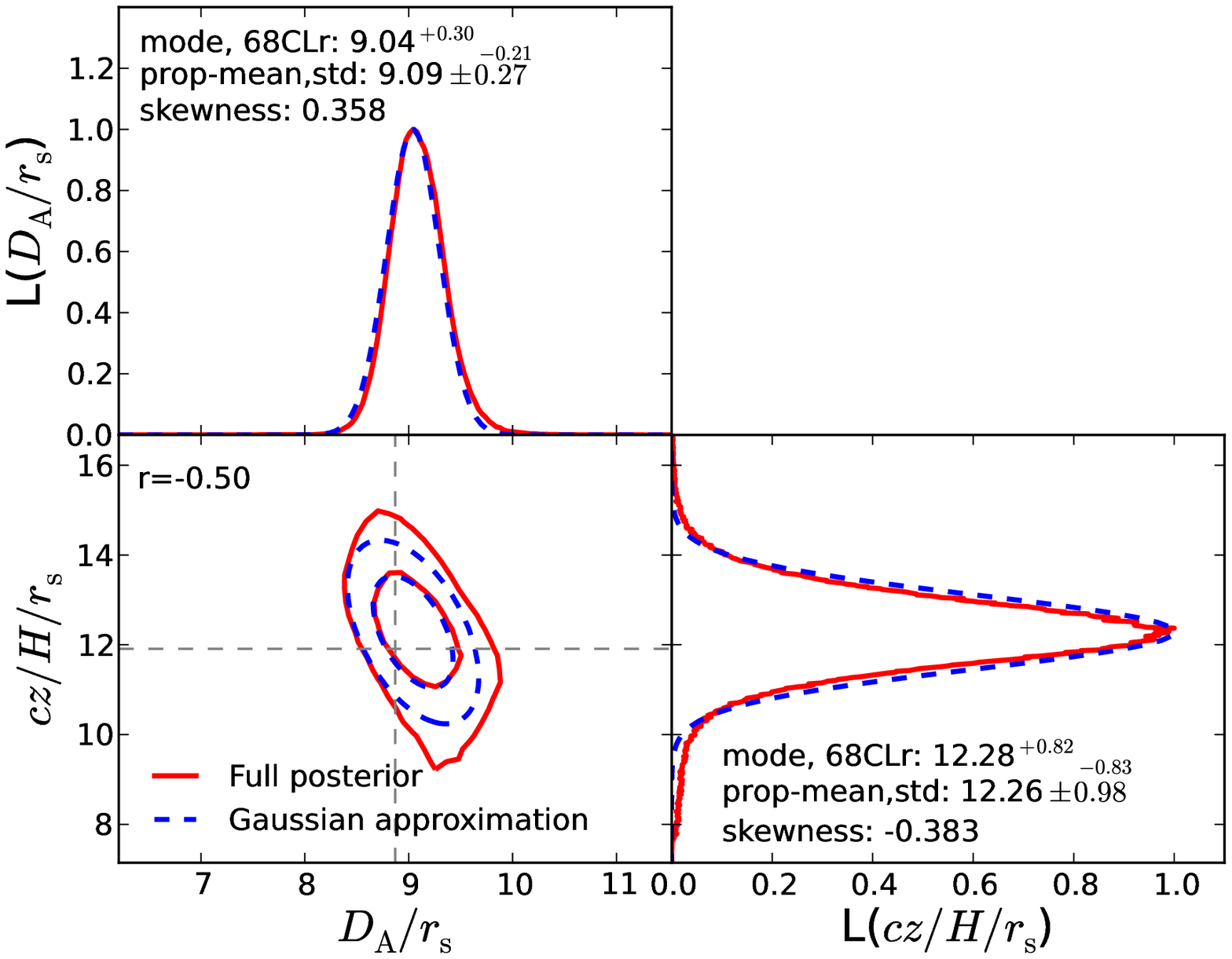} 
\caption{
CMASS results  
pre-reconstruction (left) 
and post (right). 
The marginalized results 
of \czHzrs (right panels)  
and \Dazrs (top panels), 
and the joint constraints (bottom panels). 
The solid red lines are the posterior, 
and the dashed blue lines are a Gaussian 
approximation, as described in the text. 
The panels indicate  
the modes, $68\%$ CL region boundaries, proposition-mean,  
proposition standard deviation,  
skewness and cross-correlation coefficient ($r$). 
The contours indicate the $68.27,95.45\%$ CL regions. 
For plotting purposes the post-reconstruction likelihoods 
assume a prior $|\epsilon|<0.15$. 
The gray dashed lines indicate the fiducial cosmology. 
}
\label{best_fit_figure_Hz_Daplane}  
\end{center}
\end{figure*}

Our main pre- and post-reconstruction results are summarized 
in Figures \ref{cmass_xi1d_stats_figure_postrec} and \ref{best_fit_figure_Hz_Daplane}.
Post-reconstruction we measure 
\czHzrs$= \ $\czHrscmass$\ \pm \ $\czHrsunc  (\czHrsperc accuracy; uncertainties are quoted at 68\% CL)
and 
\Dazrs$= \ $\Darscmass$\ \pm \ $\Darsunc (\Darsperc accuracy). 
The correlation coefficient betwen 
\czHzrs and \Dazrs is measured at 
\crosscorrczHrsDarsii, 
similar to that predicted by \cite{seo07}. 
The best fit model shows an excellent fit at 
$\chi^2/$dof=$0.82$ with  dof=66  degrees of freedom. 
Compared to the mocks realizations, this result is better than 
398/600 mocks. 
With the pre-reconstruction $\xi_{\Delta\mu}$ we obtain $\chi^2/$dof=$0.64$ (better than 578/600 realizations). 

Figure \ref{best_fit_figure_Hz_Daplane} compares 
the posterior results (solid red lines) 
with a Gaussian approximation (dashed blue lines), 
based on the same quoted modes, 
uncertainties and cross-correlation coefficients. 
In both the pre- and post-reconstruction cases,   
we see that the Gaussian approximation
describes the $68.27\%$ CL region fairly well, 
but clearly underestimates the $95.45\%$ CL reigon. 
We also note that the full posterior $99.73\%$ CL regions  
obtained both pre- and post-reconstruction 
are not well defined. 
These indicate the limited S/N in these measurement. 
We expect the agreement to improve with larger samples. 


In the top plot of Figure \ref{contourcomparison_plot} 
we show a direct comparison of the likelihood profiles 
pre- and post-reconstruction of the RPT-based clustering wedges. 
Both results 
appear to be similar, well within 
the $68\%$ CL region, 
although 
in the 
the post-reconstruction 
case \czHzrs is not as tightly constrained. 

For an 
average mock DR9-volume realization 
we find a mode cross-correlation of 
$r_{1/H}$,$r_{D_{\rm A}}\sim 0.35-0.4$ 
(or $\sim 0.6$ when examining the high S/N mocks; Appendix \ref{recimprove_section}) 
should be expected, 
where $r_{1/H}$ is the cross-correlation between the \czHzrs 
modes obtained when using one method (here pre-reconstruction) and 
when using a second (here post-reconstruction), 
and similar for $r_{D_{\rm A}}$, when discussing 
\Dazrs results. 
Also, although one does expect tighter constraints   
when applying reconstruction, 
the DR9 mocks indicate a $19\% $ (116/600) possibility
of not improving \czHzrsii. 
%
Using mocks with expected S/N of the final BOSS footprint 
(described in \S\ref{dr12_forecast_section}), 
this probability is reduced to $\sim 1.5\%$.

The CMASS \czHzrsii, \Dazrs  
results are summarized in Table \ref{cmass_table} 
along with various related parameters. 

\subsubsection{Comparing results of various $\xi$ methods}\label{cmass_varioustemplate_section}
The results quoted in the previous 
section are obtained   
when using the $\Delta\mu=0.5$ clustering wedges 
with the RPT-based template. 
Table \ref{cmass_table} contains the results obtained for  
eight different combinations of statistics.  

When applying the dewiggled template we 
obtain similar results to those obtained 
with RPT-based one.  
According to our mocks we expect 
$r_{1/H}$,$r_{D_{\rm A}}\sim 0.5-0.65$ 
amongst the templates both pre- and post-reconstruction.

\begin{figure}
\begin{center}
\includegraphics[width=0.49\textwidth]{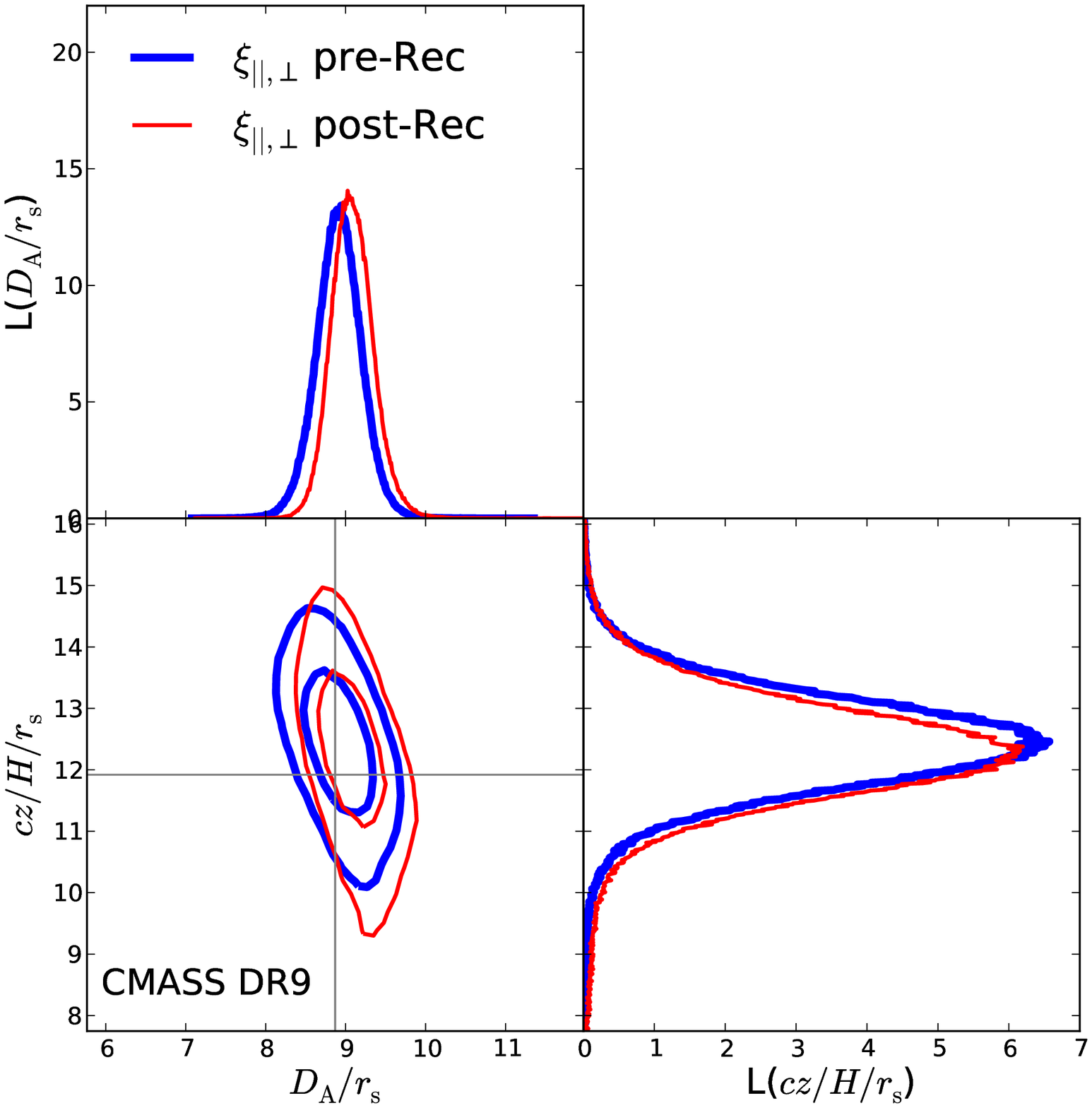}
\includegraphics[width=0.49\textwidth]{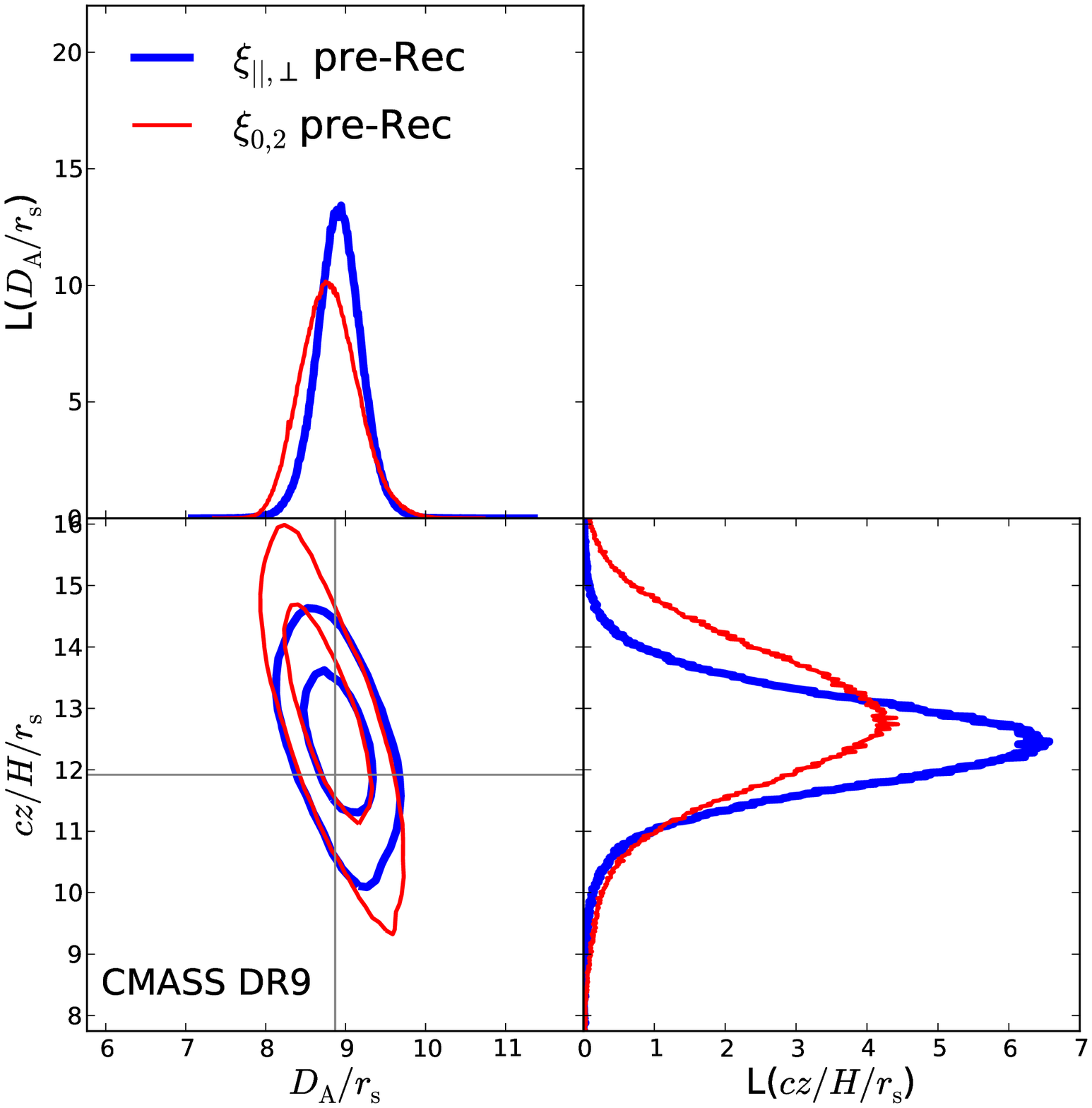}
\caption{
Comparison of the CMASS \czHzrs and \Dazrs 
results obtained with the pre-reconstruction wedges with 
alternative methods. 
The top plot shows a comparison with the post-reconstruction 
$\xi_{||,\perp}$ 
result. The Bottom plot shows a comparison with the 
pre-reconstruction clustering multipoles $\xi_0,\xi_2$. 
All methods use the RPT-based template. 
The contour plots show the $68,95\%$ CL regions. 
The solid lines are the fiducial cosmology. 
}
\label{contourcomparison_plot}  
\end{center}
\end{figure}

We apply the same test on the [$\xi_0$,$\xi_2$] multipoles 
and obtain slightly different results, 
but consistent within the $68\%$ CL regions, 
as seen in 
the bottom plot of Figure \ref{contourcomparison_plot}.  
According to the DR9 mock realizations 
we expect cross correlations between 
wedges results to multipoles by 
$r_{1/H}$,$r_{D_{\rm A}}\sim 0.4-0.45$.



\begin{figure}
\begin{center}
\includegraphics[width=0.45\textwidth]{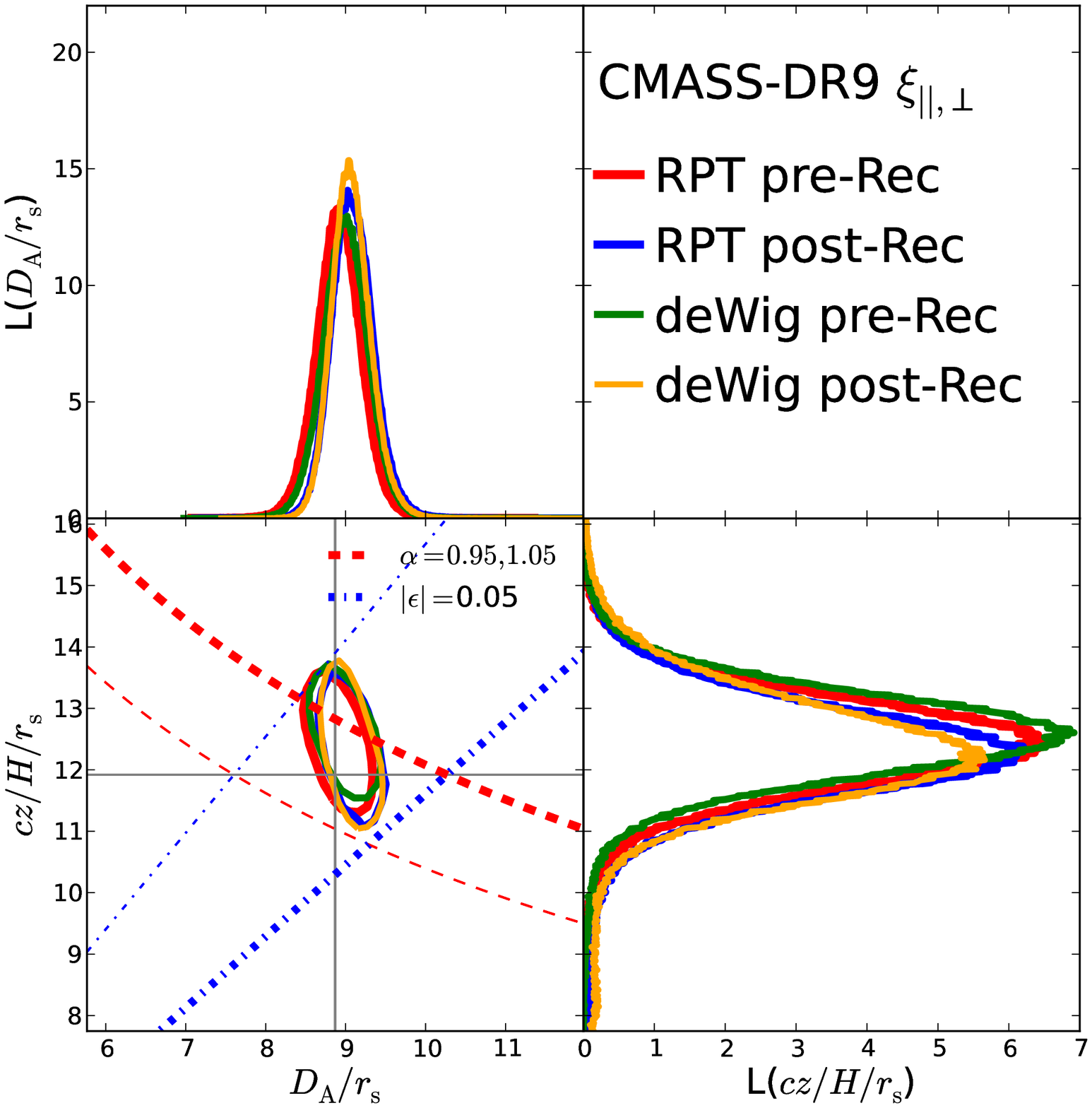}
\includegraphics[width=0.45\textwidth]{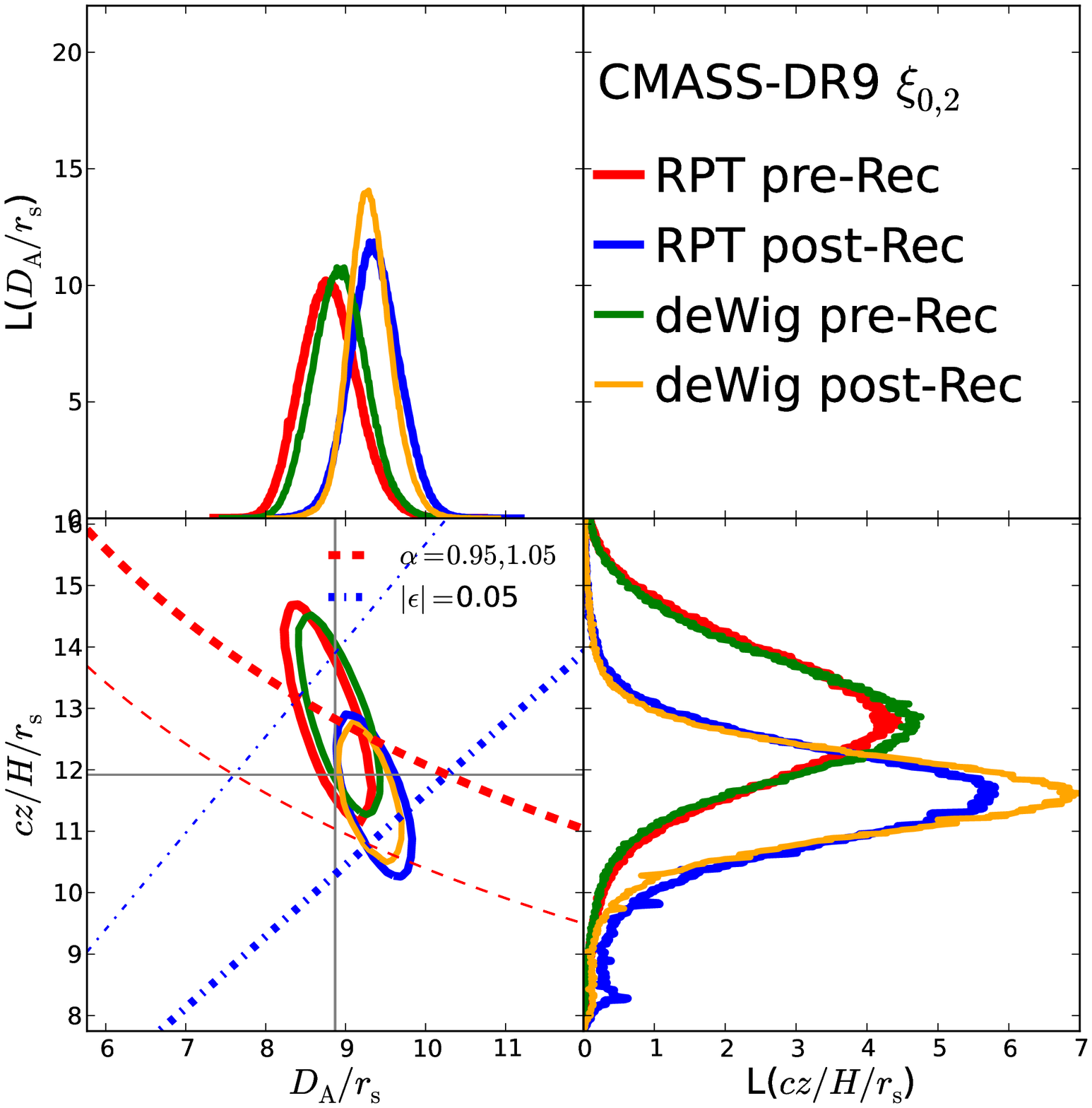}
\caption{
Comparison of the CMASS-DR9 \czHzrs and \Dazrs
marginalized profiles obtained with all the methods tested here. 
The top panel shows results when using the $\xi_{\Delta\mu}$ 
and the bottom panel when using $\xi_{0,2}$. 
The contour plots show the $68\%$ CL regions. 
The solid lines are the fiducial cosmology. 
To guide the eye we plot the regions of constant $\alpha$ and 
$\epsilon$, as indicated in the legend 
(where the thicker line of each indicates the larger value).
}
\label{all8_contourcomparison_plot}  
\end{center}
\end{figure}

Figure \ref{all8_contourcomparison_plot} displays 
\czHzrsii, \Dazrs likelihood profiles of 
all eight different methods 
analyzed here. 
The plot shows that all 
methods yield consistent results. 
The $\xi_{0,2}$ pre-rec (both RPT-based and dewiggled) 
\czHzrs profiles 
appear to be wider than the rest, 
where the 
$\xi_{0,2}$ post-rec (both RPT-based and dewiggled) 
appear to be the furthest from the rest, 
although clearly consistent within the $68-95\%$ CL regions. 
These differences are as expected based on the results 
from the mocks (for a visual of higher S/N mock results see top plot in Figure \ref{dr12forcast_figure}). 
We investigate various 
methods of shape parameters, 
and find similar results. 

\begin{table*} 
\begin{minipage}{172mm}
\caption{CMASS DR9 $\avg{z}=0.57$ results}
\label{cmass_table}
\begin{tabular}{@{}cccccccc@{}}
\hline
\hline
$\xi$&$\alpha_{||}$&\czHzrsii&$\alpha_{\perp}$&\Dazrsii&$r_{\alpha_{||},\alpha_{\perp}}$&$H$&$D_{\rm A}$\\
\hline

{\bf No prior on $\epsilon$:} & & &  & & &km$\cdot$s$^{-1}$Mpc$^{-1}$&Mpc \\
\hline
RPT-based $\xi_{||,\perp}$ pre-Rec    &1.042& 12.41$\pm 0.75\ (6.1\%)$& 1.006& 8.92$\pm 0.27\ (3.0\%)$& -0.50&  89.9$\pm 5.6$& 1367$\pm 45$  \\
RPT-based $\xi_{0,2}$ pre-Rec         &1.072& 12.77$\pm 1.15\ (9.0\%)$& 0.989& 8.77$\pm 0.36\ (4.1\%)$& -0.72&  87.4$\pm 7.9$& 1344$\pm 57$  \\
dewig $\xi_{||,\perp}$ pre-Rec        &1.055& 12.57$\pm 0.73\ (5.8\%)$& 1.014& 8.99$\pm 0.28\ (3.1\%)$& -0.57&  88.8$\pm 5.3$& 1378$\pm 46$  \\
dewig $\xi_{0,2}$ pre-Rec             &1.070& 12.74$\pm 1.06\ (8.3\%)$& 1.008& 8.94$\pm 0.33\ (3.7\%)$& -0.72&  87.5$\pm 7.4$& 1370$\pm 53$  \\
\\ RPT-based $\xi_{||,\perp}$ post-Rec&1.031& 12.28$\pm 0.83\ (6.8\%)$& 1.020& 9.05$\pm 0.25\ (2.8\%)$& -0.51&  90.8$\pm 6.2$& 1386$\pm 42$  \\ 
RPT-based $\xi_{0,2}$ post-Rec&0.974  & 11.60$\pm 1.44\ (12.4\%)$     & 1.055& 9.36$\pm 0.34\ (3.6\%)$& -0.67& 96.2$\pm 12.0$& 1434$\pm 55$  \\ 
dewig $\xi_{||,\perp}$ post-Rec&1.026 & 12.22$\pm 0.96\ (7.8\%)$      & 1.020& 9.05$\pm 0.24\ (2.7\%)$& -0.54&91.3$\pm 7.3$& 1386$\pm 41$  \\
dewig $\xi_{0,2}$ post-Rec            &0.974& 11.60$\pm 0.79\ (6.8\%)$& 1.046& 9.28$\pm 0.27\ (3.0\%)$& -0.62& 96.2$\pm 6.7$& 1422$\pm 45$  \\

\hline
\hline
{\bf prior $|\epsilon|\leq 15\%$:} & & & &  & & & \\
\hline
RPT-based $\xi_{||,\perp}$ pre-Rec     &1.042& 12.41$\pm 0.75\ (6.1\%)$& 1.006& 8.92$\pm 0.27\ (3.0\%)$& -0.50&89.9$\pm 5.6$& 1367$\pm 45$  \\
RPT-based $\xi_{0,2}$ pre-Rec          &1.072& 12.77$\pm 1.15\ (9.0\%)$& 0.989& 8.77$\pm 0.36\ (4.1\%)$& -0.75&87.4$\pm 7.9$& 1344$\pm 57$  \\
dewig $\xi_{||,\perp}$ pre-Rec         &1.055& 12.57$\pm 0.72\ (5.8\%)$& 1.014& 8.99$\pm 0.27\ (3.0\%)$& -0.53&88.8$\pm 5.2$& 1378$\pm 45$  \\
dewig $\xi_{0,2}$ pre-Rec              &1.070& 12.74$\pm 1.06\ (8.3\%)$& 1.008& 8.94$\pm 0.33\ (3.7\%)$& -0.72&87.5$\pm 7.4$& 1370$\pm 53$  \\
\\ RPT-based $\xi_{||,\perp}$ post-Rec &1.031& 12.28$\pm 0.82\ (6.7\%)$& 1.020& 9.05$\pm 0.27\ (3.0\%)$& -0.50&90.8$\pm 6.2$& 1386$\pm 45$  \\ 
RPT-based $\xi_{0,2}$ post-Rec         &0.974& 11.60$\pm 0.87\ (7.5\%)$& 1.052& 9.33$\pm 0.36\ (3.8\%)$& -0.78& 96.2$\pm 7.3$& 1430$\pm 57$  \\ 
dewig $\xi_{||,\perp}$ post-Rec        &1.026& 12.22$\pm 0.91\ (7.4\%)$& 1.020& 9.05$\pm 0.27\ (3.0\%)$& -0.53&91.3$\pm 6.9$& 1386$\pm 45$  \\
dewig $\xi_{0,2}$ post-Rec             &0.974& 11.60$\pm 0.73\ (6.3\%)$& 1.046& 9.28$\pm 0.33\ (3.6\%)$& -0.63& 96.2$\pm 6.2$& 1422$\pm 54$  \\

\hline
\hline
\end{tabular}

\medskip
 * We define $\alpha_{||}\equiv$\HrsfidHrs and $\alpha_{\perp}\equiv$\DrsDrsfidii.\\ 
 * Uncertainties $\Delta$ quoted correspond to half of the $68\%$ marginalized CL region (68CLr). In parentheses is the mean percentage of 68CLr.\\
 * All values are unitless, unless otherwise indicated. \\
 * Fiducial values used at $\avg{z}=0.57$: (\czHzrsii$)^{\rm f}=11.93$, (\Dazrsii$)^{\rm f}=8.88$, based on WMAP5 cosmology (\citealt{komatsu09a}).\\
 * The $H(0.57)$, $D_{\rm A}(0.57)$ columns assume WMAP5 result: $r_{\rm s}(z_{\rm d})=153.3\pm 2.0$ Mpc (Table 3 in \citealt{komatsu09a}). \\
 * $r_{\alpha_{||},\alpha_{\perp}}$ is the cross-correlation coefficient for \czHzrs and \Dazrsii.
\end{minipage}
\end{table*}

\subsubsection{Robustness of results to the range of fitted scales}\label{cmass_notshape_section}
As discussed in \S\ref{paramspace_section}, these measurements 
focus on the information of the anisotropic \baf 
and not from the full shape. 
As such, we do not expect dependency of our results 
on the range of scales used in the analysis. 

\begin{figure}
\begin{center}
\includegraphics[width=0.5\textwidth]{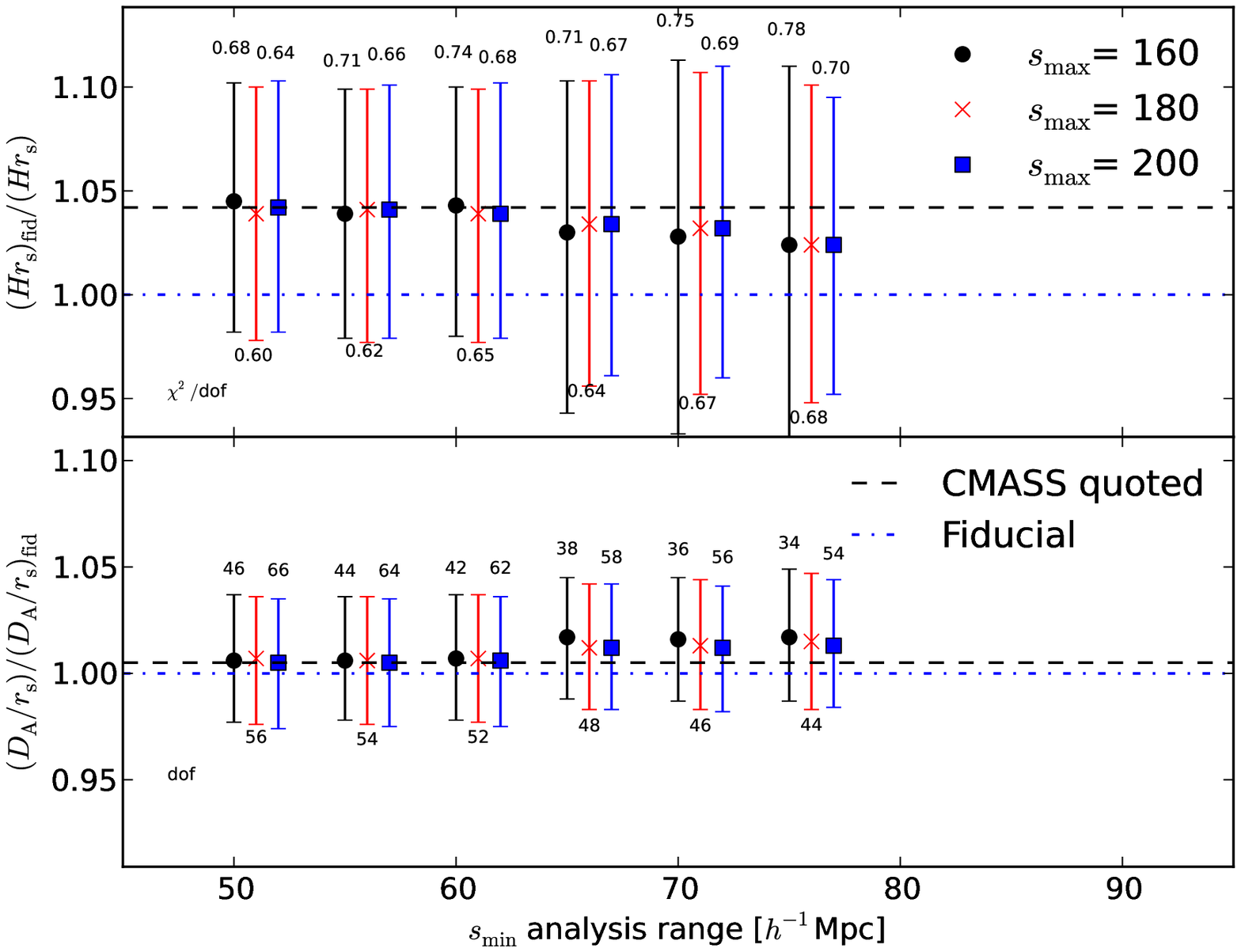} 
\caption{
This plot shows that the CMASS 
RPT-based $\xi_{||,\perp}$ pre-reconstruction results are  
insensitive to the range of analysis used 
when $s_{\rm min}<65$. 
The $x$ axes are the minimum separation used $s_{\rm min}$, 
where we compare results of $s_{\rm min}=$50,55,60,65,70,75 \hmpc
with maximum separations of 
$s_{\rm max}=$160 (black circles), 180 (red crosses), and 200\hmpc (blue squares).  
The number of degrees of freedom (dof) 
and  $\chi^2/$dof are quoted. 
All uncertainties indicate the $68\%$ CL regions.
The blue dot-dashed lines are the fiducial input values used to convert 
$z$ into comoving distances, 
and the black dashed lines are the chosen result quoted in Table \ref{cmass_table}. 
}
\label{smin_smax_H_Da_comparison_figure}  
\end{center}
\end{figure}

The results quoted in the previous sections are 
obtained when analyzing data in the region  
of separations between $[s_{\rm min}, s_{\rm max}]=[50,200]$. 
We compare the 
results obtained for various choices of $s_{\rm min}, s_{\rm max}$. 
Figure \ref{smin_smax_H_Da_comparison_figure} shows the comparison of the results.

We find that, for the most part, the range of analysis 
does not affect our main results: 
mode values, uncertainties, cross-correlation coefficient or skewness.
Regions of exception involve 
those with $s_{\rm min}\geq 65$\hmpcii, 
in which the \czHzrs uncertainties increase from 
$\sim 6\%$ to $7\%$ and even higher, 
when limiting to $s_{\rm max}$=160\hmpcii. 
This result could be explained by the fact 
that in this latter test the full dip of the \baf is not used, 
and shape parameter values that cause spurious 
dips are accepted, whereas for lower values of 
$s_{\rm min}$ they are not. 
We conclude that a more reliable result would include data 
points along the full shape, even though that information is 
marginalized over through the linear bias and $A(s)$ terms. 

We do not consider analyses with $s_{\rm min}<50$\hmpcii, 
because the templates used do not describe well the 
velocity-dispersion damping in the PTHalo mock-mean signal, 
and hence models would too heavily depend on the $A(s)$ terms.

In all ranges investigated the 
$\chi^2/{\rm dof}$ is between 
$0.6-0.8$, with the 
$s_{\rm max}=180$\hmpc 
yielding the best 
fits, 
although not significantly better ones. 

\subsubsection{Regarding the nuisance and fixed parameters}\label{cmass_nuisance_section}

As described in \S\ref{paramspace_section}, 
we use a set of ten parameters $\Phi_{10}$. 
To best understand the effects and correlations 
of these parameters amongst themselves and 
with \czHzrsii, \Dazrs we examine the results 
of both the data and the mock-mean signal. 
We perform these tests both pre- and post-reconstruction 
in both templates for $\xi_{\Delta\mu}$ and $\xi_{0,2}$. 

Overall, we do not see particular 
strong correlations between the $A(s)$ shape parameters 
with \czHzrsii, \Dazrsii, where most cross-correlations 
are $r<0.2$, but do illuminate 
a few findings of interest. 

Most of the shape parameters 
have marginalized likelihood profiles 
that are fairly symmetric (low skewness). 
We find that amplitude parameters
$a_{0 \ ||}$ and $a_{0 \ \perp}$ are uncorrelated 
with each other. 
All correlations of these parameters 
with \czHzrs and \Dazrs are $r<10\%$. 
The constant parameters ($a_3$) are uncorrelated 
to  \czHzrs and \Dazrsii, 
as expected. 
The other 
shape terms have 
weak correlations with \czHzrs and \Dazrsii, 
(at $r<0.2$).

The most important finding of the 
shape parameters, 
however, 
regards the $a_{0 \ \xi_2}$ 
(the amplitude of the quadrupole). 
In both pre- and post-reconstruction 
its marginalized likelihood profile 
is not well constrained, 
causing 
strong skewness in the joint 
likelihoods with other parameters. 
We decide to fix its value, 
which yields results similar to $\xi_{\Delta\mu}$, 
where this behaviour is not present. 

We find all the above similar 
for the data and mock-mean in the pre-reconstruction 
case. 
In the post-reconstruction case 
this is true as well,  
after we apply 
a prior $|\epsilon|<0.15$. 
Before applying the prior, 
the $99.7\%$ CL region is not well defined 
as the MCMC chains tend to accept values 
at the low limit set $\HrsfidHrsii=0.5$. 

Finally we address the question of 
the $A_{\rm MC}$ parameter in the RPT-based template 
(Equation \ref{pkrpt_equation}). 
\cite{crocce08} introduced this parameterization 
to effectively take into account the coupling 
between the $k-$modes, 
which results in a $0.5\%$ shift 
in the peak position in $\xi_0$. 
To obtain reliable templates of the 
post-reconstruction $\xi_{\Delta\mu}$ and $\xi_\ell$   
we find that a model without an $A_{\rm MC}$ term  
yields biassed results in the mocks, 
by about $\sim 1\%$ in \HrsfidHrsii. 
When analyzing the post-reconstruction CMASS $\xi_{||, \perp}$ results,  
we see 
a shift in $\alpha$ from 
1.026 ($A_{\rm MC}=2.44$) to 1.030 ($A_{\rm MC}=0$), 
a $0.4\%$ increase. 
The $1+\epsilon$ value is similar at $1.003$. 
This results in a $0.3\%$ shift in \czHzrs 
and $0.2\%$ shift in \Dazrsii, 
well below the uncertainties. 
For the post-reconstruction  $\xi_{0,2}$ 
we find similar results. 

\subsection{Final CMASS  Forecasts}\label{dr12_forecast_section}
By the conclusion of BOSS (2014), 
the survey 
will cover 
three times the area of the data 
set analyzed here, meaning 
The full CMASS 
sample will have a volume 
three times as large. 
By stacking the PTHalo mocks by 
groups of three, 
we can effectively, to first order, 
forecast the \czHzrsii, \Dazrs results of the full CMASS galaxy sample. 
Using the 600 realizations, 
we analyze here results of 200 $\xi_{||,\perp}$ stacked mocks.  

It is important to emphasize  
that the estimates yielded 
here should be considered 
{\it maximum} bounds. 
We argue this due to the 
fact that the $C_{ij}$ used 
is the same DR9 volume covariance matrix 
as in Equation (\ref{cij_equation})  
but divided by three. 
This means that we do not 
account for noisy cross-correlations 
which should be reduced with 
the actual full CMASS geometry, 
thus we expect 
the constraining power to be 
tighter when using a more reliable $C_{ij}$. 
Furthermore, 
we note 
that replicating the DR9 geometry does not improve
the reconstruction boundary effects.

\begin{figure}
\begin{center}
\includegraphics[width=0.55\textwidth]{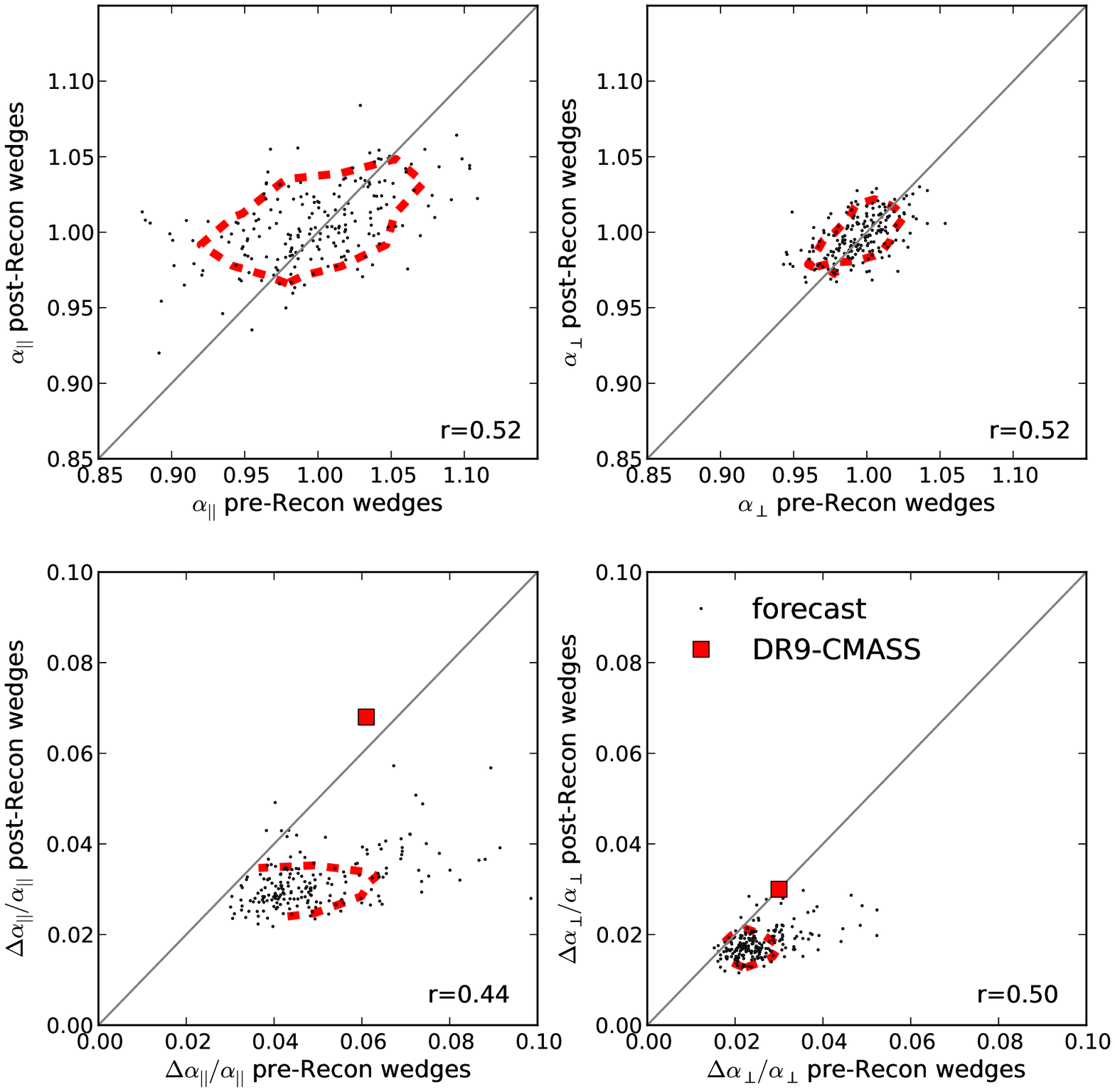} 
\includegraphics[width=0.55\textwidth]{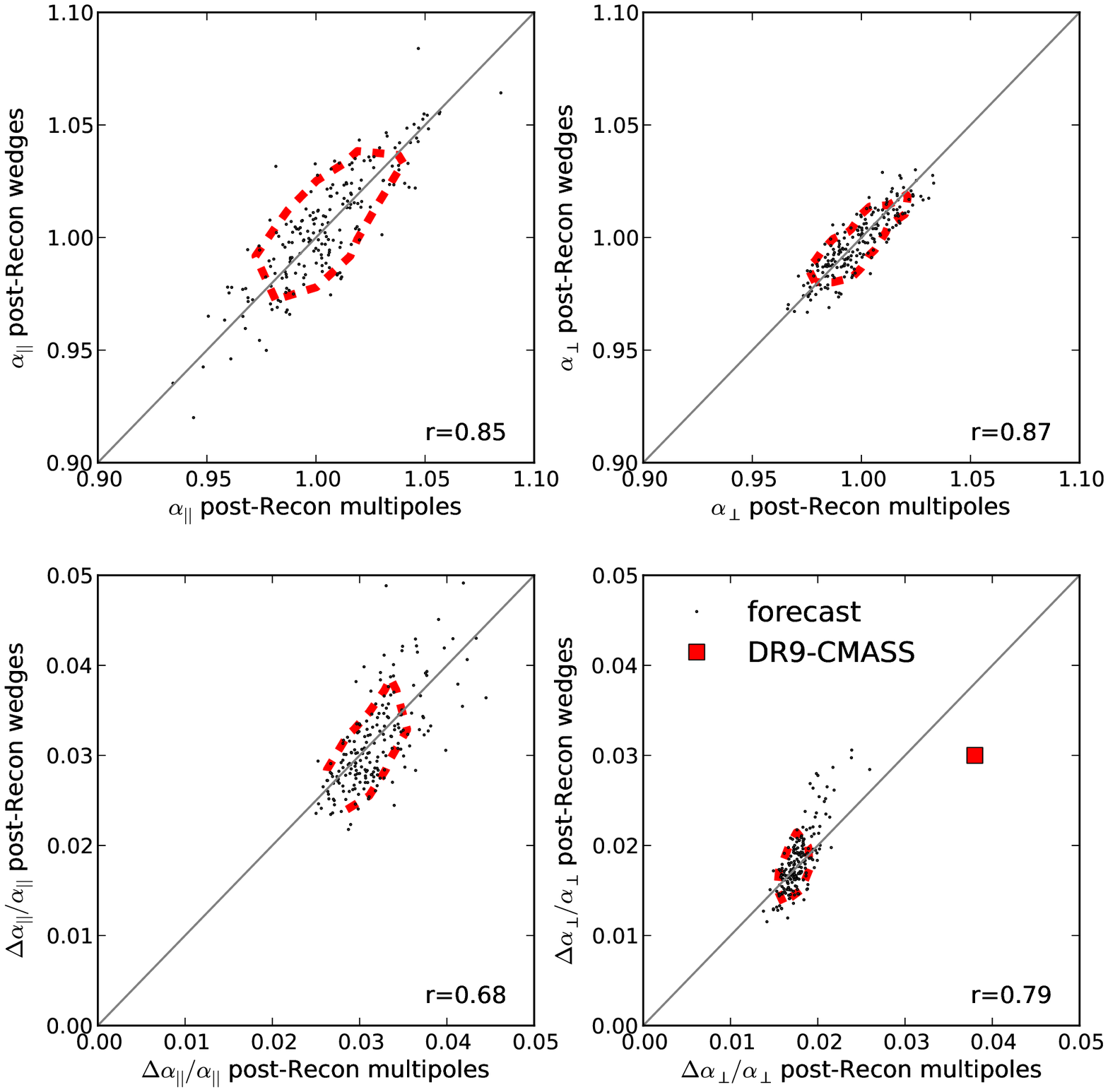} 
\caption{
 $\alpha_{||}\equiv$\HrsfidHrs and $\alpha_{\perp}\equiv$\DrsDrsfid 
 mode and fractional uncertainty  
 forecasts of 200 pseudo final BOSS CMASS volumes. 
 In all plots the 
 y-axis results are for post-reconstruction wedges. 
 In the top plot the x-axis results are for 
 pre-reconstruction wedges, on the bottom post-reconstruction 
 multipoles. In each plot the comparisons are 
 between $\alpha_{||}$ modes (top left panels), 
 $\alpha_{\perp}$ modes (top right), 
 $\Delta\alpha_{||}/\alpha_{||}$ uncertainties (bottom left),
 $\Delta\alpha_{\perp}/\alpha_{\perp}$ uncertainties (bottom right). 
 The cross-correlation in each is $r$. 
 The dashed red lines are the $68\%$ CL regions.  
 For the comparison, the red boxes are the DR9-CMASS results. 
}
\label{dr12forcast_figure}  
\end{center}
\end{figure}

Figure \ref{dr12forcast_figure} displays the 
\czHzrs and \Dazrs results obtained by means 
of the expected modes and uncertainties, 
comparing between post- and pre-reconstruction 
$\xi_{||,\perp}$ (top), 
and post-reconstruction $\xi_{||,\perp}$ to $\xi_{0,2}$ (bottom). 

When comparing \czHzrsii, \Dazrs results of the $\xi_{0,2}$ 
to the $\xi_{||,\perp}$ we find strong 
correlations where mode biases are sub $0.3\%$.  
Uncertainties show that no method 
is preferred over the other. 
When comparing pre- and post-reconstruction wedges, 
we find a $r\sim 0.52$ between the modes.

When applying reconstruction, 
the  \czHzrs uncertainties are predicted to improve 
from $0.045\pm0.017$ to $0.030\pm0.006$, 
a $33\%$ improvement. 
For \Dazrs the improvement is forecast 
to be from $0.024\pm0.007$ to $0.017\pm0.003$, 
a $\sim 30\%$ improvement. 
The mock result distributions yield 
Gaussian-like features, 
although application the K-S tests indicates  
they are not Gaussian. 
These trends are similar  
to those seen with 100 
six-stacked mocks (see Appendix \ref{stackmock_section}). 



\section{Discussion}\label{discussion_section}
The \czHzrs and \Dazrs results obtained here 
are consistent across the various techniques 
investigated:
\begin{enumerate}
{\item $\xi_{||,\perp}$, $\xi_{0,2}$}
{\item $\xi$ template: RPT-based, dewiggled}
{\item pre- and post-reconstruction}
\end{enumerate}
The likelihood profiles 
obtained with 
these eight combinations investigated, 
are shown in 
Figure \ref{all8_contourcomparison_plot} 
(as well as Figure \ref{contourcomparison_plot} and Table \ref{cmass_table}). 
Differences between the results are as expected from mock simulations.

As these posteriors are not Gaussian, 
we provide joint 2D marginalized 
likelihood profiles 
of \czHzrs and \Dazrsii, 
as well as provide the 
CMASS-DR9 $\xi_{||,\perp}$ and $C_{ij}$, $C_{ij}^{-1}$ 
on the World Wide Web.\footnote{\url{http://www.sdss3.org/science/boss_publications.php}} 
We conclude this study by using results obtained post-reconstruction  
over those yielded pre-reconstruction, 
because we show that 
mock results expect an improvement of $30\%$ in the 
marginalized constraints of \czHzrs and \Dazrsii, 
even though this is not the 
case in the data. 
We also prefer the RPT-based template 
over the dewiggled due to the larger bias in 
the mock results when using the latter. 
In the data, we find the posteriors to be similar 
regardless of choice of template 
(see Figure \ref{all8_contourcomparison_plot}).
Comparison of our results to other analyses 
of the same data set can be found in the following studies. 
\cite{sdss3dr9aniso} 
measures \czHzrs and \Dazrs
by applying a similar 
model-independent method on the $\xi_{0,2}$, 
using the same dewiggled templates. 
The main differences in analysis involve their use of  
a grid of $\alpha$ and $\epsilon$, 
where the rest of the nuisance parameters are 
determined by the least-squares method. 
We perform extensive comparisons between the methods, 
and find the \czHzrs and \Dazrs results to be fairly similar 
(see Figures 13 and 14 in \citealt{sdss3dr9aniso}). 
\cite{sdss3dr9aniso} continue to use 
results obtained in both studies to produce 
a ``consensus result" and calculate cosmological implications. 

Model-dependent analyses are performed on the full shape 
of $\xi_{||,\perp}$  (\citealt{sanchez13a}) and 
$\xi_{0,2}$ (\citealt{reid12a,chuang13a}). 
\cite{sanchez13a} shows that results amongst these 
studies are compatible. 
Figure 15 in \cite{sanchez13a} shows 
a comparison between our pre-reconstruction 
model independent 
result and their 
results from the full 
shape which are independent 
of parameter space, 
but assume $f$ follows GR predictions. 
They find an excellent agreement with our results, 
although tighter constraints as the \baf only method effectively 
accepts parameter values (e.g, $\Omega_{\rm M}$) that the full shape does not.

\section{Summary}\label{summary_section}
In this study 
we investigate the ability 
of the BOSS DR9-CMASS volume 
to constrain cosmic geometry at $z=0.57$, 
through the use of the AP technique 
applied on the anisotropic \bafii. 
We analyze the information contained in the anisotropic \bafii, for the first time, 
using a new technique called clustering wedges $\xi_{\Delta\mu}$, 
and compare results to the multipoles $\xi_{0,2}$. 

We find the anisotropic \baf to be detected in DR9-CMASS at a 
significance of $4.7\sigma$ compared to a featureless model (\S\ref{significance_section}). 
We find this level to be fairly fortunate 
(from a cosmological variance perspective), 
but consistent with that expected from mock realizations. 
The application of reconstruction 
leads to a 
significant improvement 
of detection of the peak in mock catalogues from 
$3.7\sigma\pm 0.9\sigma$ to $4.5\sigma\pm 0.9\sigma$ (median $\pm$ standard deviations; see Figure \ref{significance_plots}). 
Pre-reconstruction mocks also show that 
$23\%$ (138/600) yield a detection lower than $3\sigma$, 
whereas post-reconstruction $4.6\%$ do (28/600; Figure \ref{subsample_by_significance_plots}). 
Although we see clear improvement 
in the average mock realization, 
the significance of the detection 
of the anisotropic \baf in the data does 
not improve after applying reconstruction. 
We find this, however, consistent 
with 89/600 ($15\%$) of the mock realizations (Figure \ref{subsample_by_significance_plots}). 

To obtain geometrical constraints that are model independent, 
we use information from the post-reconstruction anisotropic \baf 
and measure 
\czHzrsii$\ = \ $\czHrscmass$\ \pm \ $\czHrsunc (\czHrsperc accuracy)
and 
\Dazrsii$\ = \ $\Darscmass$\ \pm \ $\Darsunc (\Darspercii)
with a correlation coefficient of \crosscorrczHrsDars  
(uncertainties are quoted at 68\% CL). 
In terms of constraining \czHzrs and \Dazrsii, 
the pre-reconstruction DR9-CMASS yields mutual constraints tighter than 
584/600 of the mocks, putting it in 
the fortunate top $2.5\%$. 
In the post-reconstruction case this is reduced 
to the top 444/600, meaning the top $26\%$. 
Although CMASS-DR9 results do not improve with 
reconstruction, mock catalogs 
indicate that, on average, 
one should expect an improvement 
of constraining power of $\sim 30\%$.  
Throughout this study we show that the posteriors of 
\czHzrs and \Dazrs 
from the DR9 volume are not expected to be Gaussian. 
In \S\ref{discussion_section} we explain how to use the results presented here, 
pointing out that the provided full likelihood function should be used 
instead of a Gaussian approximation.    
\cite{sdss3dr9aniso} analyze cosmological consequences of this measurement.

In our analysis of mock catalogues we 
also demonstrate that the constraining power 
of $\xi_{0,2}$ and $\xi_{||,\perp}$ are 
expected to be similar.  
With this information we conclude that the 
analysis of the clustering wedges 
and comparison to the multipoles technique, 
as performed here, is vital for testing 
systematics when measuring 
\czHzrs and \Dazrsii. 
Here we use wide clustering wedges of $\Delta\mu=0.5$, 
which are fairly correlated (see Figure \ref{NCij_50-200_figure}). 
As long as covariances can be adequately taken into account, 
this method could be generalized to narrower $\Delta\mu$ 
clustering wedges, as 
future surveys will yield better signal-to-noise ratio.

\section*{Acknowledgements}
It is a pleasure to thank Chris Blake for his insight. 
Also we thank  
David Kirkby, 
Felipe Marin,  
Cameron McBride and 
Uros Seljak 
for useful discussions. 
EK is supported by the Australian Research Council Centre of Excellence for All-sky Astrophysics (CAASTRO), through project number CE110001020.
AGS acknowledges support by the Trans-regional Collaborative Research 
Centre TR33 `The Dark Universe' of the German Research Foundation (DFG).
EK thanks Erin Sheldon for software used here.  
Numerical computations for the PTHalos mocks were done on the Sciama
High Performance Compute (HPC)
cluster which is supported by the ICG, SEPNet and the University of Portsmouth. 
Funding for SDSS-III has been provided by the Alfred P. Sloan Foundation, 
the Participating Institutions, the National Science Foundation, 
and the U.S. Department of Energy Office of Science. 
The SDSS-III web site is http://www.sdss3.org/.
SDSS-III is managed by the Astrophysical Research Consortium for the Participating Institutions of the SDSS-III Collaboration including the 
University of Arizona, 
the Brazilian Participation Group, 
Brookhaven National Laboratory, 
University of Cambridge, 
Carnegie Mellon University, 
University of Florida, 
the French Participation Group, 
the German Participation Group, 
Harvard University, 
the Instituto de Astrofisica de Canarias, 
the Michigan State/Notre Dame/JINA Participation Group, 
Johns Hopkins University, 
Lawrence Berkeley National Laboratory, 
Max Planck Institute for Astrophysics, 
Max Planck Institute for Extraterrestrial Physics, 
New Mexico State University, 
New York University, 
Ohio State University, 
Pennsylvania State University, 
University of Portsmouth, Princeton 
University, the Spanish Participation Group, 
University of Tokyo, 
University of Utah, 
Vanderbilt University, 
University of Virginia, 
University of Washington, 
and Yale University.



\begin{thebibliography}{90}
\expandafter\ifx\csname natexlab\endcsname\relax\def\natexlab#1{#1}\fi

\bibitem[{{Ahn} {et~al}\mbox{.}(2012){Ahn}, {Alexandroff}, {Allende Prieto},
  {Anderson}, {Anderton}, {Andrews}, {Aubourg}, {Bailey}, {Balbinot}, {Barnes},
  \& et~al.}]{ahn12a}
{Ahn} C.~P. {et~al.}, 2012, \apjs, 203, 21

\bibitem[{{Aihara} {et~al}\mbox{.}(2011){Aihara}, {Allende Prieto}, {An},
  {Anderson}, {Aubourg}, {Balbinot}, {Beers}, {Berlind}, {Bickerton},
  {Bizyaev}, {Blanton}, {Bochanski}, {Bolton}, {Bovy}, {Brandt}, {Brinkmann},
  {Brown}, {Brownstein}, {Busca}, {Campbell}, {Carr}, {Chen}, {Chiappini},
  {Comparat}, {Connolly}, {Cortes}, {Croft}, {Cuesta}, {da Costa}, {Davenport},
  {Dawson}, {Dhital}, {Ealet}, {Ebelke}, {Edmondson}, {Eisenstein},
  {Escoffier}, {Esposito}, {Evans}, {Fan}, {Femen{\'{\i}}a Castell{\'a}},
  {Font-Ribera}, {Frinchaboy}, {Ge}, {Gillespie}, {Gilmore}, {Gonz{\'a}lez
  Hern{\'a}ndez}, {Gott}, {Gould}, {Grebel}, {Gunn}, {Hamilton}, {Harding},
  {Harris}, {Hawley}, {Hearty}, {Ho}, {Hogg}, {Holtzman}, {Honscheid}, {Inada},
  {Ivans}, {Jiang}, {Johnson}, {Jordan}, {Jordan}, {Kazin}, {Kirkby}, {Klaene},
  {Knapp}, {Kneib}, {Kochanek}, {Koesterke}, {Kollmeier}, {Kron}, {Lampeitl},
  {Lang}, {Le Goff}, {Lee}, {Lin}, {Long}, {Loomis}, {Lucatello}, {Lundgren},
  {Lupton}, {Ma}, {MacDonald}, {Mahadevan}, {Maia}, {Makler}, {Malanushenko},
  {Malanushenko}, {Mandelbaum}, {Maraston}, {Margala}, {Masters}, {McBride},
  {McGehee}, {McGreer}, {M{\'e}nard}, {Miralda-Escud{\'e}}, {Morrison},
  {Mullally}, {Muna}, {Munn}, {Murayama}, {Myers}, {Naugle}, {Fausti Neto},
  {Cuong Nguyen}, {Nichol}, {O'Connell}, {Ogando}, {Olmstead}, {Oravetz},
  {Padmanabhan}, {Palanque-Delabrouille}, {Pan}, {Pandey}, {P{\^a}ris},
  {Percival}, {Petitjean}, {Pfaffenberger}, {Pforr}, {Phleps}, {Pichon},
  {Pieri}, {Prada}, {Price-Whelan}, {Raddick}, {Ramos}, {Reyl{\'e}}, {Rich},
  {Richards}, {Rix}, {Robin}, {Rocha-Pinto}, {Rockosi}, {Roe}, {Rollinde},
  {Ross}, {Ross}, {Rossetto}, {S{\'a}nchez}, {Sayres}, {Schlegel},
  {Schlesinger}, {Schmidt}, {Schneider}, {Sheldon}, {Shu}, {Simmerer},
  {Simmons}, {Sivarani}, {Snedden}, {Sobeck}, {Steinmetz}, {Strauss}, {Szalay},
  {Tanaka}, {Thakar}, {Thomas}, {Tinker}, {Tofflemire}, {Tojeiro}, {Tremonti},
  {Vandenberg}, {Vargas Maga{\~n}a}, {Verde}, {Vogt}, {Wake}, {Wang}, {Weaver},
  {Weinberg}, {White}, {White}, {Yanny}, {Yasuda}, {Yeche}, \&
  {Zehavi}}]{aihara11a}
{Aihara} H. {et~al.}, 2011, \apjs, 193, 29

\bibitem[{{Alcock} \& {Paczynski}(1979)}]{alcock79}
{Alcock} C., {Paczynski} B., 1979, \nat, 281, 358

\bibitem[{{Anderson} {et~al}\mbox{.}(2012){Anderson}, {Aubourg}, {Bailey},
  {Bizyaev}, {Blanton}, {Bolton}, {Brinkmann}, {Brownstein}, {Burden},
  {Cuesta}, {da Costa}, {Dawson}, {de Putter}, {Eisenstein}, {Gunn}, {Guo},
  {Hamilton}, {Harding}, {Ho}, {Honscheid}, {Kazin}, {Kirkby}, {Kneib},
  {Labatie}, {Loomis}, {Lupton}, {Malanushenko}, {Malanushenko}, {Mandelbaum},
  {Manera}, {Maraston}, {McBride}, {Mehta}, {Mena}, {Montesano}, {Muna},
  {Nichol}, {Nuza}, {Olmstead}, {Oravetz}, {Padmanabhan},
  {Palanque-Delabrouille}, {Pan}, {Parejko}, {P{\^a}ris}, {Percival},
  {Petitjean}, {Prada}, {Reid}, {Roe}, {Ross}, {Ross}, {Samushia},
  {S{\'a}nchez}, {Schlegel}, {Schneider}, {Sc{\'o}ccola}, {Seo}, {Sheldon},
  {Simmons}, {Skibba}, {Strauss}, {Swanson}, {Thomas}, {Tinker}, {Tojeiro},
  {Maga{\~n}a}, {Verde}, {Wagner}, {Wake}, {Weaver}, {Weinberg}, {White}, {Xu},
  {Y{\`e}che}, {Zehavi}, \& {Zhao}}]{anderson12a}
{Anderson} L. {et~al.}, 2012, \mnras, 427, 3435

\bibitem[{{Anderson et al.}(2013)}]{sdss3dr9aniso}
{Anderson} L. et al., 2013, ArXiv e-prints

\bibitem[{{Ballinger} {et~al}\mbox{.}(1996){Ballinger}, {Peacock}, \&
  {Heavens}}]{ballinger96a}
{Ballinger} W.~E., {Peacock} J.~A., {Heavens} A.~F., 1996, \mnras, 282, 877

\bibitem[{{Bassett} \& {Hlozek}(2010)}]{bassett10a}
{Bassett} B., {Hlozek} R., 2010, {Baryon acoustic oscillations},
  {Ruiz-Lapuente} P., ed., p. 246

\bibitem[{{Berlind} \& {Weinberg}(2002)}]{berlind02a}
{Berlind} A.~A., {Weinberg} D.~H., 2002, \apj, 575, 587

\bibitem[{{Bernardeau} {et~al}\mbox{.}(2002){Bernardeau}, {Colombi},
  {Gazta{\~n}aga}, \& {Scoccimarro}}]{bernardeau02a}
{Bernardeau} F. {et~al.}, 2002, \physrep, 367, 1

\bibitem[{{Beutler} {et~al}\mbox{.}(2011){Beutler}, {Blake}, {Colless},
  {Jones}, {Staveley-Smith}, {Campbell}, {Parker}, {Saunders}, \&
  {Watson}}]{beutler11a}
{Beutler} F. {et~al.}, 2011, \mnras, 416, 3017

\bibitem[{{Beutler} {et~al}\mbox{.}(2012){Beutler}, {Blake}, {Colless},
  {Jones}, {Staveley-Smith}, {Poole}, {Campbell}, {Parker}, {Saunders}, \&
  {Watson}}]{beutler12a}
{Beutler} F. {et~al.}, 2012, \mnras, 423, 3430

\bibitem[{{Blake} {et~al}\mbox{.}(2011{\natexlab{a}}){Blake}, {Brough},
  {Colless}, {Contreras}, {Couch}, {Croom}, {Davis}, {Drinkwater}, {Forster},
  {Gilbank}, {Gladders}, {Glazebrook}, {Jelliffe}, {Jurek}, {Li}, {Madore},
  {Martin}, {Pimbblet}, {Poole}, {Pracy}, {Sharp}, {Wisnioski}, {Woods},
  {Wyder}, \& {Yee}}]{blake11a}
{Blake} C. {et~al.}, 2011{\natexlab{a}}, \mnras, 415, 2876

\bibitem[{{Blake} {et~al}\mbox{.}(2011{\natexlab{b}}){Blake}, {Davis}, {Poole},
  {Parkinson}, {Brough}, {Colless}, {Contreras}, {Couch}, {Croom},
  {Drinkwater}, {Forster}, {Gilbank}, {Gladders}, {Glazebrook}, {Jelliffe},
  {Jurek}, {Li}, {Madore}, {Martin}, {Pimbblet}, {Pracy}, {Sharp}, {Wisnioski},
  {Woods}, {Wyder}, \& {Yee}}]{blake11b}
{Blake} C. {et~al.}, 2011{\natexlab{b}}, \mnras, 415, 2892

\bibitem[{{Blake} \& {Glazebrook}(2003)}]{blake03}
{Blake} C., {Glazebrook} K., 2003, \apj, 594, 665

\bibitem[{{Blake} {et~al}\mbox{.}(2011{\natexlab{c}}){Blake}, {Glazebrook},
  {Davis}, {Brough}, {Colless}, {Contreras}, {Couch}, {Croom}, {Drinkwater},
  {Forster}, {Gilbank}, {Gladders}, {Jelliffe}, {Jurek}, {Li}, {Madore},
  {Martin}, {Pimbblet}, {Poole}, {Pracy}, {Sharp}, {Wisnioski}, {Woods},
  {Wyder}, \& {Yee}}]{blake11d}
{Blake} C. {et~al.}, 2011{\natexlab{c}}, \mnras, 1599

\bibitem[{{Blake} {et~al}\mbox{.}(2011{\natexlab{d}}){Blake}, {Kazin},
  {Beutler}, {Davis}, {Parkinson}, {Brough}, {Colless}, {Contreras}, {Couch},
  {Croom}, {Croton}, {Drinkwater}, {Forster}, {Gilbank}, {Gladders},
  {Glazebrook}, {Jelliffe}, {Jurek}, {Li}, {Madore}, {Martin}, {Pimbblet},
  {Poole}, {Pracy}, {Sharp}, {Wisnioski}, {Woods}, {Wyder}, \&
  {Yee}}]{blake11c}
{Blake} C. {et~al.}, 2011{\natexlab{d}}, \mnras, 1598

\bibitem[{{Bolton} {et~al}\mbox{.}(2012){Bolton}, {Schlegel}, {Aubourg},
  {Bailey}, {Bhardwaj}, {Brownstein}, {Burles}, {Chen}, {Dawson}, {Eisenstein},
  {Gunn}, {Knapp}, {Loomis}, {Lupton}, {Maraston}, {Muna}, {Myers}, {Olmstead},
  {Padmanabhan}, {P{\^a}ris}, {Percival}, {Petitjean}, {Rockosi}, {Ross},
  {Schneider}, {Shu}, {Strauss}, {Thomas}, {Tremonti}, {Wake}, {Weaver}, \&
  {Wood-Vasey}}]{bolton12a}
{Bolton} A.~S. {et~al.}, 2012, \aj, 144, 144

\bibitem[{{Busca} {et~al}\mbox{.}(2012){Busca}, {Delubac}, {Rich}, {Bailey},
  {Font-Ribera}, {Kirkby}, {Le Goff}, {Pieri}, {Slosar}, {Aubourg}, {Bautista},
  {Bizyaev}, {Blomqvist}, {Bolton}, {Bovy}, {Brewington}, {Borde}, {Brinkmann},
  {Carithers}, {Croft}, {Dawson}, {Ebelke}, {Eisenstein}, {Hamilton}, {Ho},
  {Hogg}, {Honscheid}, {Lee}, {Lundgren}, {Malanushenko}, {Malanushenko},
  {Margala}, {Maraston}, {Mehta}, {Miralda-Escud{\'e}}, {Myers}, {Nichol},
  {Noterdaeme}, {Olmstead}, {Oravetz}, {Palanque-Delabrouille}, {Pan},
  {P{\^a}ris}, {Percival}, {Petitjean}, {Roe}, {Rollinde}, {Ross}, {Rossi},
  {Schlegel}, {Schneider}, {Shelden}, {Sheldon}, {Simmons}, {Snedden},
  {Tinker}, {Viel}, {Weaver}, {Weinberg}, {White}, {Y{\`e}che}, \&
  {York}}]{busca12a}
{Busca} N.~G. {et~al.}, 2012, ArXiv e-prints

\bibitem[{{Chuang} \& {Wang}(2011)}]{chuang11a}
{Chuang} C.-H., {Wang} Y., 2011, ArXiv e-prints

\bibitem[{{Chuang} \& {Wang}(2012{\natexlab{a}})}]{chuang12c}
{Chuang} C.-H., {Wang} Y., 2012{\natexlab{a}}, ArXiv e-prints

\bibitem[{{Chuang} \& {Wang}(2012{\natexlab{b}})}]{chuang12b}
{Chuang} C.-H., {Wang} Y., 2012{\natexlab{b}}, ArXiv e-prints

\bibitem[{{Chuang et al.}(2013)}]{chuang13a}
{Chuang} C.-H. et al., 2013, ArXiv e-prints 

\bibitem[{{Cole} {et~al}\mbox{.}(2005){Cole}, {Percival}, {Peacock}, {Norberg},
  {Baugh}, {Frenk}, {Baldry}, {Bland-Hawthorn}, {Bridges}, {Cannon}, {Colless},
  {Collins}, {Couch}, {Cross}, {Dalton}, {Eke}, {De Propris}, {Driver},
  {Efstathiou}, {Ellis}, {Glazebrook}, {Jackson}, {Jenkins}, {Lahav}, {Lewis},
  {Lumsden}, {Maddox}, {Madgwick}, {Peterson}, {Sutherland}, \&
  {Taylor}}]{cole05a}
{Cole} S. {et~al.}, 2005, \mnras, 362, 505

\bibitem[{{Cooray} \& {Sheth}(2002)}]{cooray02}
{Cooray} A., {Sheth} R., 2002, \physrep, 372, 1

\bibitem[{{Crocce} \& {Scoccimarro}(2008)}]{crocce08}
{Crocce} M., {Scoccimarro} R., 2008, \prd, 77, 023533

\bibitem[{{Dawson} {et~al}\mbox{.}(2013){Dawson}, {Schlegel}, {Ahn},
  {Anderson}, {Aubourg}, {Bailey}, {Barkhouser}, {Bautista}, {Beifiori},
  {Berlind}, {Bhardwaj}, {Bizyaev}, {Blake}, {Blanton}, {Blomqvist}, {Bolton},
  {Borde}, {Bovy}, {Brandt}, {Brewington}, {Brinkmann}, {Brown}, {Brownstein},
  {Bundy}, {Busca}, {Carithers}, {Carnero}, {Carr}, {Chen}, {Comparat},
  {Connolly}, {Cope}, {Croft}, {Cuesta}, {da Costa}, {Davenport}, {Delubac},
  {de Putter}, {Dhital}, {Ealet}, {Ebelke}, {Eisenstein}, {Escoffier}, {Fan},
  {Filiz Ak}, {Finley}, {Font-Ribera}, {G{\'e}nova-Santos}, {Gunn}, {Guo},
  {Haggard}, {Hall}, {Hamilton}, {Harris}, {Harris}, {Ho}, {Hogg}, {Holder},
  {Honscheid}, {Huehnerhoff}, {Jordan}, {Jordan}, {Kauffmann}, {Kazin},
  {Kirkby}, {Klaene}, {Kneib}, {Le Goff}, {Lee}, {Long}, {Loomis}, {Lundgren},
  {Lupton}, {Maia}, {Makler}, {Malanushenko}, {Malanushenko}, {Mandelbaum},
  {Manera}, {Maraston}, {Margala}, {Masters}, {McBride}, {McDonald}, {McGreer},
  {McMahon}, {Mena}, {Miralda-Escud{\'e}}, {Montero-Dorta}, {Montesano},
  {Muna}, {Myers}, {Naugle}, {Nichol}, {Noterdaeme}, {Nuza}, {Olmstead},
  {Oravetz}, {Oravetz}, {Owen}, {Padmanabhan}, {Palanque-Delabrouille}, {Pan},
  {Parejko}, {P{\^a}ris}, {Percival}, {P{\'e}rez-Fournon},
  {P{\'e}rez-R{\`a}fols}, {Petitjean}, {Pfaffenberger}, {Pforr}, {Pieri},
  {Prada}, {Price-Whelan}, {Raddick}, {Rebolo}, {Rich}, {Richards}, {Rockosi},
  {Roe}, {Ross}, {Ross}, {Rossi}, {Rubi{\~n}o-Martin}, {Samushia},
  {S{\'a}nchez}, {Sayres}, {Schmidt}, {Schneider}, {Sc{\'o}ccola}, {Seo},
  {Shelden}, {Sheldon}, {Shen}, {Shu}, {Slosar}, {Smee}, {Snedden}, {Stauffer},
  {Steele}, {Strauss}, {Streblyanska}, {Suzuki}, {Swanson}, {Tal}, {Tanaka},
  {Thomas}, {Tinker}, {Tojeiro}, {Tremonti}, {Vargas Maga{\~n}a}, {Verde},
  {Viel}, {Wake}, {Watson}, {Weaver}, {Weinberg}, {Weiner}, {West}, {White},
  {Wood-Vasey}, {Yeche}, {Zehavi}, {Zhao}, \& {Zheng}}]{dawson13a}
{Dawson} K.~S. {et~al.}, 2013, \aj, 145, 10

\bibitem[{{Drinkwater} {et~al}\mbox{.}(2010){Drinkwater}, {Jurek}, {Blake},
  {Woods}, {Pimbblet}, {Glazebrook}, {Sharp}, {Pracy}, {Brough}, {Colless},
  {Couch}, {Croom}, {Davis}, {Forbes}, {Forster}, {Gilbank}, {Gladders},
  {Jelliffe}, {Jones}, {Li}, {Madore}, {Martin}, {Poole}, {Small}, {Wisnioski},
  {Wyder}, \& {Yee}}]{drinkwater10a}
{Drinkwater} M.~J. {et~al.}, 2010, \mnras, 401, 1429

\bibitem[{{Einstein}(1916)}]{einstein16a}
{Einstein} A., 1916, Annalen der Physik, 354, 769

\bibitem[{{Eisenstein} \& {Hu}(1998)}]{eisenstein98}
{Eisenstein} D.~J., {Hu} W., 1998, \apj, 496, 605

\bibitem[{{Eisenstein} {et~al}\mbox{.}(2007{\natexlab{a}}){Eisenstein}, {Seo},
  {Sirko}, \& {Spergel}}]{eisenstein07}
{Eisenstein} D.~J. {et~al.}, 2007{\natexlab{a}}, \apj, 664, 675

\bibitem[{{Eisenstein} {et~al}\mbox{.}(2007{\natexlab{b}}){Eisenstein}, {Seo},
  \& {White}}]{eisenstein07b}
{Eisenstein} D.~J., {Seo} H.-J., {White} M., 2007{\natexlab{b}}, \apj, 664, 660

\bibitem[{{Eisenstein} {et~al}\mbox{.}(2011){Eisenstein}, {Weinberg}, {Agol},
  {Aihara}, {Allende Prieto}, {Anderson}, {Arns}, {Aubourg}, {Bailey},
  {Balbinot}, \& et~al.}]{eisenstein11a}
{Eisenstein} D.~J. {et~al.}, 2011, \aj, 142, 72

\bibitem[{Eisenstein {et~al}\mbox{.}(2005)Eisenstein {et~al.}}]{eisenstein05b}
Eisenstein D.~J., {et~al.}, 2005, \apj, 633, 560

\bibitem[{{Gazta{\~n}aga} {et~al}\mbox{.}(2009){Gazta{\~n}aga}, {Cabr{\'e}}, \&
  {Hui}}]{gaztanaga08iv}
{Gazta{\~n}aga} E., {Cabr{\'e}} A., {Hui} L., 2009, \mnras, 399, 1663

\bibitem[{{Gunn} {et~al}\mbox{.}(1998){Gunn}, {Carr}, {Rockosi}, {Sekiguchi},
  {Berry}, {Elms}, {de Haas}, {Ivezi{\'c}}, {Knapp}, {Lupton}, {Pauls},
  {Simcoe}, {Hirsch}, {Sanford}, {Wang}, {York}, {Harris}, {Annis}, {Bartozek},
  {Boroski}, {Bakken}, {Haldeman}, {Kent}, {Holm}, {Holmgren}, {Petravick},
  {Prosapio}, {Rechenmacher}, {Doi}, {Fukugita}, {Shimasaku}, {Okada}, {Hull},
  {Siegmund}, {Mannery}, {Blouke}, {Heidtman}, {Schneider}, {Lucinio}, \&
  {Brinkman}}]{gunn98a}
{Gunn} J.~E. {et~al.}, 1998, \aj, 116, 3040

\bibitem[{{Gunn} {et~al}\mbox{.}(2006){Gunn} {et~al.}}]{gunn05a}
{Gunn} J.~E., {et~al.}, 2006, \aj, 131, 2332

\bibitem[{{Guzzo} {et~al}\mbox{.}(2007){Guzzo}, {Pierleoni}, {Meneux},
  {Branchini}, {Le F{\'e}vre}, {Marinoni}, {Garilli}, {Blaizot}, {De Lucia},
  {Pollo}, {McCracken}, {Bottini}, {Le Brun}, {Maccagni}, {Picat},
  {Scaramella}, {Scodeggio}, {Tresse}, {Vettolani}, {Zanichelli}, {Adami},
  {Arnouts}, {Bardelli}, {Bolzonella}, {Bongiorno}, {Cappi}, {Charlot},
  {Ciliegi}, {Contini}, {Cucciati}, {de La Torre}, {Foucaud}, {Franzetti},
  {Gavignaud}, {Ilbert}, {Iovino}, {Lamareille}, {Marano}, {Mazure}, {Memeo},
  {Mellier}, {Merighi}, {Moscardini}, {Paltani}, {Pell{\`o}}, {Pozzetti},
  {Radovich}, {Vergani}, {Zamorani}, \& {Zucca}}]{guzzo07}
{Guzzo} L. {et~al.}, 2007, Nuovo Cimento B Serie, 122, 1385

\bibitem[{{Hamilton}(1998)}]{hamilton98a}
{Hamilton} A.~J.~S., 1998, in Astrophysics and Space Science Library, Vol. 231,
  The Evolving Universe, {D.~Hamilton}, ed., pp. 185--+

\bibitem[{{Hartlap} {et~al}\mbox{.}(2007){Hartlap}, {Simon}, \&
  {Schneider}}]{hartlap07a}
{Hartlap} J., {Simon} P., {Schneider} P., 2007, \aap, 464, 399

\bibitem[{{Hinshaw} {et~al}\mbox{.}(2012){Hinshaw}, {Larson}, {Komatsu},
  {Spergel}, {Bennett}, {Dunkley}, {Nolta}, {Halpern}, {Hill}, {Odegard},
  {Page}, {Smith}, {Weiland}, {Gold}, {Jarosik}, {Kogut}, {Limon}, {Meyer},
  {Tucker}, {Wollack}, \& {Wright}}]{hinshaw12a}
{Hinshaw} G. {et~al.}, 2012, ArXiv e-prints

\bibitem[{{Hoffman} \& {Ribak}(1991)}]{hoffman91a}
{Hoffman} Y., {Ribak} E., 1991, \apjl, 380, L5

\bibitem[{{Hu} \& {Haiman}(2003)}]{hu03a}
{Hu} W., {Haiman} Z., 2003, \prd, 68, 063004

\bibitem[{{Hu} {et~al}\mbox{.}(1997){Hu}, {Sugiyama}, \& {Silk}}]{hu97a}
{Hu} W., {Sugiyama} N., {Silk} J., 1997, \nat, 386, 37

\bibitem[{{Jones} {et~al}\mbox{.}(2009){Jones}, {Read}, {Saunders}, {Colless},
  {Jarrett}, {Parker}, {Fairall}, {Mauch}, {Sadler}, {Watson}, {Burton},
  {Campbell}, {Cass}, {Croom}, {Dawe}, {Fiegert}, {Frankcombe}, {Hartley},
  {Huchra}, {James}, {Kirby}, {Lahav}, {Lucey}, {Mamon}, {Moore}, {Peterson},
  {Prior}, {Proust}, {Russell}, {Safouris}, {Wakamatsu}, {Westra}, \&
  {Williams}}]{jones09a}
{Jones} D.~H. {et~al.}, 2009, \mnras, 399, 683

\bibitem[{{Kaiser}(1987)}]{kaiser87}
{Kaiser} N., 1987, \mnras, 227, 1

\bibitem[{{Kazin} {et~al}\mbox{.}(2010){Kazin}, {Blanton}, {Scoccimarro},
  {McBride}, \& {Berlind}}]{kazin10b}
{Kazin} E.~A. {et~al.}, 2010, \apj, 719, 1032

\bibitem[{{Kazin} {et~al}\mbox{.}(2012){Kazin}, {S{\'a}nchez}, \&
  {Blanton}}]{kazin11a}
{Kazin} E.~A., {S{\'a}nchez} A.~G., {Blanton} M.~R., 2012, \mnras, 419, 3223

\bibitem[{{Kirkby} {et~al}\mbox{.}(2013){Kirkby}, {Margala}, {Slosar},
  {Bailey}, {Busca}, {Delubac}, {Rich}, {Blomqvist}, {Brownstein}, {Carithers},
  {Croft}, {Dawson}, {Font-Ribera}, {Miralda-Escud{\'e}}, {Myers}, {Nichol},
  {Palanque-Delabrouille}, {P{\^a}ris}, {Petitjean}, {Rossi}, {Schlegel},
  {Schneider}, {Viel}, {Weinberg}, \& {Y{\`e}che}}]{kirkby13a}
{Kirkby} D. {et~al.}, 2013, ArXiv e-prints

\bibitem[{{Komatsu} {et~al}\mbox{.}(2009){Komatsu}, {Dunkley}, {Nolta},
  {Bennett}, {Gold}, {Hinshaw}, {Jarosik}, {Larson}, {Limon}, {Page},
  {Spergel}, {Halpern}, {Hill}, {Kogut}, {Meyer}, {Tucker}, {Weiland},
  {Wollack}, \& {Wright}}]{komatsu09a}
{Komatsu} E. {et~al.}, 2009, \apjs, 180, 330

\bibitem[{{Komatsu} {et~al}\mbox{.}(2011){Komatsu}, {Smith}, {Dunkley},
  {Bennett}, {Gold}, {Hinshaw}, {Jarosik}, {Larson}, {Nolta}, {Page},
  {Spergel}, {Halpern}, {Hill}, {Kogut}, {Limon}, {Meyer}, {Odegard}, {Tucker},
  {Weiland}, {Wollack}, \& {Wright}}]{komatsu11a}
{Komatsu} E. {et~al.}, 2011, \apjs, 192, 18

\bibitem[{{Landy} \& {Szalay}(1993)}]{landy93a}
{Landy} S.~D., {Szalay} A.~S., 1993, \apj, 412, 64

\bibitem[{{Lavaux} \& {Wandelt}(2012)}]{lavaux12a}
{Lavaux} G., {Wandelt} B.~D., 2012, \apj, 754, 109

\bibitem[{{Linder}(2008)}]{linder08}
{Linder} E.~V., 2008, Astroparticle Physics, 29, 336

\bibitem[{{Manera} {et~al}\mbox{.}(2012){Manera}, {Scoccimarro}, {Percival},
  {Samushia}, {McBride}, {Ross}, {Sheth}, {White}, {Reid}, {S{\'a}nchez}, {de
  Putter}, {Xu}, {Berlind}, {Brinkmann}, {Nichol}, {Montesano}, {Padmanabhan},
  {Skibba}, {Tojeiro}, \& {Weaver}}]{manera12a}
{Manera} M. {et~al.}, 2012, ArXiv e-prints

\bibitem[{{Masters} {et~al}\mbox{.}(2011){Masters}, {Maraston}, {Nichol},
  {Thomas}, {Beifiori}, {Bundy}, {Edmondson}, {Higgs}, {Leauthaud},
  {Mandelbaum}, {Pforr}, {Ross}, {Ross}, {Schneider}, {Skibba}, {Tinker},
  {Tojeiro}, {Wake}, {Brinkmann}, \& {Weaver}}]{masters11a}
{Masters} K.~L. {et~al.}, 2011, \mnras, 418, 1055

\bibitem[{{Mehta} {et~al}\mbox{.}(2011){Mehta}, {Seo}, {Eckel}, {Eisenstein},
  {Metchnik}, {Pinto}, \& {Xu}}]{mehta11a}
{Mehta} K.~T. {et~al.}, 2011, \apj, 734, 94

\bibitem[{{Noh} {et~al}\mbox{.}(2009){Noh}, {White}, \& {Padmanabhan}}]{noh09a}
{Noh} Y., {White} M., {Padmanabhan} N., 2009, \prd, 80, 123501

\bibitem[{{Nuza} {et~al}\mbox{.}(2012){Nuza}, {Sanchez}, {Prada}, {Klypin},
  {Schlegel}, {Gottloeber}, {Montero-Dorta}, {Manera}, {McBride}, {Ross},
  {Angulo}, {Blanton}, {Bolton}, {Favole}, {Samushia}, {Montesano}, {Percival},
  {Padmanabhan}, {Steinmetz}, {Tinker}, {Skibba}, {Schneider}, {Guo}, {Zehavi},
  {Zheng}, {Bizyaev}, {Malanushenko}, {Malanushenko}, {Oravetz}, {Oravetz}, \&
  {Shelden}}]{nuza12a}
{Nuza} S.~E. {et~al.}, 2012, ArXiv e-prints

\bibitem[{{Okumura} {et~al}\mbox{.}(2008){Okumura}, {Matsubara}, {Eisenstein},
  {Kayo}, {Hikage}, {Szalay}, \& {Schneider}}]{okumura08}
{Okumura} T. {et~al.}, 2008, \apj, 676, 889

\bibitem[{Padmanabhan \& White(2008)}]{padmanabhan08a}
Padmanabhan N., White M., 2008, Physical Review D, 77, 123540, (c) 2008: The
  American Physical Society

\bibitem[{Padmanabhan {et~al}\mbox{.}(2009)Padmanabhan, White, \&
  Cohn}]{padmanabhan09a}
Padmanabhan N., White M., Cohn J.~D., 2009, Physical Review D, 79, 63523, (c)
  2009: The American Physical Society

\bibitem[{{Padmanabhan} {et~al}\mbox{.}(2012){Padmanabhan}, {Xu}, {Eisenstein},
  {Scalzo}, {Cuesta}, {Mehta}, \& {Kazin}}]{padmanabhan12a}
{Padmanabhan} N. {et~al.}, 2012, ArXiv e-prints

\bibitem[{{Peacock} \& {Smith}(2000)}]{peacock00a}
{Peacock} J.~A., {Smith} R.~E., 2000, \mnras, 318, 1144

\bibitem[{{Peebles} \& {Yu}(1970)}]{peebles70a}
{Peebles} P.~J.~E., {Yu} J.~T., 1970, \apj, 162, 815

\bibitem[{{Perlmutter} {et~al}\mbox{.}(1999){Perlmutter}
  {et~al.}}]{perlmutter99a}
{Perlmutter} S., {et~al.}, 1999, \apj, 517, 565

\bibitem[{{Phillipps}(1994)}]{phillipps94a}
{Phillipps} S., 1994, \mnras, 269, 1077

\bibitem[{{Reid} {et~al}\mbox{.}(2012){Reid}, {Samushia}, {White}, {Percival},
  {Manera}, {Padmanabhan}, {Ross}, {S{\'a}nchez}, {Bailey}, {Bizyaev},
  {Bolton}, {Brewington}, {Brinkmann}, {Brownstein}, {Cuesta}, {Eisenstein},
  {Gunn}, {Honscheid}, {Malanushenko}, {Malanushenko}, {Maraston}, {McBride},
  {Muna}, {Nichol}, {Oravetz}, {Pan}, {de Putter}, {Roe}, {Ross}, {Schlegel},
  {Schneider}, {Seo}, {Shelden}, {Sheldon}, {Simmons}, {Skibba}, {Snedden},
  {Swanson}, {Thomas}, {Tinker}, {Tojeiro}, {Verde}, {Wake}, {Weaver},
  {Weinberg}, {Zehavi}, \& {Zhao}}]{reid12a}
{Reid} B.~A. {et~al.}, 2012, \mnras, 426, 2719

\bibitem[{{Riess} {et~al}\mbox{.}(1998){Riess}, {Filippenko}, {Challis},
  {Clocchiatti}, {Diercks}, {Garnavich}, {Gilliland}, {Hogan}, {Jha},
  {Kirshner}, {Leibundgut}, {Phillips}, {Reiss}, {Schmidt}, {Schommer},
  {Smith}, {Spyromilio}, {Stubbs}, {Suntzeff}, \& {Tonry}}]{riess98}
{Riess} A.~G. {et~al.}, 1998, \aj, 116, 1009

\bibitem[{{Samushia} {et~al}\mbox{.}(2012){Samushia}, {Percival}, \&
  {Raccanelli}}]{samushia11a}
{Samushia} L., {Percival} W.~J., {Raccanelli} A., 2012, \mnras, 420, 2102

\bibitem[{{Samushia} {et~al}\mbox{.}(2013){Samushia}, {Reid}, {White},
  {Percival}, {Cuesta}, {Lombriser}, {Manera}, {Nichol}, {Schneider},
  {Bizyaev}, {Brewington}, {Malanushenko}, {Malanushenko}, {Oravetz}, {Pan},
  {Simmons}, {Shelden}, {Snedden}, {Tinker}, {Weaver}, {York}, \&
  {Zhao}}]{samushia13a}
{Samushia} L. {et~al.}, 2013, \mnras, 429, 1514

\bibitem[{{S{\'a}nchez} {et~al}\mbox{.}(2008){S{\'a}nchez}, {Baugh}, \&
  {Angulo}}]{sanchez08}
{S{\'a}nchez} A.~G., {Baugh} C.~M., {Angulo} R., 2008, \mnras, 390, 1470

\bibitem[{{S{\'a}nchez} {et~al}\mbox{.}(2012){S{\'a}nchez}, {Sc{\'o}ccola},
  {Ross}, {Percival}, {Manera}, {Montesano}, {Mazzalay}, {Cuesta},
  {Eisenstein}, {Kazin}, {McBride}, {Mehta}, {Montero-Dorta}, {Padmanabhan},
  {Prada}, {Rubi{\~n}o-Mart{\'{\i}}n}, {Tojeiro}, {Xu}, {Maga{\~n}a},
  {Aubourg}, {Bahcall}, {Bailey}, {Bizyaev}, {Bolton}, {Brewington},
  {Brinkmann}, {Brownstein}, {Gott}, {Hamilton}, {Ho}, {Honscheid}, {Labatie},
  {Malanushenko}, {Malanushenko}, {Maraston}, {Muna}, {Nichol}, {Oravetz},
  {Pan}, {Ross}, {Roe}, {Reid}, {Schlegel}, {Shelden}, {Schneider}, {Simmons},
  {Skibba}, {Snedden}, {Thomas}, {Tinker}, {Wake}, {Weaver}, {Weinberg},
  {White}, {Zehavi}, \& {Zhao}}]{sanchez12a}
{S{\'a}nchez} A.~G. {et~al.}, 2012, \mnras, 425, 415

\bibitem[{{S{\'a}nchez et al.}(2013)}]{sanchez13a}
{S{\'a}nchez} A. et al., 2013, ArXiv e-prints

\bibitem[{{Scoccimarro} {et~al}\mbox{.}(2001){Scoccimarro}, {Sheth}, {Hui}, \&
  {Jain}}]{scoccimarro01a}
{Scoccimarro} R. {et~al.}, 2001, \apj, 546, 20

\bibitem[{{Seljak}(2000)}]{seljak00a}
{Seljak} U., 2000, \mnras, 318, 203

\bibitem[{{Seo} \& {Eisenstein}(2007){Seo}, \& {Eisenstein}}]{seo07}
{Seo} H., {Eisenstein} D.~J., 2007, \apj, 665, 14


\bibitem[{{Seo} {et~al}\mbox{.}(2010){Seo}, {Eckel}, {Eisenstein}, {Mehta},
  {Metchnik}, {Padmanabhan}, {Pinto}, {Takahashi}, {White}, \& {Xu}}]{seo10a}
{Seo} H. {et~al.}, 2010, \apj, 720, 1650

\bibitem[{{Shoji} {et~al}\mbox{.}(2009){Shoji}, {Jeong}, \&
  {Komatsu}}]{shoji09a}
{Shoji} M., {Jeong} D., {Komatsu} E., 2009, \apj, 693, 1404

\bibitem[{{Slosar} {et~al}\mbox{.}(2013){Slosar}, {Ir{\v s}i{\v c}}, {Kirkby},
  {Bailey}, {Busca}, {Delubac}, {Rich}, {Bhardwaj}, {Blomqvist}, {Bolton},
  {Bovy}, {Brownstein}, {Carithers}, {Croft}, {Dawson}, {Font-Ribera}, {Le
  Goff}, {Ho}, {Honscheid}, {Lee}, {Margala}, {McDonald}, {Medolin},
  {Miralda-Escud{\'e}}, {Myers}, {Nichol}, {Noterdaeme}, {P{\^a}ris},
  {Petitjean}, {Pieri}, {Roe}, {Ross}, {Rossi}, {Schlegel}, {Schneider},
  {Sheldon}, {Seljak}, {Viel}, {Weinberg}, \& {Y{\`e}che}}]{slosar13a}
{Slosar} A. {et~al.}, 2013, ArXiv e-prints

\bibitem[{{Smee} {et~al}\mbox{.}(2012){Smee}, {Gunn}, {Uomoto}, {Roe},
  {Schlegel}, {Rockosi}, {Carr}, {Leger}, {Dawson}, {Olmstead}, {Brinkmann},
  {Owen}, {Barkhouser}, {Honscheid}, {Harding}, {Long}, {Lupton}, {Loomis},
  {Anderson}, {Annis}, {Bernardi}, {Bhardwaj}, {Bizyaev}, {Bolton},
  {Brewington}, {Briggs}, {Burles}, {Burns}, {Castander}, {Connolly},
  {Davenport}, {Ebelke}, {Epps}, {Feldman}, {Friedman}, {Frieman}, {Heckman},
  {Hull}, {Knapp}, {Lawrence}, {Loveday}, {Mannery}, {Malanushenko},
  {Malanushenko}, {Merrelli}, {Muna}, {Newman}, {Nichol}, {Oravetz}, {Pan},
  {Pope}, {Ricketts}, {Shelden}, {Sandford}, {Siegmund}, {Simmons}, {Smith},
  {Snedden}, {Schneider}, {Strauss}, {SubbaRao}, {Tremonti}, {Waddell}, \&
  {York}}]{smee12a}
{Smee} S. {et~al.}, 2012, ArXiv e-prints

\bibitem[{{Taruya} {et~al}\mbox{.}(2010){Taruya}, {Nishimichi}, \&
  {Saito}}]{taruya10a}
{Taruya} A., {Nishimichi} T., {Saito} S., 2010, \prd, 82, 063522

\bibitem[{{Taruya} {et~al}\mbox{.}(2009){Taruya}, {Nishimichi}, {Saito}, \&
  {Hiramatsu}}]{taruya09a}
{Taruya} A. {et~al.}, 2009, \prd, 80, 123503

\bibitem[{{Taruya} {et~al}\mbox{.}(2011){Taruya}, {Saito}, \&
  {Nishimichi}}]{taruya11a}
{Taruya} A., {Saito} S., {Nishimichi} T., 2011, \prd, 83, 103527

\bibitem[{{Wagner} {et~al}\mbox{.}(2008){Wagner}, {M{\"u}ller}, \&
  {Steinmetz}}]{wagner08a}
{Wagner} C., {M{\"u}ller} V., {Steinmetz} M., 2008, \aap, 487, 63

\bibitem[{{Weinberg} {et~al}\mbox{.}(2012){Weinberg}, {Mortonson},
  {Eisenstein}, {Hirata}, {Riess}, \& {Rozo}}]{weinberg12a}
{Weinberg} D.~H. {et~al.}, 2012, ArXiv e-prints

\bibitem[{{White} {et~al}\mbox{.}(2011){White}, {Blanton}, {Bolton},
  {Schlegel}, {Tinker}, {Berlind}, {da Costa}, {Kazin}, {Lin}, {Maia},
  {McBride}, {Padmanabhan}, {Parejko}, {Percival}, {Prada}, {Ramos}, {Sheldon},
  {de Simoni}, {Skibba}, {Thomas}, {Wake}, {Zehavi}, {Zheng}, {Nichol},
  {Schneider}, {Strauss}, {Weaver}, \& {Weinberg}}]{white10a}
{White} M. {et~al.}, 2011, \apj, 728, 126

\bibitem[{{Xu} {et~al}\mbox{.}(2012{\natexlab{a}}){Xu}, {Cuesta},
  {Padmanabhan}, {Eisenstein}, \& {McBride}}]{xu12b}
{Xu} X. {et~al.}, 2012{\natexlab{a}}, ArXiv e-prints

\bibitem[{{Xu} {et~al}\mbox{.}(2012{\natexlab{b}}){Xu}, {Padmanabhan},
  {Eisenstein}, {Mehta}, \& {Cuesta}}]{xu12a}
{Xu} X. {et~al.}, 2012{\natexlab{b}}, ArXiv e-prints

\bibitem[{{York} {et~al}\mbox{.}(2000){York} {et~al.}}]{york00a}
{York} D.~G., {et~al.}, 2000, \aj, 120, 1579

\bibitem[{{Zaroubi} {et~al}\mbox{.}(1995){Zaroubi}, {Hoffman}, {Fisher}, \&
  {Lahav}}]{zaroubi95a}
{Zaroubi} S. {et~al.}, 1995, \apj, 449, 446

\end{thebibliography}

\appendix
\section{The AP mapping in practice}\label{ap_inpractice_section}
Here we describe the geometrical correction mapping (or AP shifting) 
of $1$D statistics as 
$\xi_{||,\perp},\xi_{0,2}$.

As we compare a $\xi$ template 
to data which are affected by 
geometrical distortions we 
must distinguish between two sets 
of coordinate systems, which are, 
ultimately, related through 
$H$ and $D_{\rm A}$.

In \S\ref{geometry_section} we define the
geometric distortions 
of the components of ${\bf s}$. In the final product, though,  
we use its absolute value and $\mu$ related by: 
\beq\label{smu_defintions_eq}
s\equiv |{\bf s}|=\sqrt{s_{||}^2+s_{\perp}^2}, \ \mu=\frac{s_{||}}{s},
\eeq
where $s_{||}$ is the line-of-sight separation component.  

The template, from which the model is constructed, is calculated in a ``true" 
or ``test" 
coordinate system $s_{\rm t}, \mu_{\rm t}$, 
where the data are in shifted axes based on the fiducial cosmology, 
hence we define its
separations and angles $s_{\rm f}, \mu_{\rm f}$. 
Because we apply the model to the data, the model, 
which is based on the template, should be in the fiducial coordinate system, as well,  
hence the AP shifting of the template to $\xi^{\rm template}(s_{\rm f},\mu_{\rm f})$.  


Using Equations (\ref{strv_shift_eq})-(\ref{alphalos_eq}) along 
with Equations (\ref{smu_defintions_eq}) 
we obtain 
\beq\label{s_shifted_equation}
s_{\rm t} = s_{\rm f} \cdot \sqrt{   \alpha_{||}^2\mu_{\rm f}^2  +  \alpha_{\perp}^2(1-\mu_{\rm f}^2)}, 
\eeq
and 
\beq\label{mu_shifted_equation}
\mu_{\rm t} = \mu_{\rm f}  \frac{\alpha_{||}}{ \sqrt{   \alpha_{||}^2\mu_{\rm f}^2  +  \alpha_{\perp}^2(1-\mu_{\rm f}^2)}}. 
\eeq
After $\xi^{\rm template}(s_{\rm f},\mu_{\rm f})$  is produced 
(see below for details of its construction), we calculate:
\begin{dmath}\label{apshift_wedges_equation}
\xi_{\Delta\mu}^{\rm {\mathcal AP} \ template}(s_{\rm f})= \frac{1}{\Delta\mu}\int_{\mu_{\rm f \ min}}^{\mu_{\rm f \ min }+\Delta\mu} 
\xi^{\rm template}(s_{\rm f},\mu_{\rm f}) d\mu_{\rm f}
\end{dmath}
for the clustering wedges.
For the multipoles we calculate:
\begin{dmath}\label{apshift_multipoles_equation}
\xi_{\ell}^{\rm {\mathcal AP} \ template}(s_{\rm f})= (2\ell+1)\int_{0}^{1} 
\xi^{\rm template}(s_{\rm f},\mu_{\rm f}) \mathcal{L_{\ell}}(\mu_{\rm f})d\mu_{\rm f}.
\end{dmath}

To calculate $\xi^{\rm template}(s_{\rm f},\mu_{\rm f})$ in practice we apply the following steps:
\begin{enumerate}
{\item At every point of the integration we use Equations (\ref{s_shifted_equation})-(\ref{mu_shifted_equation}) 
to convert the fiducial $s_{\rm f},\mu_{\rm f}$ into the template true coordinates $s_{\rm t},\mu_{\rm t}$}. 
{\item We interpolate stored arrays of a pre-calculated $\xi_{0},\xi_2$ templates to the resulting $s_{\rm t}$ value.
For details regarding the templates used see \S\ref{xitemplates_section}.}
{\item We calculate $\xi(s_{\rm f},\mu_{\rm f})$ by interpolation of $\xi\left(s_{\rm t}\left(s_{\rm f},\alpha_{||},\alpha_{\perp},\mu_{\rm f}\right),\mu_{\rm t}\left(\alpha_{||},\alpha_{\perp},\mu_{\rm f}\right)\right)$=$\xi_0(s_{\rm t})+{\mathcal L_2(\mu_{\rm t})}\xi_2(s_{\rm t})$}
\end{enumerate}

Note that to calculate $\xi_{2}^{\rm {\mathcal AP} \ template}$ (Equation \ref{apshift_multipoles_equation})  
we need to calculate $\mathcal{L}_2(\mu_{\rm f})$,  
where for the $\xi_{\Delta\mu}$ (Equation \ref{apshift_wedges_equation}) 
this is not needed. 
We test our algorithm by applying it on mock catalogues 
in which we assume an incorrect fiducial cosmology,  
and apply the above algorithm and obtain the true $1/H$ and $D_{\rm A}$ values. 

In this method we make two main assumptions. 
First, 
the AP shifting is based on 
a template that consists of multipoles $\ell=0,2$. 
This template can be easily expanded to higher orders of $\ell$, 
although at scales of interest $\ell\geq 4$ components 
should be fairly weak. 
Second, we assume the plane parallel approximation for each pair. 
\cite{wagner08a} show that light-coning yields minimal 
effects at $z=1,3$, as do \cite{kazin11a} at $z=0.35$. 

\section{Linear vs. Non-linear AP effect}\label{linear_ap_appendix}
\begin{figure*}
\begin{center}
\includegraphics[width=0.49\textwidth]{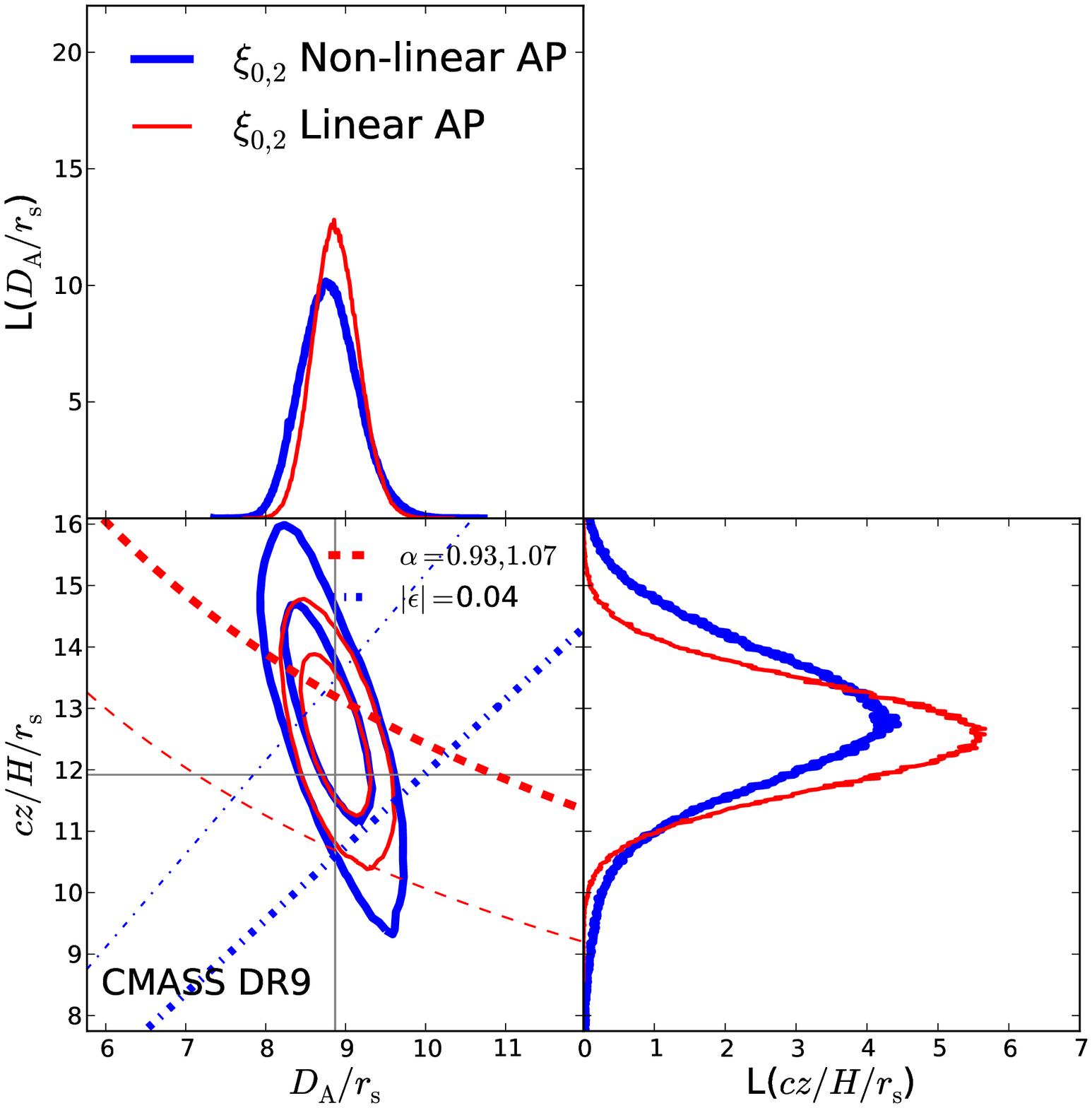}
\includegraphics[width=0.49\textwidth]{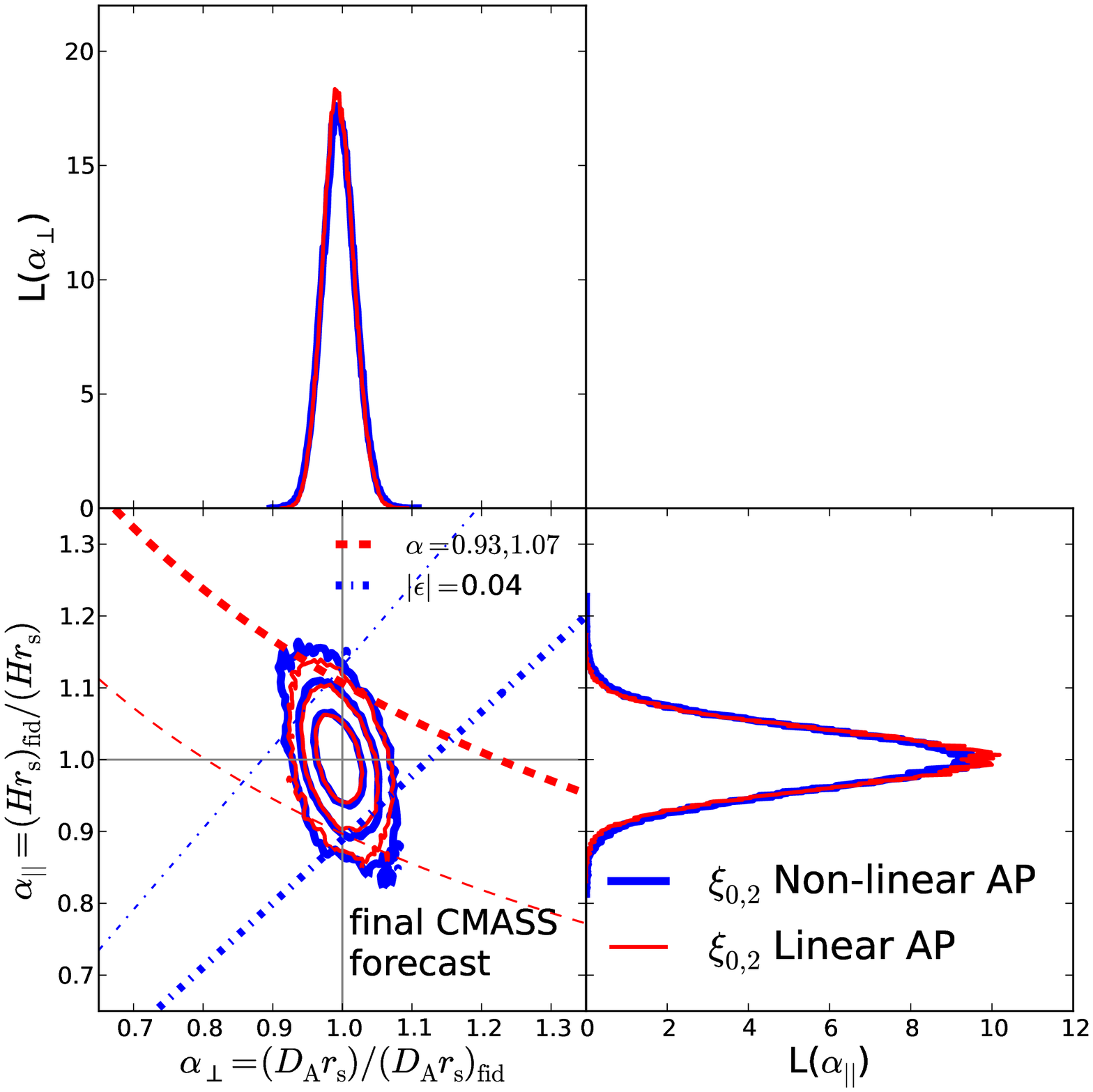}
\caption{
We use the RPT-based multipoles pre-reconstruction to test the linear (thin red)
AP correction against the non-linear (thick blue) 
in constraining \czHzrs  
and \Dazrsii. 
The left plot shows results for CMASS DR9 investigated here 
(contours are $68,95\%$ CL regions), 
and the right plot for projections of the final BOSS CMASS footprint (contours are $68,95,99.7\%$ CL regions). 
(As mentiond in \S\ref{dr12_forecast_section}, this forecast should be considered 
overestimated constraints as the $C_{ij}$ used is noiser  
than that expected of a the full CMASS volume.). 
To guide the eye we plot the regions of constant $\alpha$ and 
$\epsilon$, as indicated in the legend 
(where the thicker line of each indicates the larger value). 
}
\label{linearAP_figure}  
\end{center}
\end{figure*}
Throughout this analysis we apply 
the non-linear AP correction 
as described in Appendix \ref{ap_inpractice_section}. 
In this section we investigate differences with the 
linear AP effect as used in \cite{xu12a}. 
This linear approach was introduced in \cite{padmanabhan08a} 
in the $P(k)$ formulation, 
and analyzed in $\xi$ in \cite{kazin11a}. 
However, as pointed out by \cite{padmanabhan08a}, 
this linear approach breaks down when $|\epsilon|>2\%$, 
which is clearly the case in the DR9-CMASS for a large 
part of the $95\%$ CL region. 

The linear AP correction, when applied on the 
clustering multipoles, is as follows:

\begin{eqnarray}
\label{mono_equation}
\xi_0(s_{\rm t})&=&\xi_0(\alpha s_{\rm f})+\nonumber\\ 
&&\epsilon \left( \frac{2}{5}\frac{d\xi_2(x)}{d\ln(x)}\bigg|_{x=\alpha s_{\rm f}}+\frac{6}{5} \xi_2(\alpha s_{\rm f})\right), \\
\label{quad_equation}
\xi_2(s_{\rm t})&=&\left( 1+\frac{6}{7}\epsilon \right)\xi_2(\alpha s_{\rm f}) + \frac{4}{7}\epsilon \frac{d\xi_2(x)}{d\ln(x)}\bigg|_{x=\alpha s_{\rm f}}+\nonumber\\ 
&&2\epsilon\frac{d\xi_0(x)}{d\ln(x)}\bigg|_{x=\alpha s_{\rm f}}. 
\end{eqnarray}
Here we neglect terms of order ${\mathcal O}(\epsilon^2)$, 
as well as $\xi_4$ terms. 
(For a discussion of higher order terms see \S2.2.4 in \citealt{kirkby13a}.)

The left plot of Figure \ref{linearAP_figure} shows  
the results obtained when applying the 
non-linear AP (thick blue) 
and the linear correction (thin red) as 
to the CMASS-DR9 $\xi$.  
The dotted and dashed lines convey constant 
values of $\alpha$ and $\epsilon$, respectively. 

The results clearly show that the 
linear correction under-estimates the uncertainties of   
\czHzrs and \Dazrs by 
$\sigma^{\rm linear}_{H}/\sigma^{\rm non-linear}_{H}=7.2/9.6$ and 
$\sigma^{\rm linear}_{D_{\rm A}}/\sigma^{\rm non-linear}_{D_{\rm A}}=3.2/3.9$,  
where $\sigma^{\rm method}_{\rm X}$ is the 68CLr of X=$H, D_{\rm A}$. 
The method results agree fairly well where 
$\epsilon$ is small (and regardless of $\alpha$), 
but differ as $\epsilon$ grows.
These differences should vary with the choice of the fiducial model, 
as well as the volume investigated. 

We apply a similar comparison for a mock-mean signal 
(of 600 mocks) with the $C_{ij}$ divided by three 
(as in \S\ref{dr12_forecast_section}) and plot the results 
in the right of Figure \ref{linearAP_figure}. 
In this higher S/N test we clearly see that 
the two methods agree with each other extremely  well, 
due to the fact that $\epsilon$ is  low. 
There is a slight under-estimation of the linear approximation 
at the $95\%$ CL region. 
Note that here we test the 
case where the fiducial $H$ and $D_{\rm A}$ 
correspond to the mock true values ($\epsilon=0$, $\alpha=1$), 
whereas 
if 
we would 
apply a geometric distoriton of 
$|\epsilon|>2\%$ 
we should expect larger differences.  

In conclusion,  
the non-linear AP correction should be applied 
to avoid potential estimation biases. 

\section{Testing the algorithm on high S/N mocks}\label{stackmock_section}
We test our methodology by applying it on 
a set of 100 mocks 
with higher S/N 
than those used in the final mock DR9 analysis. 
The motivation for this procedure is to separate between potential 
systematics and effects due to weak \baf signals. 

The higher S/N mocks, called ``stacked-mocks", 
are built by stacking the 600 PTHalo DR9-volume mocks 
by groups of six, providing us with one hundred realizations. 
For purposes of this analysis we divide the 
DR9 $C_{ij}$ (see \S\ref{cij_section}) by a factor of six.

\begin{figure*}
\begin{center}
\includegraphics[width=0.49\textwidth]{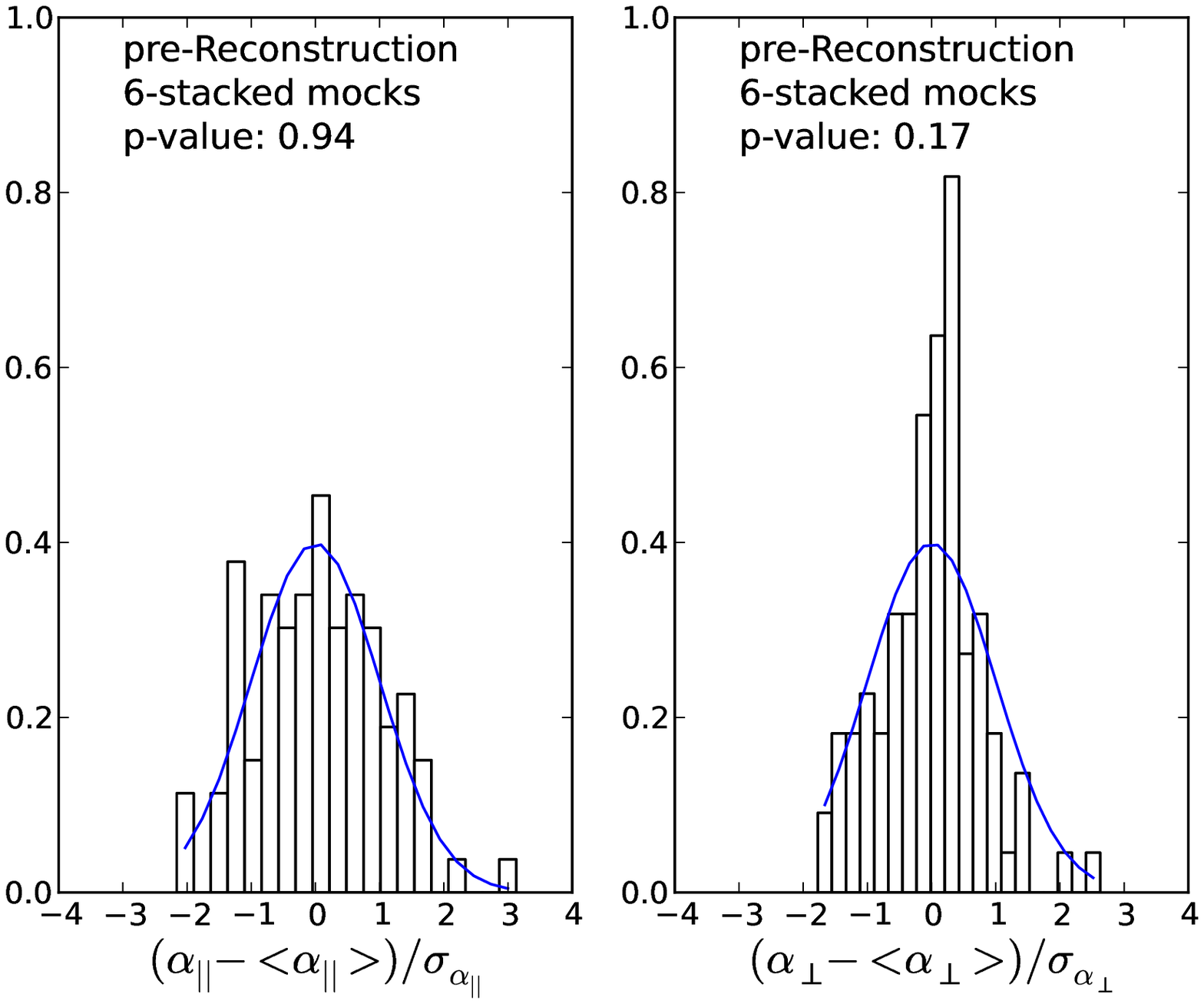}
\includegraphics[width=0.49\textwidth]{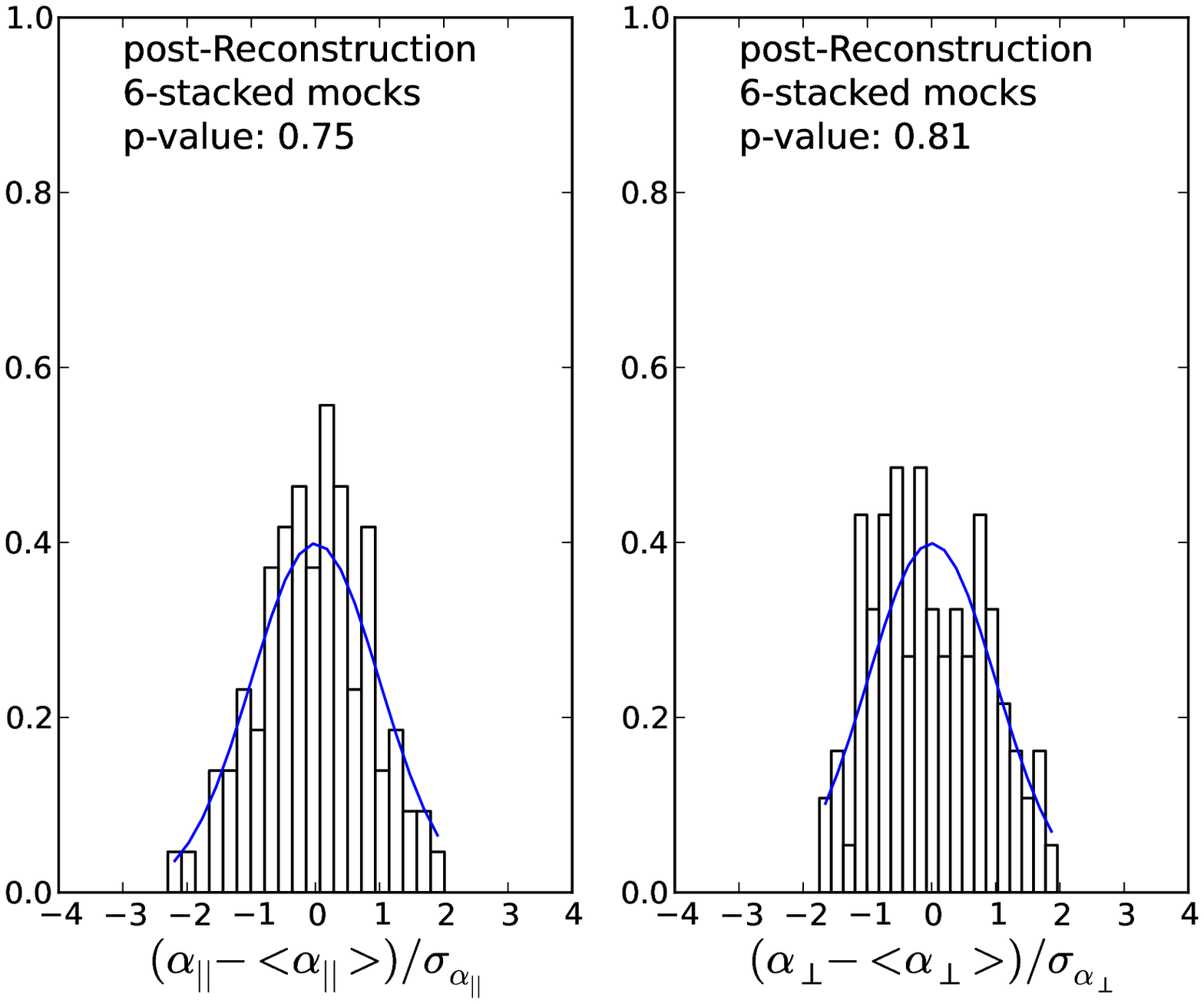}
\includegraphics[width=0.49\textwidth]{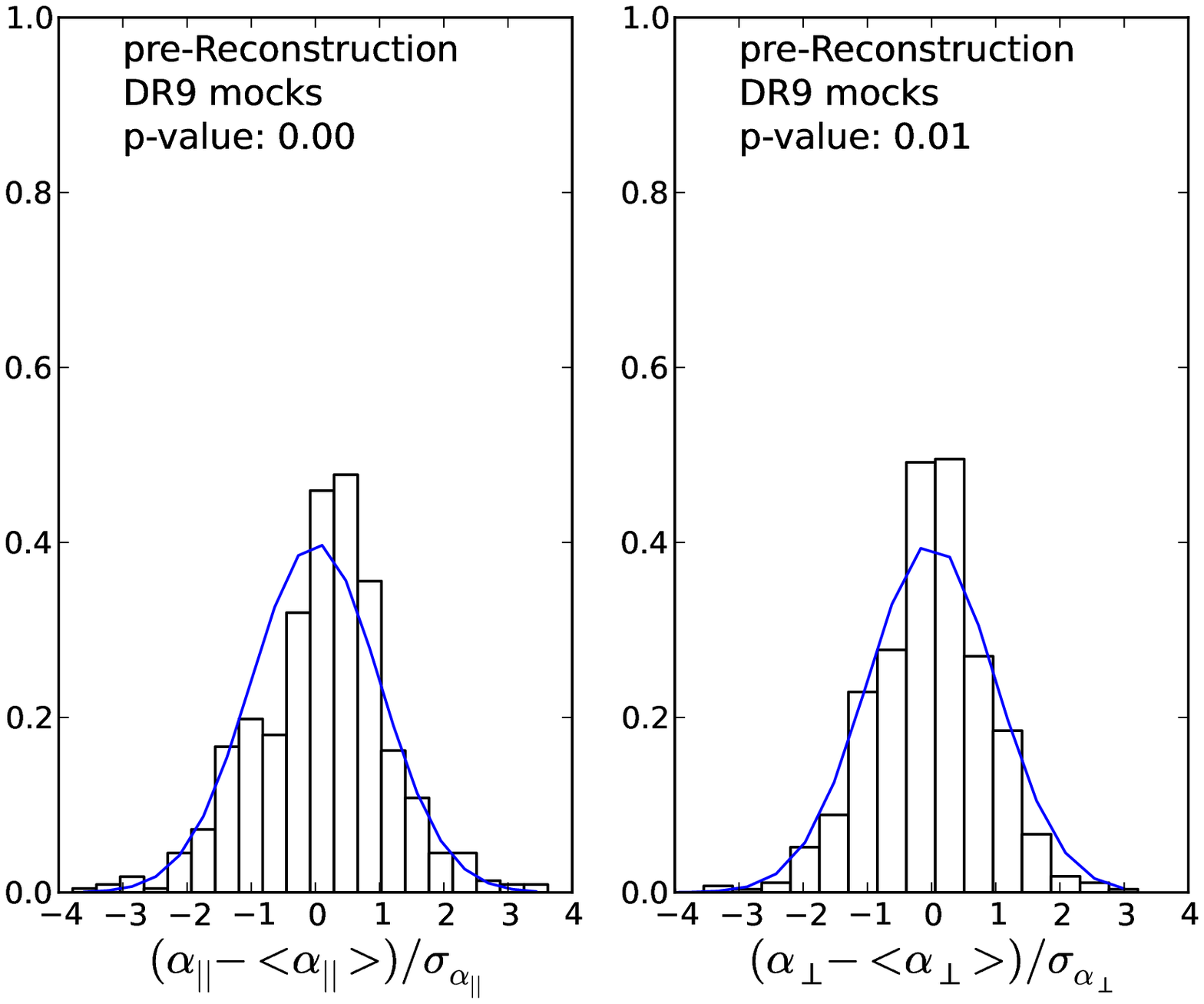}
\includegraphics[width=0.49\textwidth]{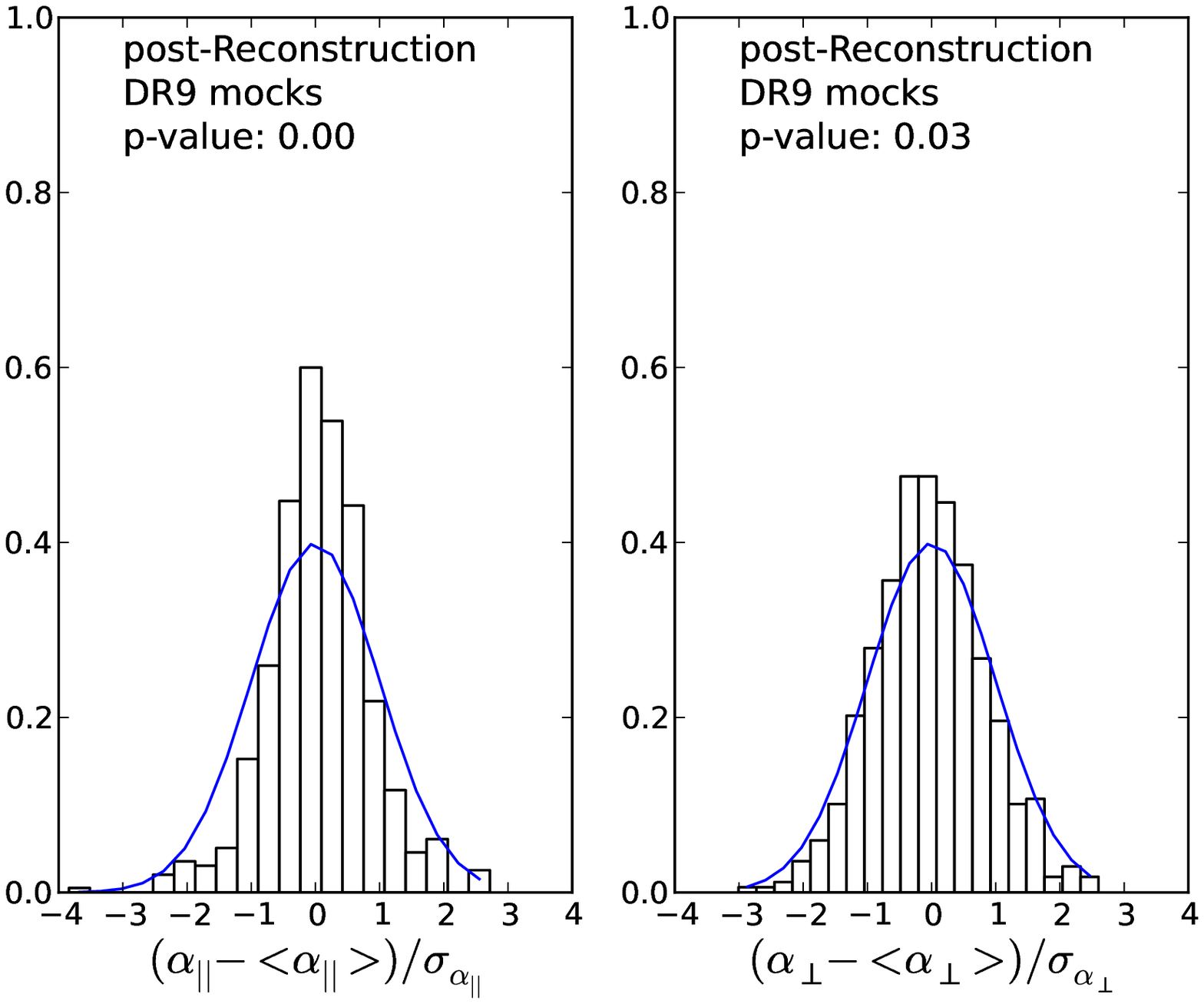}
\caption{
($\alpha_{||}$-$\avg{\alpha_{||}}$)/$\sigma_{\alpha_{||}}$ results 
(and similar for $\alpha_{\perp}$) 
of 100 6-stacked-mocks (Top) 
and 600 DR9 mocks (Bottom) pre- (Left) and post-reconstruction (Right). 
Results are for RPT-based clustering wedges. 
The p-values reflect K-S tests when comparing 
to a Gaussian distribution (blue lines). 
The p-values vary by template (RPT-based, dewiggled), 
and $\xi$ statistic (clustering wedges, multipoles) used. 
The stacked-mocks yield p-values between $20-95\%$, 
where the DR9 mocks results have negligible p-values. 
}
\label{ks_plots}  
\end{center}
\end{figure*}

Figure \ref{ks_plots} shows distributions of 
($\alpha_{||}$-$\avg{\alpha_{||}}$)/$\sigma_{\alpha_{||}}$
and 
($\alpha_{\perp}$-$\avg{\alpha_{\perp}}$)/$\sigma_{\alpha_{\perp}}$
for the stacked mocks (top) 
and the DR9 mocks (bottom) 
both pre- (left) and post-reconstruction (right).
The quoted p-values are obtained when performing 
the standard Kolmogorov-Smirnov test between the distributions 
and a Gaussian one. 

We find that the stacked mock 
results yield various Gaussian (or symmetric) 
attributes not found in the DR9 mock results. 
First, in the stacked mocks the 
means of the MCMC propositions are similar 
to the mode values, the standard deviations 
of the MCMC propositions are similar to the 68CLr 
and they yield low skewness values 
of the marginalized 1D likelihood distributions.
As discussed in \S\ref{testingmethodology_section}
in the DR9-volume mocks we find large skewness causing differences 
in these statistics. Using the DR9 mocks, we find in the 
that the modes and 68CLr are more reliable, 
as they are better defined. 

One of the most important Gaussian-like features 
found in the stacked-mock \HzfidHzii, \DazDazfid results  
is that the scatter in the modes 
is similar to the mean of the uncertainties. 
This is not the case for the DR9-volume mocks, 
probably due to weak anisotropic \baf detections. 

Finally, 
the stacked mock results (modes and uncertainties) 
are similar to those yielded 
when applying the same $C_{ij}^{-1}$ 
on the mock-mean signal (i.e., the mean signal of all 600 mocks). 
We find this to be true for all 
eight combinations investigated: 
clustering wedges, multipoles; RPT-based, dewiggled templates;  
pre-, post-reconstruction. 
All results are presented in Table \ref{hda_highsn_table}. 


\begin{table*} 
\begin{minipage}{172mm}
\caption{High S/N  (6-stacked) mock results}
\label{hda_highsn_table}
\begin{tabular}{@{}ccccc@{}}
\hline
\hline
$\xi$ ($\#$ of realizations) & $\alpha_{||}$ & $\Delta\alpha_{||}/\alpha_{||}$ &$\alpha_{\perp}$ &  $\Delta\alpha_{\perp}/\alpha_{\perp}$ \\
\hline
RPT-based wedges pre-rec (100)&$ 1.002\pm 0.033$&$ 0.031\pm 0.006$&$ 0.996\pm 0.013$&$ 0.017\pm 0.002$  \\
RPT-based multipoles pre-rec (100)&$ 1.002\pm 0.033$&$ 0.030\pm 0.004$&$ 0.994\pm 0.013$&$ 0.016\pm 0.001$  \\
RPT-based wedges post-Rec (100)&$ 1.005\pm 0.018$&$ 0.021\pm 0.003$&$ 0.996\pm 0.010$&$ 0.012\pm 0.002$  \\
RPT-based multipoles post-rec (100)&$ 1.003\pm 0.016$&$ 0.022\pm 0.002$&$ 0.997\pm 0.010$&$ 0.012\pm 0.001$  \\

\\  dewiggled wedges pre-rec (100)&$ 1.014\pm 0.034$&$ 0.032\pm 0.006$&$ 1.003\pm 0.014$&$ 0.017\pm 0.002$  \\
dewiggled multipoles pre-rec (100)&$ 1.009\pm 0.032$&$ 0.029\pm 0.004$&$ 1.003\pm 0.013$&$ 0.016\pm 0.001$  \\
dewiggled wedges post-rec (100)&$ 1.008\pm 0.019$&$ 0.020\pm 0.003$&$ 1.000\pm 0.011$&$ 0.012\pm 0.002$  \\
dewiggled multipoles post-rec (100)&$ 1.007\pm 0.014$&$ 0.017\pm 0.001$&$ 1.001\pm 0.009$&$ 0.010\pm 0.001$  \\
\hline
\hline
\end{tabular}

\medskip
 * The $\alpha_{||}$ and $\alpha_{\perp}$ columns show the median and rms of the modes. \\
 * The $\Delta \alpha_{||}/\alpha_{||}$ and $\Delta \alpha_{\perp}$/$\alpha_{\perp}$ columns show the median and rms of the fractional uncertainties.
\end{minipage}
\end{table*}

\subsection{RPT-based vs. dewiggled templates}\label{rptdiwiggled_section}

As for preference of template (RPT-based vs. dewiggled) 
for constraining \HzfidHz and \DazDazfidii, 
when using the stacked mocks we find strong 
cross correlation coefficients of $r\sim 0.9-1$ 
in both modes and uncertainties. 
This comparison shows no difference in uncertainties. 
The only oddity we find 
is that the dewiggled pre-reconstruction 
wedges and multipoles yield 
median (mean) biases of $1.4,0.9\%$ ($0.9,1.0\%$) 
in \HzfidHz modes, respectively, which is reduced 
post-reconstruction to $0.8,0.7\%$ ($0.7\%$). 
These \HzfidHz biases, 
when using the dewiggled model, 
do not appear when applied to the DR9-mocks. 
In those mocks, we find that the 
pre-reconstruction dewiggled model yields 
a bias of $\sim 1\%$ on determining \DazDazfidii. 


In all four RPT-based cases 
(wedges, multipoles; pre-, post-reconstruction) 
the mean biases 
of \HzfidHz \DazDazfid 
are $\leq 0.5\%$. 
\cite{sanchez08} thoroughly analyze differences 
between RPT-based and dewiggled $\xi_0$ and report  
that, when using the latter, 
one should expect systematic shifts in $\alpha$ 
due to the lack of a $k$-mode coupling term. 
In \S\ref{postrec_templates_section} we demonstrate that 
the post-reconstruction mocks do not prefer a template with 
$A_{\rm MC}=0$, and hence suggest templates require a mode coupling term. 

For all the reasons above our choice of preference is the 
RPT-based template. 

\subsection{Clustering wedges vs. multipoles}\label{wedgesmultipoles_section}
The stacked mocks show  
no significant difference regarding 
the constraining power of $\xi_{||,\perp}$ and $\xi_{0,2}$ 
on 
\HzfidHz or \DazDazfidii; 
post-reconstruction RPT-based yields sub $0.1\%$ differences. 
The cross correlation between the uncertainties of 
\HzfidHz are found to be $r\sim 0.6, \ 0.7$ (dewiggled, RPT-based), 
and $0.88, \ 0.83$ for \DazDazfidii. 
The pre-reconstruction templates yield similar 
results. 

We then ask if  multipoles and wedges  yield 
similar mode results. The post-reconstruction stacked mocks 
indicate $r\sim 0.80$ for \HzfidHz and $r\sim 0.85$ for \DazDazfid
in both RPT-based and dewiggled templates. 
Pre-reconstruction results yield similar correlations. 

For a visual of the results of the 3-stacked mocks, 
please refer to the bottom plot 
of Figure \ref{dr12forcast_figure}, 
which is described in \S\ref{dr12_forecast_section}.

\subsection{Improvement due to reconstruction}\label{recimprove_section}
According to the stacked mock $\xi_{||,\perp}$ (and hence also $\xi_{0,2}$), 
we find the 
uncertainty of \HzfidHz 
improves 
by $32\%$ and that for 
\DazDazfid by $30\%$. 

The stacked mocks show that the  
\HzfidHz modes should have a moderate 
correlation of $r\sim 0.5-0.55$ and 
\DazDazfid of $0.5-0.6$. 
For a visual of results from the 3-stacked mocks, 
please refer to the top plot 
of Figure \ref{dr12forcast_figure}, 
which is described in \S\ref{dr12_forecast_section}.

Another value of interest is the 
cross-correlation between \HzfidHz and \DazDazfidii. 
With the stacked mocks we find this correlation to be of 
order $r\sim -0.55$ pre-reconstruction and 
$r\sim -0.35$ post-reconstruction. 
Also we find no correlation between 
$\alpha$ and $\epsilon$ modes, as expected.


\end{document}